\documentclass[12pt]{article}

\newtheorem{Proposition}{Proposition}

\newcommand{\imag}{\mathrm{i}\,}

\title{Unified treatment and classification of  superintegrable
systems with integrals quadratic in momenta
on a two dimensional manifold
\footnote{Extended version of the paper published in J. Math. Phys. \textbf{47}, 042904 (2006)} }

\author{C. Daskaloyannis\thanks{e:mail address: daskalo@math.auth.gr}\\
{\it Mathematics Department}\\
{and}\\
{K. Ypsilantis}\\
                  {\it  Physics Department,}\\
        {\it Aristotle University of Thessaloniki,}\\
                {\it  54124 Thessaloniki, Greece}
        }

\date{May 2006}

\begin{document}

 \maketitle

\newpage

\begin{abstract}
In this paper we prove that the two dimensional superintegrable
systems with quadratic integrals of motion on a manifold can be
classified by using the Poisson algebra of the integrals of
motion. There are six general fundamental classes of
superintegrable systems. Analytic formulas for the involved
integrals are
 calculated in all the cases. All the known
superintegrable systems are  classified as special cases of these
six general  classes.
\end{abstract}

\vfill

Running title: Classification of two dimensional superintegrable
systems

PACS Numbers: 03.65.Fd; 02.10.Tq; 45.20.Jj;

\newpage

\section{Introduction}\label{sec:Alg}

In classical mechanics, an integrable system on a manifold of $N$
dimensions, is a system which has $N$ integrals of motion in
evolution. Superintegrable  (or maximally integrable) system is a
system possessing the maximum number of constants of motion, i.e.
$2N-1$ integrals of motion.

 The simplest case is the two
dimensional superintegrable problems with integrals of motion,
which are linear and quadratic functions of momenta. The
investigation of such superintegrable systems on a two dimensional
manifold is a quite old problem, dated on 19th century. Initially
the  problem was formulated as a geometry problem. The challenge
was to find two dimensional manifolds whose the geodesics are
curves which possess additional integrals than the free
Hamiltonian. This problem was studied in the four volume treatise
of Darboux  \textit{Le\c{c}ons sur la Th\'{e}orie G\'{e}n\'{e}rale
des Surfaces} \cite{Darboux}. The main result of this study is
that, there are five classes of general forms of metrics,  whose
the geodesics have three integrals of motion (the Hamiltonian and
two additional functionally independent integrals). These metrics
are called "formes essentielles" and they depend on four
parameters. All the metrics having more than two integrals of
motion can be obtained as partial cases of these "formes essentielles"
by choosing appropriate values of the four parameters. The five classes of metrics
are tabulated in "Tableau VII" by Koenigs\cite[vol IV  p.385]{Darboux}

In Classical Mechanics language, Darboux and Koenigs results can
be translated as searching   manifolds, where the free Hamiltonian
accepts quadratic integrals of motion. The evolution of this
problem  is to find superintegrable systems, whose  Hamiltonian is
the free Hamiltonian plus a potential and these systems possess
additional quadratic integrals of motion.

  The simplest integrable and
superintegrable systems are the systems defined on  the real plane.
A comprehensive review of the real two-dimensional integrable
classical systems on the plane is given in Ref
\cite{Fris,Hietarinta87}. The complex superintegrable systems with
quadratic moments on a flat space were  recently catalogued by
Kalnins, Miller, Pogosyan et al
\cite{KaMiPogo96,KaMiPogo00,KaKrPogo01,KaMiPogo02_PAN}. The flat
space is a two dimensional manifold with curvature zero. The Drach
potentials are also systems defined on a manifold with curvature
zero. The Drach systems with quadratic integrals of motion were
investigated by Ra\~{n}ada \cite{Ranada97}. The superintegrable
systems on the hyperbolic plane were studied in
\cite{RaSan99,RaSan02}.  These systems were studied separately,
while they are connected by obvious coordinate transformations.
Therefore they are naturally connected, i.e. there is a common
classification scheme of all these systems \cite{RaSan99,RaSan02}.

The case of non flat space is under current intensive
investigation.  The superintegrable systems on the sphere were
studied in \cite{Ranada97,RaSan99,KaMiPogo00b} and they were
classified in \cite{KaKrPogo01}. The sphere is a special case of
manifold with constant curvature. In refs \cite{RaSan99,RaSan02}
the superintegrable systems on the sphere and on the hyperbolic
plane were studied using an unified formulation.

In the case of manifolds of non-constant curvature, the known
examples of superintegrable systems are those which are defined on
manifolds which are surfaces of revolution.  The corresponding
problem of the geodesics with three quadratic integrals of motion
on surfaces of revolution was treated by Koenigs in \cite[n${ }^o$
5, vol IV, p. 377]{Darboux}. Recently Kalnins and collaborators
classified the superintegrable systems on a surface of revolution
\cite{KaKrWin02,KaKrMiWin03} using the manifolds which were
provided by Koenigs.

In two recent papers  Kalnins, Kress and Miller \cite{KaKrMi05a,KaKrMi05b}
give a comprehensive study of the two-dimensional superintegrable sytems.
In \cite{KaKrMi05b}, they prove that the general Koenigs essential forms of
metrics correspond to the most general forms of superintegrable systems.
Also they have shown that every two-dimensional superintegrable system
is St\"{a}ckel equivalent to a two-dimensional non degenerate superintegrable
system on a constant curvature space.

Kress \cite{Kress} in collaboration with Kalnins and Miller studied
the St\"{a}ckel equivalence classes of the superintegrable systems
on the spaces of constant curvature, and they have shown that there
are six equivalence classes. The general classes of Koenigs
classification given by the Table VII in \cite{Darboux} are five.
That means that there is a sixth class which should be added in the
Koenigs classification scheme. In this paper we investigate this
sixth class, which completes the Koenigs classification. This class
is the nondegenerate superintegrable system generated by the case
VI${}_6$ of Koenigs.

An interesting  question is, whether there could be
 a general classification scheme of superintegrable systems with
quadratic integrals of motion, which contains all the equivalence classes
of superintegrable systems on a manifold with constant curvature and the
general classes of manifolds , which were introduced by Koenigs.
The  classification schemes are
based on the Darboux relations derived for the invariants, which
are defined on a specific manifold.
 In this paper we propose a
classification scheme based on the properties of the Poisson algebra
of the integrals of motion. Then we show that there is indeed  such
a classification scheme, which determines the supporting manifold
metric. We must notice that the Kress \cite{Kress} equivalence
classes are derived by classifying  the Poisson algebra of integrals
of motion. The proposed classes of superintegrable systems in this
paper correspond to the equivalent classes studied in ref
\cite{Kress}. Analytic formulas for the metrics of the  permitted
manifolds,  the potentials and the integrals of motion are
calculated.

This paper is organized as follows: In section
\ref{sec:Integrable} the general form of the integrable two
dimensional system with one
 quadratic integral of motion is derived.
  The results of this section  correspond to the Darboux treatise
  paragraphs \cite[n${ }^o$ 593, vol. III,
p.30]{Darboux}, \cite[n${ }^o$ 593, vol. III, p.31]{Darboux}, but
we give a brief modern derivation of the formulas including the
potentials in our discussion.  These formulas will be
  used in the following sections.  The carrying manifold is a Liouville or
 a Lie surface. Therefore there are two classes of integrable systems.
In a specific coordinate system, which is called Liouville (or
Lie) system, the analytic expressions
  of the potential and the integrals of motion are given and the action is calculated.
In Section \ref{prop:Quadrquadr} the Poisson algebra of the
integrals of a superintegrable two dimensional system is
discussed. This algebra is a quadratic algebra, the coefficients
of the quadratic terms are characteristic of the carrying
manifold. In Section \ref{sec:PoissonSuper} we prove that the
coefficients of the Poisson algebra impose the classification of
the superintegrable systems with two quadratic integrals in six
fundamental classes. The method of analytic calculation of form of
the permitted carrying manifolds, the potentials and the integrals
of motion are discussed. In this section we prove that the
general form of the
superintegrable potential   can be written as a fraction $V=w(x,y)/g(x,y)$ and
the two functions $w(x,y)$ and the metric $g(x,y)$ are two solutions of the same
partial differential equation.
 The existence of the Poisson algebra was
assumed as obvious by several authors
\cite{KaMiPogo96}--\cite{KaMiPogo02_PAN},
\cite{Vinet}--\cite{GLZ-R3}. In Appendix \ref{sec:Poly} we give a
proof of the existence of the Poisson algebra, for two dimensional
superintegrable systems with quadratic integrals of motion.
 In
Section \ref{sec:Classification} the analytic formulas of the
manifolds and integrals are given for all the six fundamental
classes of superintegrable systems. From these analytic formulas
we can show that there are new superintegrable systems, because
they are defined on manifolds which have not constant curvature
and are not surfaces of revolution.
In Section \ref{sec:Koenigs}  the superintegrable systems corresponding
to the Koenigs essential forms of Table VII are given.
In Section \ref{sec:RevolQQ}
the superintegrable systems on a surface of revolution are
studied. We find that there is a new system which was not revealed
by the other classification schemes. In Section \ref{sec:QQzero}
the superintegrable systems on a manifold with curvature zero are
studied and in Section \ref{sec:ConstantCurvQQ} the systems on a
manifold with constant curvature are listed. In Section
\ref{sec:LinearQuadratic} the systems with a linear and a
quadratic integral of motion are discussed.   Finally,  in Section
\ref{sec:Discussion} the results of the paper are summarized.

\section{Integrable systems on a two dimensional manifold}\label{sec:Integrable}
Let us consider an  integrable system defined on a two dimensional
mani\-fold with metric:
\[
ds^2= E(u,v) du^2 +2 F(u,v) du dv +G(u,v) dv^2
\]
There is a conformal coordinate system where the metric can be
written as:
\begin{equation}
ds^2=g(x,y) dx dy \label{eq:metric}
\end{equation}
The passage from the original coordinate system $(u,\,v)$ to the
conformal one $(x,y)$ can be realized by using the Beltrami
partial differential equation. We must notice that the choice of
the conformal coordinate system
 is not unique, i.e. there are several conformal coordinate systems for a given
 metric, these systems are conformally equivalent.

In a conformal  coordinate system the general form of the Hamiltonian is:
\begin{equation}
H= \frac{p_x \,p_y}{g(x,y)} +V(x,y)\label{eq:ClassHamiltonian}
\end{equation}
where the Hamiltonian is a quadratic form of the momenta.

 Let us
consider an integral of motion which is quadratic in momenta. The
most general form  can be written as:
\begin{equation}
I= A(x,y) p_x^2 + B(x,y) p_y^2 - 2 p_x  p_y
\frac{\beta(x,y)}{g(x,y)}+Q(x,y) \label{eq:IntegralOfMotion}
\end{equation}
By definition the Poisson bracket between the Hamiltonian and the
integral of motion is zero:
\begin{equation}
\label{eq:PB}
 \left\{ I,\,  H\right\}_P=\frac{\partial
I}{\partial x}\frac{\partial H}{\partial p_x} -\frac{\partial
I}{\partial p_x}\frac{\partial H}{\partial x}+ \frac{\partial
I}{\partial y}\frac{\partial H}{\partial p_y}-\frac{\partial
I}{\partial p_y}\frac{\partial H}{\partial y}=0
\end{equation}
The above equality implies  restrictions on the functions involved
in equations (\ref{eq:ClassHamiltonian}) and
(\ref{eq:IntegralOfMotion}).
The left hand side of the above equation is an odd function
of cubic order in the momenta.  The coefficients of $p^3_x$ and $p^3_y$ must be
zero:
\begin{equation}
\label{eq:AB}
\begin{array}{lcl}
\displaystyle \frac{\partial A}{\partial y}=0 & \Rightarrow & A=A(x)\\
\\
\displaystyle \frac{\partial B}{\partial x}=0& \Rightarrow &
B=B(y)
\end{array}
\end{equation}
The coefficients of $p^2_x p_y$ and $p^2_y p_x$ in (\ref{eq:PB}) must be indeed
zero:
\begin{eqnarray}
\label{eq:beta}\displaystyle \frac{\partial \beta}{\partial y}
 =  \displaystyle  A(x) \frac{\partial g}{\partial x}+
 \frac{g}{2}A'(x)=a(x) \frac{ \partial \left( a(x) g(x,y)\right)}{\partial x}\\[0.1in]
\label{eq:beta_1}
\displaystyle  \frac{\partial \beta}{\partial x}
 =\displaystyle B(y)\frac{\partial g}{\partial y}
 +\frac{g}{2}B'(y)=b(y) \frac{ \partial \left( b(y) g(x,y) \right)}{\partial y}
\end{eqnarray}
where
\[
A(x)=a^2(x), \quad B(y)=b^2(y)
\]
The partial $x$-derivative of the right hand side in
(\ref{eq:beta}) is equal to the $y$-derivative of the right hand
side in equation (\ref{eq:beta_1}),
 therefore:
\begin{equation}
\label{eq:Diff_beta1} (A''(x)-B''(y))g(x,y) +3 A'(x)\frac{\partial
g}{\partial x} - 3 B'(y)\frac{\partial g}{\partial y}+ 2
A(x)\frac{\partial^2 g}{\partial x^2}-2 B(y) \frac{\partial^2
g}{\partial y^2}=0
\end{equation}
or
\begin{equation}
\label{eq:Diff_beta2} \frac{\partial }{\partial x} \left(
a(x)\frac{\partial}{\partial x}\left( a(x) g(x,y)\right) \right)=
\frac{\partial }{\partial y} \left( b(y)\frac{\partial}{\partial
y}\left( b(y) g(x,y)\right) \right)
\end{equation}

 The coefficients of $p_x$ and $p_y$ in (\ref{eq:PB}) must be zero:
\begin{equation}
\label{eq:Coeff_px_py}
\begin{array}{l}
\displaystyle \frac{\partial Q}{\partial y}=2 A(x)\frac{\partial
V}{\partial x}+2 \beta(x,y) \frac{\partial V}{\partial y} \\
\\
\displaystyle \frac{\partial Q}{\partial x}=2 B(y)\frac{\partial
V}{\partial y}+2 \beta(x,y) \frac{\partial V}{\partial y}
\end{array}
\end{equation}
The above relations imply:
\begin{equation}
\begin{array}{r}
g(x,y)\left(2 A(x)\frac{\partial^2 V}{\partial x^2}-2 B(y)
\frac{\partial^2 V}{\partial y^2}+3 A'(x)\frac{\partial
V}{\partial x} - 3 B'(y)\frac{\partial V}{\partial y}\right) +\\
+4\left( A(x)\frac{\partial g}{\partial x}\frac{\partial
V}{\partial x}- B(y) \frac{\partial g}{\partial y}\frac{\partial
V}{\partial y}\right)=0 \label{eq:DiffQ}
\end{array}
\end{equation}
At this point we must to distinguish two cases. In the first case
$A(x)$ and $B(y)$ are both different from zero, whereas in the
second case $B(y)$ is assumed to be zero.

\noindent \underline{\textbf{Class I:  $A(x) B(y) \ne 0$}}\\
Following the method given in Koenigs original paper, we can choose
a new coordinate system
\[
\xi= \int \frac{dx}{\sqrt{A(x)}} \quad \mbox{and} \quad \eta=\int
\frac{dy}{\sqrt{B(y)}}
\]
where the associated momenta are:
\[
p_\xi= \sqrt{A(x)}\, p_x, \quad \mbox{and} \quad p_\eta=
\sqrt{B(y)}\, p_y
\]
 In this case
the metric is written
\[
ds^2= \widehat{g}(\xi,\eta) \, d\xi\, d\eta, \quad \mbox{where}
\quad \widehat{g}(\xi,\eta)= g(x,y)\,\sqrt{A(x)}\,\sqrt{B(y)}
\]
In these new coordinate system $(\xi,\,\eta)$ all the above
equations are considerably simplified. We can easily show that the
formulas are the same by replacing $x\rightarrow\xi,\;y\rightarrow
\eta$ and  fixing $A(x)=1$ and $B(y)=1$. For simplicity reasons we
omit the hat on the metric $\widehat{g}(\xi,\eta)\rightarrow
{g}(\xi,\eta)$. Equations (\ref{eq:ClassHamiltonian}) and
(\ref{eq:IntegralOfMotion}) are written:
\begin{equation}
\begin{array}{c}
\displaystyle  H= \frac{p_\xi p_\eta}{g(\xi,\eta)}+ V(\xi,\eta)\\
\displaystyle I= p_\xi^2+ p_\eta^2 - 2 p_\xi  p_\eta
\frac{\beta(\xi,\eta)}{g(\xi,\eta)}+Q(\xi,\eta)
\end{array}
\label{eq:Aeq1Beq1}
\end{equation}
We call these specific coordinates $(\xi,\,\eta)$
\underline{Liouville coordinates}. In Liouville coordinates the
Hamiltonian $H$ \underline{and} the integral $I$ are written as it
was given in the above equation (\ref{eq:Aeq1Beq1}).

 In the Liouville coordinates,  equation
(\ref{eq:Diff_beta2}) is considerably simplified:
\[
\frac{\partial^2 g}{\partial \xi^2}-\frac{\partial^2 g}{\partial
\eta^2}=0
\]
the general solution is:
\[
g(\xi,\eta)= F(\xi+\eta)+G(\xi-\eta)
\]
where $F(u)$ and $G(v)$ are arbitrary functions. The above metric
characterizes a surface which is called Liouville surface in the
geometry textbooks.  Therefore we have shown that a system is
integrable in Class I only when the surface is a Liouville
surface. The surfaces of constant curvature or the rotation
surfaces are Liouville surfaces, but there are Liouville surfaces
which have not constant curvature and they are not rotation
surfaces.

 The function
$\beta(\xi,\eta)$ can be calculated from equations (\ref{eq:beta})
\[
\frac{\partial \beta}{\partial \xi}=\frac{\partial g}{\partial
\eta}, \qquad \frac{\partial \beta}{\partial \eta} =\frac{\partial
g}{\partial \xi}
\]
then
\begin{equation}
\beta(\xi,\eta)=F(\xi+\eta)-G(\xi-\eta)  \label{eq:betasol}
\end{equation}
The potential $V(\xi,\eta)$ in Liouville coordinates is the
solution of equation (\ref{eq:DiffQ})
\[
\begin{array}{l}
\left(  F(\xi+\eta)+G(\xi-\eta) \right) \left( \frac{\partial^2
V}{\partial
\xi^2}-\frac{\partial^2 V}{\partial \eta^2} \right)+\\
 +2 F'(\xi+\eta)
\left( \frac{\partial V}{\partial \xi}-\frac{\partial V}{\partial
\eta} \right)+ 2 G'(\xi+\eta) \left( \frac{\partial V}{\partial
\xi}+\frac{\partial V}{\partial \eta} \right)=0
\end{array}
\]
The general solution of the above equation in Liouville
coordinates is
\begin{equation}
V(\xi,\eta)= \frac{
f(\xi+\eta)\,+\,g(\xi-\eta)}{F(\xi+\eta)\,+\,G(\xi-\eta)}
\label{eq:solV}
\end{equation}
The functions $f(u)$ and $g(v)$ are arbitrary functions. The
function $Q(\xi,\eta)$ is determined from equations
(\ref{eq:Coeff_px_py}) and the solution is easily calculated:
\begin{equation}
Q(\xi,\eta)= 4\frac{f(\xi+\eta)\,G(\xi-\eta)\,-\,g(\xi-\eta)\,
F(\xi+\eta)}{F(\xi+\eta)\,+\,G(\xi-\eta)} \label{eq:solQ}
\end{equation}
Usually is more convenient the use of the coordinates $u,\; v$,
which are defined by:
\[
\xi=u+i v, \quad \eta= u - i v, \quad p_\xi=\frac{p_u- i p_v}{2},
\quad \mbox{and}\quad p_\eta=\frac{p_u+ i p_v}{2}
\]
The Hamiltonian $H$ and the integral can be written as:
\[
H= \frac{{p_u}^2 + {p_v}^2 + 4\,(f(u) + g(v))}{4\,(F(u) + G(v))}
\]
\[
I=\frac{ {p_u}^2 G(v) - {p_v}^2 F(u)}{F(u) + G(v)} + 4\,\frac{f(u)
G(v) - g(v)F(u)}{F(u) + G(v)}
\]
The above formula has been investigated in a different context in
\cite{KaKrMiPo02}.  In these coordinates the action $S(u,v)$
satisfy the following equations:
\begin{equation}
\label{eq:Action_eq}
\begin{array}{c}
E=H(u,v,p_u,p_v),\quad   J=I(u,v,p_u,p_v)\\
p_u=\frac{\partial S}{\partial u}, \quad p_v=\frac{\partial
S}{\partial v}
\end{array}
\end{equation}
and we can find the action $S(u,v)$ by separation of variables.
\[
S=-E t + \int{\sqrt{4 \, E \, F(u)+J -4\, f(u)}\; du}+
\int{\sqrt{4\, E \,G(v)-J -4 \,g(v)}\; dv}
\]

\noindent \underline{\textbf{CLass II:  $ B(y) = 0$}}\\
We can choose a new coordinate system
\[
\xi= \int \frac{dx}{\sqrt{A(x)}} \quad \mbox{and} \quad \eta=y
\]
the associated momenta are:
\[
p_\xi= \sqrt{A(x)}\, p_x, \quad \mbox{and} \quad p_\eta= \, p_y
\]
 In this case
the metric is written
\[
ds^2= \widehat{g}(\xi,\eta) \, d\xi\, d\eta, \quad \mbox{where}
\quad \widehat{g}(\xi,\eta)= g(x,y)\,\sqrt{A(x)}
\]
 We can easily show that the
formulas are the same by replacing $x\rightarrow\xi$,
$y\rightarrow \eta$ and  fixing $A(x)=1$ and $B(y)=0$. For
simplicity reasons we omit the hat on the metric
$\widehat{g}(\xi,\eta)\rightarrow {g}(\xi,\eta)$. Equations
(\ref{eq:ClassHamiltonian}) and (\ref{eq:IntegralOfMotion}) are
written:
\begin{equation}
\begin{array}{c}
\displaystyle  H= \frac{p_\xi p_\eta}{g(\xi,\eta)}+ V(\xi,\eta)\\
\displaystyle I= p_\xi^2 - 2 p_\xi  p_\eta
\frac{\beta(\xi,\eta)}{g(\xi,\eta)}+Q(\xi,\eta)
\end{array}
\label{eq:Aeq1Beq0}
\end{equation}
We call these specific coordinates $(\xi,\,\eta)$ \underline{Lie
coordinates}. In Lie coordinates the Hamiltonian $H$
\underline{and} the integral $I$ are written as it is given in the
above equation (\ref{eq:Aeq1Beq0}).

 In Lie coordinates equation
(\ref{eq:Diff_beta2}) is written:
\[
\frac{\partial^2 g}{\partial \xi^2}=0
\]
The general solution is:
\[
g(\xi,\eta)= F(\eta)\xi+G(\eta)
\]
where $F(\eta)$ and $G(\eta)$ are arbitrary functions. The above
metric characterizes a surface which will be called Lie surface.
Therefore we have shown that a system is integrable in Case 2 only
when the surface is a Lie  surface.

Equations (\ref{eq:beta}) imply:
\begin{equation}
\left.
\begin{array}{l}
\displaystyle \frac{\partial \beta }{\partial \eta}=
\frac{\partial g
}{\partial \xi}\\
\\
\displaystyle \frac{\partial \beta }{\partial \xi}=0
\end{array}
\right\}
 \Rightarrow
\left\{
\begin{array}{l}
g(\xi,
\eta)= F(\eta) \xi + G(\eta)\\
\\
\beta(\xi,\eta)={\int}{\, F(\eta)\, d\eta}
\end{array}
\right. \label{eq:gsol2}
\end{equation}
Equation (\ref{eq:DiffQ}) is written:
\[
\left(  F(\eta)\,\xi +G(\eta) \right)\, \frac{\partial^2
V}{\partial \xi  ^2} +2 F(\eta)\, \frac{\partial V}{\partial
\xi}=0
\]
and the general solution of the above equation is:
\begin{equation}
V(\xi,\eta)= \frac{f(\eta)\,\xi +g(\eta)}{F(\eta)\,\xi +G(\eta)}
\label{eq:solV1}
\end{equation}
The functions $F(\eta),\; G(\eta),\;f(\eta)$ and $g(\eta)$ are
arbitrary functions. In this case the solution of the system of
equations (\ref{eq:Coeff_px_py}) is given by:
\begin{equation}\label{eq:DiffQ1}
Q(\xi,\eta)= -2 \frac{\left(f(\eta)\,\xi +g(\eta)\right) \int\,
F(\eta)\, d\eta}{F(\eta)\,\xi +G(\eta)}+2 \int\, f(\eta)\, d\eta
\end{equation}
The action integral in this case can be easily calculated:
\[
S=\xi  \sqrt{J-2  \left(\int\, f(\eta)\,d\eta\;- E \int\, F(\eta)
\, d\eta\right)}-\int\, d\eta\,\frac{g(\eta)  - E
G(\eta)}{\sqrt{J-2 \left(\int\, f(\eta)\,d\eta\;- E \int\, F(\eta)
\, d\eta\right)}}
\]
where
\[
\begin{array}{c}
E=H(\xi,\eta,p_\xi,p_\eta),\quad   J=I(\xi,\eta,p_\xi,p_\eta)\\
p_\xi=\frac{\partial S}{\partial \xi}, \quad p_\eta=\frac{\partial
S}{\partial \eta}
\end{array}
\]
The above findings are summarized in the following
Proposition.\\
\begin{Proposition}\label{prop:Integrability1}
 A Hamiltonian $H$ quadratic in momenta, which is defined on a
 two dimensional manifold possesses  an integral of motion $I$
  quadratic in momenta, only  in two cases:\\
  \underline{\textbf{Class I:}}\\
The manifold is a Liouville surface, i.e.   there is a coordinate
system $(\xi,\; \eta)$
  where the metric can be written:
  \[
  ds^2=g(\xi,\eta ) d\xi \eta\quad
  \mbox{and} \quad  g(\xi,\eta)=F(\xi+\eta)+G(\xi-\eta)
  \]
  \[
  H= \frac{p_\xi\, p_\eta}{F(\xi+\eta)+G(\xi-\eta)}+
  \frac{f(\xi+\eta)+g(\xi-\eta)}{F(\xi+\eta)+G(\xi-\eta)}
\]
and simultaneously the integral of motion is written
\[
I=p_\xi^2+p_\eta^2 - 2 p_\xi p_\eta
\frac{F(\xi+\eta)-G(\xi-\eta)}{F(\xi+\eta)+G(\xi-\eta)}+
4\frac{f(\xi+\eta)\,G(\xi-\eta)\,-\,g(\xi-\eta)\,
F(\xi+\eta)}{F(\xi+\eta)\,+\,G(\xi-\eta)}
\]
  \\
\underline{\textbf{Class II:}}\\
The manifold is a Lie surface, i.e.   there is a coordinate system
$(\xi,\; \eta)$
  where the metric can be written:
\[
ds^2=g(\xi,\eta) d\xi d\eta \quad \mbox{and} \quad
g(\xi,\eta)=F(\eta)\, \xi +G(\eta)
\]
\[
H=\frac{p_\xi p_\eta}{F(\eta)\, \xi +G(\eta)}+\frac{f(\eta)\,\xi
+g(\eta)}{F(\eta)\,\xi +G(\eta)}
\]
and simultaneously
\[
I=p_\xi^2- 2 p_\xi p_\eta \frac{\int\,
F(\eta)\,d\eta}{F(\eta)\,\xi +G(\eta)} -2 \frac{\left(f(\eta)\,\xi
+g(\eta)\right) \int\, F(\eta)\, d\eta}{F(\eta)\,\xi +G(\eta)}+2
\int\, f(\eta)\, d\eta
\]
The above specific choice of coordinate system $(\xi,\, \eta)$
will be called Lie coordinate system.
\end{Proposition}
The  integrable systems  which belong in Class I and II are well
known integrable systems see Ref. \cite{KaKrWin02}, the supporting
manifold is a Liouville or a Lie surface.

We must notice that the Hamiltonian and the integral of an
integrable system determine uniquely the Liouville (or Lie)
coordinate system. Therefore the use of this privileged system is
imposed by the notion of integrability. In the next sections we
shall work in this special coordinate system, which is denoted
exclusively  by the coordinates $\xi$ and $\eta$. In Class I
integrable systems the system is separable, while in Class II
systems there is no separation of variables in general.

\section{Poisson algebra of superintegrable systems
with two quadratic integrals of motion}\label{prop:Quadrquadr}

If a system is superintegrable on a two dimensional manifold, that
means that there are three functionally independent integrals of
motion $H,\,A$ and $B$.  In this section, we assume that these
integrals of motion are quadratic functions of the momenta and
there are no other integrals of motion, which are linear functions
of the momenta. Regarding the Hamiltonian $H$ and the first
integral $A$, we can choose the Liouville coordinate system and in
this system:
\[
H=\frac{p_\xi  p_\eta}{g(\xi,\eta)}+V(\xi,\eta)
\]
As we have shown in the previous section the system is integrable
with a square integral of motion in two cases. The integral of
motion $A$ in the Liouville coordinate system is written:
\[
\begin{array}{rl}
\displaystyle A= p_\xi^2 + b \, p_\eta^2 - 2 p_\xi p_\eta
\frac{\sigma(\xi,\eta)}{g(\xi,\eta)}+\Theta(\xi,\eta)
\end{array}
\]
where
\[
b=\left\{
\begin{array}{llll}
           1& \mbox{in Class I, where}& g(\xi,\eta)=F(\xi+\eta)+G(\xi-\eta)&\mbox{  (Liouville system)}\\
           \\
           0& \mbox{in Class II, where}&g(\xi,\eta)=F(\eta)\,\xi  +G(\eta)&\mbox{  (Lie system)}
\end{array}
   \right.
\]
The integral of motion $B$ is assumed to be indeed a quadratic
function of momenta, thus the general form in Liouville
coordinates is
\[
\begin{array}{rl}
B= A(\xi) p_\xi^2 + B(\eta) p_\eta^2 -2 p_{\xi} p_{\eta}
\frac{\beta(\xi,\eta)}{g(\xi,\eta)} + Q(\xi,\eta)
\end{array}
\]
 By definition the following
relations are satisfied:
\begin{equation}
\left\{ H, A\right\}_{P}= \left\{ H, B\right\}_{P}=0
\label{eq:PBIntegrals}
\end{equation}

From the integrals of motion $A$  and $B$, we can construct the
integral of motion:
\begin{equation}
C=\left\{ A, B\right\}_{P}
 \label{eq:ClassicalC}
 \end{equation}
The integral of motion $C$ is not a new independent integral of
motion, which is a cubic function of the momenta. The integral $C$
is not functionally independent from the integrals $H,\;A$ and $B$
as it will be shown later. The fact that, the integral $C$ is a
cubic function of momenta, implies the impossibility of expressing
$C$ as a polynomial function of the other integrals, which are
quadratic functions of momenta.  We shall prove that the square of
this cubic polynomial is indeed a cubic combination of the
integrals.
 Starting from the integral of motion $C$, we can
construct the (non independent) integrals $\left\{ A,
C\right\}_{P}$ and $\left\{ B, C\right\}_{P}$. These integrals are
quartic functions of the momenta, i.e. functions of fourth order.
In Appendix \ref{sec:Poly}   we show that these integrals can be
expressed as quadratic combinations of the integrals $H,\;A$, and
$B$. Therefore the following relations are  valid:
\begin{equation}
\left\{ A, C\right\}_{P}= \alpha A^2 + \beta B^2 + 2 \gamma A B +
\delta A + \epsilon B +\zeta \label{eq:AC}
\end{equation}
and
\begin{equation}
 \left\{ B , C \right\}_P = a A^2 + b B ^2 + 2 c A B
+ d A + e B + z \label{eq:BC}
\end{equation}
By taking   an appropriate  rotation  of the integrals $A$ and $B$
 we can always consider the case
$\beta =0$.

The existence of the Poisson algebra (\ref{eq:AC})-(\ref{eq:BC})
is not evident. The above form
  was assumed as obvious by several authors
\cite{KaMiPogo96}--\cite{KaMiPogo02_PAN},
\cite{Vinet}--\cite{GLZ-R3}. In Classical mechanics the Poisson
algebra was not considered as an important point, but in Quantum
Mechanics the existence of Poisson algebra permits the algebraic
treatment of the superintegrable system \cite{Das00}. In this
paper we prove that the superintegrable systems can be classified
using the properties of the Poisson algebra. The
superintegrability is a global property of the system, and  this
fact is reflected in the Poisson algebra structure, which is
indeed a global property. The mathematical proof of the existence
of the algebra (\ref{eq:AC})-(\ref{eq:BC}) for the two dimensional
superintegrable systems can be found in the Appendix
\ref{sec:Poly}.

 The Jacobi equality for the Poisson
brackets induces the relation
\[
 \left\{ A, \left\{ B, C \right\}_P \right\}_P =
  \left\{ B, \left\{ A, C
\right\}_P \right\}_P
\]
 The following  relations
 $$ b = - \gamma , \quad c= - \alpha
\quad \mbox{and} \quad e= - \delta $$
 must be satisfied.

 The integrals $A,\;B$ and $C$ satisfy the  quadratic Poisson
 algebra:
\begin{equation}
\begin{array}{rcl}
\left\{ A, B\right\}_{P}&=&C \\
 \left\{ A , C \right\}_P &=& \alpha A^2
+2 \gamma  A B  + \delta A + \epsilon B + \zeta\\
 \left\{ B , C \right\}_P &=& a A^2 - \gamma B ^2 - 2\alpha  A B + d A
-\delta B + z
\end{array}
\label{eq:PoissonAlgebra}
\end{equation}
where $\alpha,\;\gamma,\;a$ are constants and
\[
\begin{array}{l}
\delta=\delta(H)= \delta_0 + \delta_1 H \\
 \epsilon=\epsilon(H)=\epsilon_0 + \epsilon_1 H\\
\zeta= \zeta(H) = \zeta_0 + \zeta_1 H + \zeta_2 H^2\\ d=d(H)=
d_0+d_1 H \\ z= z(H) = z_0 + z_1 H + z_2 H^2
\end{array}
\]
where $\delta_i,\;\epsilon_i,\;\zeta_i,\;d_i$ and $z_i$ are
constants. The associative algebra, whose the generators satisfy
equations (\ref{eq:PoissonAlgebra}), is the general form of the
closed Poisson algebra of the integrals of superintegrable systems
with integrals quadratic in momenta.

The quadratic Poisson algebra (\ref{eq:PoissonAlgebra}) possess a
Casimir which is a  function of momenta of degree 6 and it is
given by:
\begin{equation}
\begin{array}{rl}
 K=& C^2-2\alpha A^2 B - 2 \gamma A B^2 -2 \delta A B-\\
 &- \epsilon B^2 -2 \zeta B+\frac{2}{3}a A^3 + d A^2 +2 z A=\\
 =& k_0 + k_1 H + k_2 H^2+k_3 H^3
\end{array}
 \label{eq:ClassicalCasimir}
\end{equation}
Obviously
\[
\left\{ K,A \right\}_P = \left\{ K,B \right\}_P =\left\{ K,C
\right\}_P =0
\]
Therefore the integrals of motion of a superintegrable two
dimensional system with quadratic integrals of motion close a
constrained classical quadratic Poisson algebra
(\ref{eq:PoissonAlgebra}), corresponding to a Casimir equal at
most to a cubic function of the Hamiltonian
(\ref{eq:ClassicalCasimir}).

In the general case of a superintegrable system the integrals are
not necessarily quadratic functions of the momenta, but rather
polynomial functions of the momenta. The case of the systems with
a quadratic and a cubic integral of motion are  studied by
Tsiganov \cite{Tsyg00_TMP,Tsyg00_JPA}.

\section{Classes of superintegrable
systems on a two dimensional manifold with quadratic integrals of
motion}\label{sec:PoissonSuper}

 The main result of the previous section is that the definition of the Casimir
of the Poisson algebra, given by equation
 (\ref{eq:ClassicalCasimir})  determines
 uniquely the Poisson algebra. This Poisson algebra is specific
 for each superintegrable system,  therefore it can be used for the
 classification of the possible superintegrable systems. Usually
 the proposed classifications of superintegrable systems assumed
 the definition of the  manifold metric and the superintegrable systems were
 fixed for the given
 metric. In this
 paper we propose a classification which is based on the Poisson
 algebra.
Let us consider a superintegrable system, which is described by a
Hamiltonian $H$ and two integrals of motion $A$ and $B$. The
integrability of the system imposes several choices which are
determined two by  Classes I and II of integrable systems, as it
has been shown in section \ref{sec:Integrable}. These classes of
super integrable systems are:
\begin{description}
\item[Class I]\label{ClassI} This class contains  superintegrable
systems, whose manifold metric and integrals of motion are written
 in a specific coordinate:
\begin{equation}\label{eq:Smetric1}
ds^2=g(\xi,\eta) d\xi d\eta, \qquad g(\xi,\eta)=F(\xi+\eta)+G(\xi-\eta)
\end{equation}
\begin{equation}\label{eq:SHamiltonian1}
H=\frac{ p_\xi p_\eta}{F(\xi+\eta)+G(\xi-\eta)}
+\frac{f(\xi+\eta)+g(\xi-\eta)}{F(\xi+\eta)+G(\xi-\eta)}
\end{equation}
and
\begin{equation}\label{eq:SA1}
\begin{array}{rl}
A=&p_\xi^2+p_\eta^2 - 2 p_\xi p_\eta \frac{F(\xi+\eta)-G(\xi-\eta)}{F(\xi+\eta)+G(\xi-\eta)} +\\
+& 4\frac{f(\xi+\eta)\,G(\xi-\eta)\,-\,g(\xi-\eta)\, F(\xi+\eta)}{F(\xi+\eta)\,+\,G(\xi-\eta)}
\end{array}
\end{equation}
 The second integral of
motion has the general form:
\begin{equation}\label{eq:SB1}
B= A(\xi) p_\xi^2 + B(\eta) p_\eta^2 - 2 p_\xi p_\eta
\frac{\beta(\xi,\eta)}{F(\xi+\eta)+G(\xi-\eta)}+Q(\xi,\eta)
\end{equation}
where $A(\xi)$ and $B(\eta)$ are not zero.
\\
\item[Class II]\label{ClassII} This class contains
superintegrable systems, whose manifold metric and integrals of
motion are written
 in a specific coordinate:
\begin{equation}\label{eq:Smetric2}
ds^2=g(\xi,\eta) d\xi d\eta, \qquad g(\xi,\eta)=F(\eta)\xi+G(\eta)
\end{equation}
\begin{equation}\label{eq:SHamiltonian2}
H=\frac{p_\xi p_\eta}{F(\eta)\, \xi +G(\eta)}+\frac{f(\eta)\,\xi +g(\eta)}{F(\eta)\,\xi
+G(\eta)}
\end{equation}
and
\begin{equation}\label{eq:SA2}
A=p_\xi^2- 2 p_\xi p_\eta \frac{\int\, F(\eta)\,d\eta}{F(\eta)\,\xi +G(\eta)} -2
\frac{\left(f(\eta)\,\xi +g(\eta)\right) \int\, F(\eta)\, d\eta}{F(\eta)\,\xi
+G(\eta)}+2 \int\, f(\eta)\, d\eta
\end{equation}
 The second integral of
motion has the general form:
\begin{equation}\label{eq:SB2}
B= A(\xi) p_\xi^2 + B(\eta) p_\eta^2 - 2 p_\xi p_\eta \frac{\beta(\xi,\eta)}{F(\eta)\xi
+G(\eta)}+Q(\xi,\eta)
\end{equation}
where $A(\xi)$ and $B(\eta)$ are not zero.
\end{description}
The existence of a second integral of motion $B$ implies that there
is another Liouville coordinate system $(X,Y)$ corresponding to the
pair of integrals $H$ and $B$. In this system there are analytic
formulas for the second integral of motion by using the results of
Proposition \ref{prop:Integrability1}. The analytic calculation of
these formulas will be given in the following subsections.

 The Class I  superintegral system is described by  the
integrals of motion which are given by equations
(\ref{eq:Smetric1}--\ref{eq:SB1}). Correspondingly, the Class II
superintegrable system  is described by  the integrals of motion
which are given by equations (\ref{eq:Smetric2}  -- \ref{eq:SB2}).
 These integrals of motion
should satisfy the relation (\ref{eq:ClassicalCasimir}), which is
an identity relation between  polynomials of sixth degree for the
momenta. The coefficients of $p_\xi^6$ and $p_\eta^6$ in
(\ref{eq:ClassicalCasimir}) vanish and  the following identities
are true:
\begin{equation}
\begin{array}{l}
6 \left( A'(\xi)\right) ^2=3 \gamma\, A^2(\xi) \,+\,3 \alpha\, A(\xi) -a\\
6 \left(B'(\eta)\right)^2=3 \gamma\, B^2 (\eta)\,+\,3 \alpha\,
B(\eta) -a
\end{array}
\label{eq:AB_Identities}
\end{equation}
 In equation (\ref{eq:SB1}), the coefficients of
$p_\xi^2$ and $p_\eta^2$ in the integral $B$ are determined by the
solution of the above equations (\ref{eq:AB_Identities}).

The superintegrable systems on a manifold can be classified by the
possible solutions of equations (\ref{eq:AB_Identities}).  The
 integral of motion $A$ has a standard form, given by
equation (\ref{eq:SA1}), this form is the Liouville form and the
coefficients of $p_x^2$ and $p_y^2$ are equal to $1$. The second
integral of motion $B$ can be replaced by any combination of the
form:
\[
B\longrightarrow q B+ r H + s A
\]
where $q, r, s$ are arbitrary constants. From this fact we can show
that there are six subclasses of possible solutions of equation
(\ref{eq:AB_Identities}).

\begin{itemize}
\item  \underline{\textbf{Subclass I${}_1$}} This class
corresponds to
\[
\gamma=0,\quad \alpha =0, \quad a\ne 0
\]
if we choose $a=-6$, then
\[
A(\xi)=\xi,\quad B(\eta)=\eta
\]

 \item  \underline{\textbf{Subclass I${}_2$}} This class corresponds to
\[
\gamma=0,\quad \alpha \ne 0, \quad a= 0
\]
if we choose $\alpha=8$, then
\[
A(\xi)=\xi^2,\quad B(\eta)=\eta^2
\]

\item  \underline{\textbf{Subclass I${}_3$}} This class
corresponds to
\[
\gamma\ne 0,\quad \alpha =0, \quad a\ne 0
\]
if we choose $\gamma=2$, then we can show that all cases are
equivalent to the choice
\[
A(\xi)=\left(e^{\xi}+e^{-\xi}\right)^2,\quad
B(\eta)=\left(e^{\eta} + e^{-\eta}\right)^2
\]

\end{itemize}
and
\begin{itemize}
\item  \underline{\textbf{Subclass II${}_1$}} This class
corresponds to
\[
\gamma=0,\quad \alpha =0, \quad a=0
\]
then
\[
A(\xi)=1,\quad B(\eta)=1
\]

 \item  \underline{\textbf{Subclass II${}_2$}} This class corresponds to
\[
\gamma=0,\quad \alpha=0 , \quad a\ne 0
\]
if we choose $a=-6$, then
\[
A(\xi)=\xi,\quad B(\eta)=\eta
\]

\item  \underline{\textbf{Subclass II${}_3$}} This class
corresponds to
\[
\gamma= 0 ,\quad \alpha \ne 0, \quad a=0
\]
if we choose $\alpha=8$, then we can show that
\[
A(\xi)=\xi^2,\quad B(\eta)=\eta^2
\]

\end{itemize}
These classes of solutions will be studied in detail in
\textit{Section \ref{sec:Classification}}

\subsection{ Class I superintegrable systems}

 Starting from the
definition of functions $A(\xi)$ and $B(\eta)$ we can solve
equations (\ref{eq:beta}--\ref{eq:DiffQ}) and the superintegrable
system is fully determined. This procedure will be sketched in
detail in the next paragraphs.

Let us start by the known solutions $A(\xi),\; B(\eta)$ of equations
(\ref{eq:AB_Identities}).

Equation (\ref{eq:Diff_beta1}) is written
\begin{equation}\label{eq:Diff_betaI}
\begin{array}{l} \left( A''(\xi)-B''(\eta) \right) \left( F(\xi+\eta)+ G(\xi-\eta) \right)
+\\
+ 3 A'(\xi) \left( F'(\xi+\eta)+ G'(\xi-\eta) \right) -3 B'(\eta) \left( F'(\xi+\eta)-
G'(\xi-\eta) \right)+\\+2  \left(A(\xi)-B(\eta)\right) \left( F''(\xi+\eta)+
G''(\xi-\eta) \right)=0
\end{array}
\end{equation}
In the next paragraphs we will show that, the above equation can
be separated in two second order differential equations for the
involved functions $F(u)$ and $G(v)$. The general solution of
these equations are given by:
\begin{equation}\label{eq:FGsol}
\begin{array}{l}
F(u)= \lambda_1 \,F_1(u)\,+\,\lambda_2 \,F_2(u)\, +\, \lambda_3\, F_3(u)\,+\, \lambda_4\, F_4(u)\\
G(v)=\ell_1 \,G_1(v)\,+\,\ell_2 \,G_2(v)\, +\, \ell_3\,
G_3(v)\,+\, \ell_4\, G_4(v)
\end{array}
\end{equation}
where $F_k(u)$ and $G_k(u)$ are functions which are not generally  partial independent solutions of
two second order differential equation (\ref{eq:Diff_beta1}). Between the eight parameters $\lambda_k$  and
 $\ell_k$ for $k\,=\,1,\,2,\,3,\,4$ only four among
of them are independent.

 After the calculation of the functions $F(u)$ and $G(v)$ we can
 calculate the function $\beta(x,y)$ from equation
 (\ref{eq:beta}). The general form of the potential $V(\xi,\eta)$ is
 given by equation (\ref{eq:solV}).

 After some elementary (but
 rather lengthy) algebraic calculation we can show that
  equation (\ref{eq:DiffQ}) leads to a differential equation for
  the functions $f(u)$ and $g(v)$ which are involved in the
  definition  (\ref{eq:solV}) of the potential:
\[
\begin{array}{l}
f(\xi + \eta)\,\{ -3\,B'(\eta)\,
      \left( F'(\xi + \eta) - G'(\xi - \eta) \right)  +
     3\,A'(\xi)\,\left( F'(\xi + \eta) + G'(\xi - \eta) \right)  +\\
     +
     2\,\left( A(\xi) - B(\eta) \right) \,
      \left( F''(\xi + \eta) + G''(\xi - \eta) \right) \}  +\\
 + g(\xi - \eta)\,\{ -3\,B'(\eta)\,
      \left( F'(\xi + \eta) - G'(\xi - \eta) \right)  +
     3\,A'(\xi)\,\left( F'(\xi + \eta) + G'(\xi - \eta) \right)  +\\
    + 2\,\left( A(\xi) - B(\eta) \right) \,
      \left( F''(\xi + \eta) + G''(\xi - \eta) \right) \}  -\\
      -
  \left( F(\xi + \eta) + G(\xi - \eta) \right) \,
  \{ -3\,B'(\eta)\,
      \left( f'(\xi + \eta) - g'(\xi - \eta) \right)  +\\
     +3\,A'(\xi)\,\left( f'(\xi + \eta) +
        g'(\xi - \eta) \right)  +
     2\,\left( A(\xi) - B(\eta) \right) \,
      \left( f''(\xi + \eta) + g''(\xi - \eta) \right)
     \}=0
\end{array}
\]
We can eliminate in the above equation the functions $F(u)$ and
$G(v)$, which satisfy equation (\ref{eq:Diff_betaI}) and finally
the functions involved in the definition of the potential satisfy
the following equation:
\begin{equation}\label{eq:DiffQI}
\begin{array}{l} \left( A''(\xi)-B''(\eta) \right) \left( f(\xi+\eta)+ g(\xi-\eta) \right)
+\\
+ 3 A'(\xi) \left( g'(\xi+\eta)+ g'(\xi-\eta) \right) -3 B'(\eta) \left( g'(\xi+\eta)-
g'(\xi-\eta) \right)+\\+2  \left(A(\xi)-B(\eta)\right) \left( f''(\xi+\eta)+
g''(\xi-\eta) \right)=0
\end{array}
\end{equation}
This equation is indeed the same as (\ref{eq:Diff_betaI}) and the
general solution has been given by equation (\ref{eq:FGsol}). Then
we have proved the following general proposition:
\begin{Proposition}\label{prop:V_ClassI}
The general form of the potential $V(\xi,\eta)$ in Liouville
coordinates for a superintegrable system of Class I is given by
the general formula:
\[
V(\xi,\eta)=
\frac{f(\xi+\eta)+g(\xi-\eta)}{F(\xi+\eta)+G(\xi-\eta)}
\]
where both the pairs of functions $f(u),\, g(v)$ and $F(u),\,
G(v)$ satisfy the same differential equation (\ref{eq:Diff_betaI})
and (\ref{eq:DiffQI}).
\end{Proposition}
Therefore the solutions of equation (\ref{eq:DiffQI}) are given
by:
\begin{equation}\label{eq:fgsol}
\begin{array}{l}
f(u)= \rho_1 \,F_1(u)\,+\,\rho_2 \,F_2(u)\, +\, \rho_3\, F_3(u)\,+\, \rho_4\, F_4(u)\\
g(v)=r_1 \,G_1(v)\,+\,r_2 \,G_2(v)\, +\, r_3\, G_3(v)\,+\, r_4\,
G_4(v)
\end{array}
\end{equation}
The equations of motion don't depend on the shifts of the
potential by one constant, this fact explains physically the
identity of differential equations (\ref{eq:Diff_betaI}) and
(\ref{eq:DiffQI}).

The calculation of the second integral of motion $B$ is now
straightforward. If the functions $A(x)$ and $B(y)$ are fixed as
solutions of the characteristic equations (\ref{eq:AB_Identities})
then we can introduce the new coordinates:
\begin{equation} \label{eq:NewCoordinates}
X= \int \frac{dx}{\sqrt{A(x)}} \quad \mbox{and} \quad Y=\int
\frac{dy}{\sqrt{B(y)})}
\end{equation}
In these coordinates the metric of the manifold can be written
\begin{equation}
\label{eq:NewMetric}
 ds^2\,= \,g(x,y)\, dx dy \,= \,\widetilde{g}(X,Y) \,dX dY
\end{equation}
where
\[
\widetilde{g}(X,Y) = g(x,y)\,\sqrt{A(x)\,B(y)}
\]
and
\[
H=\frac{p_x p_y}{g(x,y)} +V(x,y) =\frac{p_X
p_Y}{\widetilde{g}(X,Y)} +\widetilde{V}(X,Y)
\]
The integrability of the above Hamiltonian implies:
\begin{equation}\label{eq:NewMetric1sum}
\widetilde{g}(X,Y)=\widetilde{F}(X+Y)+\widetilde{G}(X-Y)
\end{equation}
The above relation implies:
\[
\begin{array}{lcl}
\widetilde{g}\left(\frac{U+V}{2},\frac{U-V}{2}\right)&=&\widetilde{F}(U)+\widetilde{G}(V)\\
\widetilde{g}\left(\frac{U+d}{2},\frac{U-d}{2}\right)&=&\widetilde{F}(U)+\widetilde{G}(d)\\
\widetilde{g}\left(\frac{c+V}{2},\frac{c-V}{2}\right)&=&\widetilde{F}(c)+\widetilde{G}(V)\\
\widetilde{g}\left(\frac{c+d}{2},\frac{c-d}{2}\right)&=&\widetilde{F}(c)+\widetilde{G}(d)
\end{array}
\]
where $c$ and $d$ are two arbitrary constants. Therefore the
functions $\widetilde{F}(U)$ and $\widetilde{G}(V)$ are calculated
up to one constant by:
\begin{equation}\label{eq:FGTilda}
\begin{array}{l}
\widetilde{F}(U)=\widetilde{g}\left(\frac{U+d}{2},\frac{U-d}{2}\right)
-\frac{1}{2} \widetilde{g}\left(\frac{c+d}{2},\frac{c-d}{2}\right)+\mu\\
\widetilde{G}(V)=\widetilde{g}\left(\frac{c+V}{2},\frac{c-V}{2}\right)
-\frac{1}{2}
\widetilde{g}\left(\frac{c+d}{2},\frac{c-d}{2}\right)-\mu
\end{array}
\end{equation}
where $\mu$ can be an arbitrary chosen constant. Starting from the
well known form of the potential:
\[
V(x,y)=\frac{w(x,y)}{g(x,y)}, \qquad w(x,y)=f(x+y)+g(x-y)
\]
we can show that
\begin{equation}\label{eq:NewPotential}
\begin{array}{c}
\displaystyle
V(x,y)=\widetilde{V}(X,Y)=\frac{\widetilde{w}(X,Y)}{\widetilde{g}(X,Y)},
\\
\displaystyle
\widetilde{w}(X,Y)=\widetilde{f}(X+Y)+\widetilde{g}(X-Y)=w(x,y)\,\sqrt{A(x)\,B(y)}
\end{array}
\end{equation}
The functions $\widetilde{f}(U)$ and $\widetilde{g}(V)$ are
calculated by:
\[
\begin{array}{l}
\widetilde{f}(U)=\widetilde{w}\left(\frac{U+d}{2},\frac{U-d}{2}\right)
-\frac{1}{2} \widetilde{w}\left(\frac{c+d}{2},\frac{c-d}{2}\right)\\
\widetilde{g}(V)=\widetilde{w}\left(\frac{c+V}{2},\frac{c-V}{2}\right)
-\frac{1}{2} \widetilde{w}\left(\frac{c+d}{2},\frac{c-d}{2}\right)
\end{array}
\]
Then the second integral of motion in the coordinates $X,Y$ is:
\begin{equation}\label{eq:NewB}
\begin{array}{rl}
B=&p_X^2+p_Y^2 - 2 p_X p_Y
\frac{\widetilde{F}(X+Y)-\widetilde{G}(X-Y)}{\widetilde{F}(X+Y)+\widetilde{G}(X-Y)}+\\
+& 4
\frac{\widetilde{f}(X+Y)\widetilde{G}(X-Y)-\widetilde{g}(X-Y)\widetilde{F}(X+Y)}{\widetilde{F}(X+Y)+\widetilde{G}(X-Y)}
\end{array}
\end{equation}
After calculating the above integral in the coordinates $X,Y$, we
can compute analytically the second integral $B$ in the original
coordinates $x,y$.

\subsection{ Class II superintegrable systems}

Equation (\ref{eq:Diff_beta1}) is written
\begin{equation}\label{eq:Diff_betaII}
\begin{array}{l} \left( A''(x)-B''(y) \right) \left( F(y)x + G(y) \right)
+\\
+ 3 A'(x) F(y) -3 B'(y) \left( F'(y) x+ G'(y) \right)+\\+2
\left(A(x)-B(y)\right) \left( F''(y)  x+ G''(y) \right)=0
\end{array}
\end{equation}
 The general solutions of
 equation (\ref{eq:Diff_betaII}) are given by:
\begin{equation}\label{eq:FGsol1}
\begin{array}{l}
F(y)= \lambda_1 \,F_1(y)\,+\,\lambda_2 \,F_2(y)\, +\, \lambda_3\, F_3(y)\,+\, \lambda_4\, F_4(y)\\
G(y)=\ell_1 \,G_1(y)\,+\,\ell_2 \,G_2(y)\, +\, \ell_3\,
G_3(y)\,+\, \ell_4\, G_4(y)
\end{array}
\end{equation}
where $F_k(y)$ and $G_k(y)$ are partial independent solutions of
two second order differential equations with several constant
parameters. Between the eight parameters $\lambda_k$  and
 $\ell_k$ for $k\,=\,1,\,2,\,3,\,4$ only four among
them are linearly independent.

 After the calculation of the functions $F(y)$ and $G(y)$ we can
 calculate the function $\beta(x,y)$ from equation
 (\ref{eq:gsol2}). The general form of the potential $V(x,y)$ is
 given by equation (\ref{eq:solV1}). After some elementary (but
 rather complicated) algebraic calculation we can show that,
  equation (\ref{eq:DiffQ1}) leads to an differential equation for
  the functions $f(y)$ and $g(y)$, which are involved in the
  definition  (\ref{eq:solV}) of the potential:
\begin{equation}\label{eq:DiffQII}
\begin{array}{l} \left( A''(x)-B''(y) \right) \left( f(y)x + g(y) \right)
+\\
+ 3 A'(x)  f(y)-3 B'(y) \left( f'(y)x+ g'(y) \right)+\\+2
\left(A(x)-B(y)\right) \left( f''(y) x+ g''(y) \right)=0
\end{array}
\end{equation}
Then
we have proved the following general proposition:
\begin{Proposition}\label{prop:V_ClassII}
The general form of the potential $V(\xi,\eta)$ in Liouville
coordinates for a superintegrable system of Class II is given by
the general formula:
\[
V(\xi,\eta)=
\frac{f(\eta)\,\xi +g(\eta)}{F(\eta)\,\xi+G(\eta)}
\]
where both the pairs of functions $f(u),\, g(v)$ and $F(u),\,
G(v)$ satisfy the same differential equation (\ref{eq:Diff_betaII})
and (\ref{eq:DiffQII}).
\end{Proposition}
Therefore the solutions of equation (\ref{eq:DiffQII}) are given
by:
\begin{equation}\label{eq:fgsol1}
\begin{array}{l}
f(y)= \rho_1 \,F_1(y)\,+\,\rho_2 \,F_2(y)\, +\, \rho_3\, F_3(y)\,+\, \rho_4\, F_4(y)\\
g(y)=r_1 \,G_1(y)\,+\,r_2 \,G_2(y)\, +\, r_3\, G_3(y)\,+\, r_4\,
G_4(y)
\end{array}
\end{equation}
The equations of motion don't depend on the shifts of the
potential by one constant, this fact explains physically the
identity of differential equations (\ref{eq:Diff_betaI}) (or (\ref{eq:Diff_betaII}) ) and
(\ref{eq:DiffQI}) (or (\ref{eq:DiffQII})).

The calculation of the second integral of motion $B$ is now
straightforward, we can use the same procedure of solution as it
has been described by equations
(\ref{eq:NewCoordinates}--\ref{eq:NewB})

\section{Classification of two dimensional superintegrable
systems with two quadratic integrals of
motion}\label{sec:Classification}

In this section we give the analytical solutions for the different
classes of superintegrable systems. As we have shown there are two
general classes of superintegrable systems, each class has 3
subclasses.

\subsection{Class I superintegrable systems}

\subsubsection{Subclass I${}_1$ of superintegrable systems }
\[
A(\xi)=\xi, \qquad B(\eta)=\eta
\]
\begin{equation}\label{eq:FGfgVa1}
\begin{array}{ll}
 \displaystyle F(u)=4 \lambda \,u^2\,+ \,\kappa \,u
\,+\,{\nu}/{2}, &
\displaystyle G(v)=-\lambda \,v^2\, + \,{\mu}/{ v^2 }\,+\,{\nu}/{2}\\

 \displaystyle f(u)=4 \ell \,u^2 \,+\, k\, u
+{n}/{2}, & \displaystyle g(v)=-\ell\, v^2\, + \,{m}/{ v^2}\,
+\,{n}/{2}\\
\end{array}
\end{equation}

\[
\begin{array}{ll}
ds^2=g(\xi,\eta)\, d\xi\, d\eta, &
g(\xi,\eta)= F(\xi+\eta)+G(\xi-\eta)\\
\\
\displaystyle H=\frac{p_\xi p_\eta}{g(\xi,\eta)}+V(\xi,\eta)
&\displaystyle V(\xi,\eta)=\frac{w(\xi,\eta)}{g(\xi,\eta)}, \quad
w(\xi,\eta)=f(\xi+\eta)+g(\xi-\eta)
\end{array}
\]
The other integral of motion is:
\[
\begin{array}{rl}
\displaystyle A=p_\xi^2+p_\eta^2 - &\displaystyle 2 p_\xi p_\eta \frac{F(\xi+\eta)-G(\xi-\eta)}{F(\xi+\eta)+G(\xi-\eta)} +\\
+& \displaystyle 4\frac{f(\xi+\eta)\,G(\xi-\eta)\,-\,g(\xi-\eta)\,
F(\xi+\eta)}{F(\xi+\eta)\,+\,G(\xi-\eta)}
\end{array}
\]
We introduce the functions:
\begin{equation}\label{eq:FGfgtildeVa1}
\begin{array}{l}\displaystyle \widetilde{F}(u)=\frac{\,\ \lambda u^6
}{256}+\frac{\,\ \kappa u^4 }{128} + \frac{\,\ \nu u^2}{16}
 -\frac{\mu }{u^2}
 \\
\\
  \displaystyle
\widetilde{G}(v)=-
  \frac{\,\ \lambda v^6}{256}-\frac{\,\ \kappa v^4}{128}  -
  \frac{\,\ \nu v^2}{16}+ \frac{\mu }{v^2}
\\
\\
\displaystyle \widetilde{f}(u)=\frac{\,\ l\,u^6}{256}+ \frac{\,\
k\,u^4}{128}+
\frac{\,\ n\,u^2}{16}-\frac{m}{u^2} \\
\\
     \displaystyle
\widetilde{g}(v)=- \frac{\,\ l\,v^6}{256} -
  \frac{\,\ k\,v^4}{128} -
\frac{\,\ n\,v^2}{16}+\frac{m}{v^2}
\end{array}
\end{equation}
The second integral of motion is:
\[
\begin{array}{rl}
\displaystyle  B=& \displaystyle  p^2_X+p^2_Y - 2 p_X p _Y \,
\frac{\widetilde{F}(X+Y)-\widetilde{G}(X-Y)}{\widetilde{F}(X+Y)+\widetilde{G}(X-Y)}+\\
&\displaystyle   +
4 \frac{\widetilde{f}(X+Y) \widetilde{G}(X-Y)-
\widetilde{g}(X-Y)\widetilde{F}(X+Y)}{\widetilde{F}(X+Y)+\widetilde{G}(X-Y)}
\end{array}
\]
where
\[
X=2 \sqrt{\xi}, \quad p_X= \sqrt{\xi} \,p_\xi, \quad Y=2
\sqrt{\eta}, \quad p_Y=\sqrt{\eta} \,p_\eta
\]
The constants of the Poisson algebra are:
\[
\begin{array}{l}
\alpha=0 \quad\beta=0 \quad \gamma=0 \quad
\delta=16 (\kappa H -k) \quad
 \epsilon=256(  \lambda  H-\ell )\\
\zeta=-32  (\kappa H -k)  (\nu  H -n) \quad
a=-6 \qquad d=8 ( \nu H-n) \\
z=8  ( \nu  H -n)^2- 128 (\lambda H -\ell) ( \mu H - m)
\end{array}
\]
\[
K=32
 (\nu  H-n)^3  + 512 (\lambda H -\ell)
 ( \mu H -m) ( \nu H -n) -64 (\kappa H  -k)^2 (\mu H -m)
\]

\subsubsection{Subclass I${}_2$ of superintegrable systems }
\[
A(\xi)=\xi^2, \qquad B(\eta)=\eta^2
\]
\begin{equation}\label{eq:FGfgVa2}
\begin{array}{ll}
\displaystyle F(u)=
  \lambda\,u^2\, +\frac{\kappa }{u^2}  + \frac{\nu }{2}, &
\displaystyle G(v)= - \lambda \,v^2\,   +
  \frac{\mu }{v^2} + \frac{\nu }{2}\\
\\
\displaystyle f(u)=
  \ell\,u^2\, +\frac{k }{u^2}  + \frac{n }{2}, &
\displaystyle g(v)= - \ell\,v^2\,  +
  \frac{m }{v^2} + \frac{n }{2}
\end{array}
\end{equation}
\[
\begin{array}{ll}
ds^2=g(\xi,\eta)\, d\xi\, d\eta, &
g(\xi,\eta)= F(\xi+\eta)+G(\xi-\eta)\\
\\
\displaystyle H=\frac{p_\xi p_\eta}{g(\xi,\eta)}+V(\xi,\eta)
&\displaystyle V(\xi,\eta)=\frac{w(\xi,\eta)}{g(\xi,\eta)}, \quad
w(\xi,\eta)=f(\xi+\eta)+g(\xi-\eta)
\end{array}
\]
The other integral of motion is:
\[
\begin{array}{rl}
\displaystyle A=p_\xi^2+p_\eta^2 - &\displaystyle 2 p_\xi p_\eta \frac{F(\xi+\eta)-G(\xi-\eta)}{F(\xi+\eta)+G(\xi-\eta)} +\\
+& \displaystyle 4\frac{f(\xi+\eta)\,G(\xi-\eta)\,-\,g(\xi-\eta)\,
F(\xi+\eta)}{F(\xi+\eta)\,+\,G(\xi-\eta)}
\end{array}
\]
We introduce the functions:
\begin{equation}\label{eq:FGfgtildeVa2}
\begin{array}{l}
\displaystyle \widetilde{F}(u)=4\,\lambda\,e^{2\,u}+\nu\,e^u ,
\quad
  \displaystyle
\widetilde{G}(v)=\frac{\kappa\,\ e^v}
   {{\left( 1 + e^v \right) }^2} +
  \frac{\mu\,\ e^v }
   {{\left( -1 + e^v \right) }^2}
\\
\\
\displaystyle \widetilde{f}(u)=4\,\ell\, e^{2\,u} +n\, e^u ,
\quad
     \displaystyle
\widetilde{g}(v)=\frac{k\, e^v} {{\left(1 + e^v \right)}^2} +
  \frac{m\, e^v}
{{\left(-1 + e^v \right)}^2}
\end{array}
\end{equation}
The second integral of motion is:
\[
\begin{array}{rl}
\displaystyle  B=& \displaystyle  p^2_X+p^2_Y - 2 p_X p _Y \,
\frac{\widetilde{F}(X+Y)-\widetilde{G}(X-Y)}{\widetilde{F}(X+Y)+\widetilde{G}(X-Y)}+\\
&\displaystyle   +
4 \frac{\widetilde{f}(X+Y) \widetilde{G}(X-Y)-
\widetilde{g}(X-Y)\widetilde{F}(X+Y)}{\widetilde{F}(X+Y)+\widetilde{G}(X-Y)}
\end{array}
\]
where
\[
X= \ln \, \xi,   \quad p_X= \xi\,p_\xi, \quad Y=\ln\, \eta, \quad p_Y=\eta\, p_\eta
\]
The constants of the Poisson algebra are:
\[
\begin{array}{l}
\alpha=8 \quad\beta=0 \qquad \gamma=0\quad
\delta=0\quad
 \epsilon=256(   \lambda  H-\ell)\\
\zeta=-32 (\nu H -n)^2+256 ( \lambda H - \ell) \left(  (\mu -\kappa) H - (m-k) \right)
\\
a=0 \qquad d=0 \qquad
z=32\,\left( (\kappa  + \mu ) H - (k+m)  \right) \,
   (\nu \,H-n)
\end{array}
\]
\[
K= 256\,(\lambda \, H -\ell)
    \left( (\kappa  + \mu) H - (k+m)  \right) ^2 +
    128 \left(( \kappa  - \mu ) H -(k-m)  \right) \,
      {(\nu  H -n) }^2
\]
\subsubsection{Subclass I${}_3$ of superintegrable systems }
\[
A(\xi)=(e^{\xi}+e^{-\xi})^2, \qquad B(\eta)=(e^{\eta}+e^{-\eta})^2
\]
\begin{equation}\label{eq:FGfgVa3}
\begin{array}{ll}
 \displaystyle F(u)=\frac{\kappa\,e^{2\,u} }{{\left( -1 + e^{2\,u} \right) }^2} +
  \frac{\,\lambda\,e^u\,\left( 1 + e^{2\,u} \right) }
   {{\left( -1 + e^{2\,u} \right) }^2}, &
\displaystyle G(v)=\frac{\mu\,e^{2\,v} }{{\left( -1 + e^{2\,v}
\right) }^2} +
  \frac{\,\nu\,e^v\,\left( 1 + e^{2\,v} \right) }
   {{\left( -1 + e^{2\,v} \right) }^2}\\  \\

 \displaystyle f(u)=\frac{k\,e^{2\,u} }{{\left( -1 + e^{2\,u} \right) }^2} +
  \frac{\,\ell\,e^u\,\left( 1 + e^{2\,u} \right) }
   {{\left( -1 + e^{2\,u} \right) }^2}, & \displaystyle g(v)=\frac{m\,e^{2\,v} }{{\left( -1 + e^{2\,v}
\right) }^2} +
  \frac{\,n\,e^v\,\left( 1 + e^{2\,v} \right) }
   {{\left( -1 + e^{2\,v} \right) }^2}\\
\end{array}
\end{equation}
\[
\begin{array}{ll}
ds^2=g(\xi,\eta)\, d\xi\, d\eta, &
g(\xi,\eta)= F(\xi+\eta)+G(\xi-\eta)\\
\\
\displaystyle H=\frac{p_\xi p_\eta}{g(\xi,\eta)}+V(\xi,\eta)
&\displaystyle V(\xi,\eta)=\frac{w(\xi,\eta)}{g(\xi,\eta)}, \quad
w(\xi,\eta)=f(\xi+\eta)+g(\xi-\eta)
\end{array}
\]
The other integral of motion is:
\[
\begin{array}{rl}
\displaystyle A=p_\xi^2+p_\eta^2 - &\displaystyle 2 p_\xi p_\eta \frac{F(\xi+\eta)-G(\xi-\eta)}{F(\xi+\eta)+G(\xi-\eta)} +\\
+& \displaystyle 4\frac{f(\xi+\eta)\,G(\xi-\eta)\,-\,g(\xi-\eta)\,
F(\xi+\eta)}{F(\xi+\eta)\,+\,G(\xi-\eta)}
\end{array}
\]
We introduce the functions:
\begin{equation}\label{eq:FGfgtildeVa3}
\begin{array}{l}
\displaystyle \widetilde{F}(u)=\frac{\left( \kappa  + 2\,\lambda
\right)}{4}\,\tan^2  u  +
  \frac{2\,\nu\,-\mu }{4}\,\cot^2 u + \frac{\lambda  + \nu }{2}
 \\
\\
  \displaystyle
\widetilde{G}(v)=\frac{\left(2\,\lambda -\kappa   \right) }{4}
\,\tan^2 v + \frac{\mu  + 2\,\nu }{4}\,\cot^2  v+
  \frac{\lambda  + \nu }{2}\\
\\
\displaystyle \widetilde{f}(u)=\frac{\left(k+ 2\,\ell \right)
}{4}\,\tan^2  u +
  \frac{2\,n\,- m }{4}\,\cot^2u + \frac{\ell + n }{2}\\
\\
     \displaystyle
\widetilde{g}(v)=\frac{\left(2\,\ell - k \right) }{4}\,\tan^2 v +
\frac{m + 2\,n }{4}\,\cot^2 v+
  \frac{\ell + n }{2}
\end{array}
\end{equation}
The second integral of motion is:
\[
\begin{array}{rl}
\displaystyle  B=& \displaystyle  p^2_X+p^2_Y - 2 p_X p _Y \,
\frac{\widetilde{F}(X+Y)-\widetilde{G}(X-Y)}{\widetilde{F}(X+Y)+\widetilde{G}(X-Y)}+\\
&\displaystyle   +
4 \frac{\widetilde{f}(X+Y) \widetilde{G}(X-Y)-
\widetilde{g}(X-Y)\widetilde{F}(X+Y)}{\widetilde{F}(X+Y)+\widetilde{G}(X-Y)}
\end{array}
\]

where
$$
X=\arctan (e^{\xi}) ,\quad p_X= (e^{\xi}+e^{-\xi})  \, p_\xi, \quad    Y=\arctan (e^{\eta}),\quad
p_Y= (e^{\eta}+e^{-\eta})\, p_\eta
$$

The constants of the Poisson algebra are:
\[
\begin{array}{l}
\alpha=-32 \quad\beta=0 \quad \gamma=8 \quad
\delta=0 \quad
 \epsilon=0\quad
\zeta=-32 \, \left( \lambda H - \ell \right) \left( \nu H -n\right)
\\
a=0 \qquad d=64\,\left( k - m \right)-64\,\left( \kappa  -
    \mu  \right)\,H \\
z=32\, {\left( (\lambda  -
        \nu) H - (\ell-n)   \right) }^2 -32\, \left( (\kappa   H -k\right)  \,
       \left(\mu H - m  \right)
\end{array}
\]
\[
K=64\,\left( \kappa H -k\right) \,{\left(\nu   H - n \right)}^2-
64 {\left( \lambda H -\ell\right) }^
        2\,\left(\mu H -m \right)
\]
\subsection{Class II superintegrable systems}
\subsubsection{Subclass II$_{1}$ of superintegrable systems }
\[
A(\xi)=1, \qquad B(\eta)=1
\]
\begin{equation}\label{eq:FGfgVb1}
\begin{array}{ll}
 \displaystyle F(\eta)=\kappa\,\eta + \lambda, &
\displaystyle G(\eta)=\mu\, \eta  + \nu\\
 \displaystyle f(\eta)=k\,\eta + \ell, & \displaystyle g(\eta)=m\, \eta + n\\
\end{array}
\end{equation}
\[
\begin{array}{ll}
ds^2=g(\xi,\eta)\, d\xi\, d\eta, &
g(\xi,\eta)=\xi\, F(\eta)+G(\eta)\\
\\
\displaystyle H=\frac{p_{\xi}
p_{\eta}}{g(\xi,\eta)}+V(\xi,\eta)\,, &\displaystyle
V(\xi,\eta)=\frac{w(\xi,\eta)}{g(\xi,\eta)}\,, \quad
w(\xi,\eta)=\xi\, f(\eta)+g(\eta)
\end{array}
\]
The other integral of motion is:
\[
\begin{array}{rl}
\displaystyle A = p^2_\xi - \frac{2\,{p_{\xi} }\,{p_{\eta} }\,
     \int F(\eta )\,d\eta }{g(\xi ,\eta )}  -
  \frac{2\,\left( \xi \,f(\eta )+ g(\eta )
       \right) \,\int F(\eta )\,d\eta }{g(\xi ,
     \eta )} +  2\,\int f(\eta )\,d\eta
\end{array}
\]
We introduce the functions:
\begin{equation}\label{eq:FGfgtildeVb1}
\begin{array}{l}
\displaystyle \widetilde{F}(u)=\frac{\kappa \,u^2}{4} +
  \frac{\left( \lambda  + \mu  \right)\,u }{2} +
  \frac{\nu }{2}
 \\
\\
  \displaystyle
\widetilde{G}(v)=-\frac{\kappa\, v^2 }{4} + \frac{\left( \lambda -
\mu \right) \,v }{2} +
  \frac{\nu }{2}\\
\\
\displaystyle \widetilde{f}(u)=\frac{k\,u^2}{4} +
  \frac{\left( \ell + m  \right)\,u }{2} +
  \frac{n}{2}\\
\\
     \displaystyle
\widetilde{g}(v)=-\frac{k\, v^2 }{4} + \frac{\left( \ell - m
\right) \,v }{2} +
  \frac{n }{2}
\end{array}
\end{equation}
The second integral of motion is:
\[
\begin{array}{rl}
\displaystyle B=&\displaystyle p^2_\xi + p^2_\eta-
 2\, {p_{ \xi }}\,{p_{ \eta }}\, \frac{
     \widetilde{F}(\xi + \eta ) -
       \widetilde{G}( \xi - \eta  ) }{\widetilde{F}(\xi + \eta ) + \widetilde{G}( \xi -\eta )}\, +\\  \\
& \displaystyle +\, 4\, \frac{  \widetilde{f}(\xi+ \eta
)\,\widetilde{G}(\xi -\eta )-\widetilde{g}( \xi -\eta
)\,\widetilde{F}(\xi + \eta )
  }
     {\widetilde{F}(\xi + \eta ) + \widetilde{G}(\xi -\eta )}
\end{array}
\]

The constants of the Poisson algebra are:
\[
\begin{array}{l}
\alpha=0 \quad\beta=0 \quad \gamma=0 \quad \delta=8 (k-\kappa\, H)
\quad
 \epsilon=0\quad
\zeta=8 \,\left( \lambda H - \ell  \right)^2
\\
a=0 \qquad d=16 (k-\kappa\, H )\quad z=8  \left( \lambda H - \ell
\right)^2-\left( \mu H - m \right)^2
\end{array}
\]
\[
K=16\, \left( \nu H - n  \right)^2 \left( \kappa H - k \right)
-32\, \left( \lambda H - \ell  \right) \left( \mu H - m
\right)\left( \nu H - n  \right)
\]

\subsubsection{Subclass II$_{2}$ of superintegrable systems }
\[
A(\xi)=\xi, \qquad B(\eta)=\eta
\]
\begin{equation}\label{eq:FGfgVb2}
\begin{array}{ll}
 \displaystyle F(\eta)=\frac{\kappa }{{\sqrt{\eta }}} + \lambda, &
\displaystyle G(\eta)=3\,\kappa\,{\sqrt{\eta }}  + \lambda\,\eta +
\frac{\mu }{{\sqrt{\eta }}} + \nu \\ \\
 \displaystyle f(\eta)=\frac{k }{{\sqrt{\eta }}} + \ell, &
 \displaystyle g(\eta)=3\,k\,{\sqrt{\eta }}  + \ell\,\eta
+  \frac{m }{{\sqrt{\eta }}} + n
\end{array}
\end{equation}
\[
\begin{array}{ll}
ds^2=g(\xi,\eta)\, d\xi\, d\eta, &
g(\xi,\eta)=\xi\, F(\eta)+G(\eta)\\
\\
\displaystyle H=\frac{p_{\xi}
p_{\eta}}{g(\xi,\eta)}+V(\xi,\eta)\,, &\displaystyle
V(\xi,\eta)=\frac{w(\xi,\eta)}{g(\xi,\eta)}\,, \quad
w(\xi,\eta)=\xi\, f(\eta)+g(\eta)
\end{array}
\]
The other integral of motion is:
\[
\begin{array}{rl}
\displaystyle A = p^2_\xi-
  \frac{2\,{p_{\xi} }\,{p_{\eta} }\,
     \int F(\eta )\,d\eta }{g(\xi ,\eta )}  -
  \frac{2\,\left( \xi \,f(\eta )+ g(\eta )
       \right) \,\int F(\eta )\,d\eta }{g(\xi ,
     \eta )} +  2\,\int f(\eta )\,d\eta
\end{array}
\]
We introduce the functions:
\begin{equation}\label{eq:FGfgtildeVb2}
\begin{array}{l}
\displaystyle \widetilde{F}(u)= \frac{\,\lambda\, u^4 }{128}
+\frac{\,\kappa\,u^3 }{16} +\frac{\,\nu\, u^2 }{16}+
  \frac{\,\mu\, u\ }{4}

 \\
\\
  \displaystyle
\widetilde{G}(v)=-
  \frac{\,\lambda\,v^4 }{128}+\frac{\,\kappa\,v^3 }{16}  +
  \frac{\,\mu\,v }{4} -
  \frac{\,\nu\,v^2 }{16}\\
\\
\displaystyle \widetilde{f}(u)=\frac{\,\ell\, u^4 }{128}
+\frac{\,k\,u^3 }{16} +\frac{\,n\, u^2 }{16}+
  \frac{\,m\, u\ }{4}
\\   \\
     \displaystyle
\widetilde{g}(v)=-
  \frac{\,\ell\,v^4 }{128}+\frac{\,k\,v^3 }{16}  +
  \frac{\,m\,v }{4} -
  \frac{\,n\,v^2 }{16}
\end{array}
\end{equation}
The second integral of motion is:
\[
\begin{array}{rl}
\displaystyle  B=& \displaystyle  p^2_X+p^2_Y - 2 p_X p _Y \,
\frac{\widetilde{F}(X+Y)-\widetilde{G}(X-Y)}{\widetilde{F}(X+Y)+\widetilde{G}(X-Y)}+\\
&\displaystyle   +
4 \frac{\widetilde{f}(X+Y) \widetilde{G}(X-Y)-
\widetilde{g}(X-Y)\widetilde{F}(X+Y)}{\widetilde{F}(X+Y)+\widetilde{G}(X-Y)}
\end{array}
\]
where
\[
X=2 \sqrt{\xi}, \quad p_X= \sqrt{\xi}\, p_\xi, \quad Y=2
\sqrt{\eta}, \quad p_Y=\sqrt{ \eta}\, p_\eta
\]

The constants of the Poisson algebra are:
\[
\begin{array}{l}
\alpha=0 \quad\beta=0 \quad \gamma=0 \quad
\delta=4\,(\ell - \lambda\, H) \qquad
 \epsilon=0\quad
\zeta=8 {\left(\kappa  H - k \right)}^2
\\
a=-\, 6 \qquad d=\, 8\,(\nu\, H -n )\quad
z=  -8\, \left(\kappa\,H-k\right) \left( \mu H -m \right)  -2   {\,
\left(\nu  H -n  \right)}^2
\end{array}
\]
\[
K=8\,\left( \lambda\,H - \ell \right)  \left( \mu H -m \right) ^2-
16\, \left(\kappa\,H - k)  (\mu\ H -m\right) \left( \nu\, H -n\right)
\]

\subsubsection{Subclass II$_{3}$ of superintegrable systems }
\[
A(\xi)={\xi}^2, \qquad B(\eta)={\eta}^2
\]
\begin{equation}\label{eq:FGfgVb3}
\begin{array}{ll}
 \displaystyle F(\eta)=\lambda\, \eta + \frac{\kappa }{{\eta }^3} , &
\displaystyle G(\eta)=\nu + \frac{\mu }{{\eta }^2}  \\  \\

 \displaystyle f(\eta)=\ell\, \eta + \frac{k}{{\eta }^3}, &
 \displaystyle g(\eta)=n + \frac{m}{{\eta }^2}
\end{array}
\end{equation}
\[
\begin{array}{ll}
ds^2=g(\xi,\eta)\, d\xi\, d\eta, &
g(\xi,\eta)=\xi\, F(\eta)+G(\eta)\\
\\
\displaystyle H=\frac{p_{\xi}
p_{\eta}}{g(\xi,\eta)}+V(\xi,\eta)\,, &\displaystyle
V(\xi,\eta)=\frac{w(\xi,\eta)}{g(\xi,\eta)}\,, \quad
w(\xi,\eta)=\xi\, f(\eta)+g(\eta)
\end{array}
\]
The other integral of motion is:
\[
\begin{array}{rl}
\displaystyle A = p^2_\xi -
  \frac{2\,{p_{\xi} }\,{p_{\eta} }\,
     \int F(\eta )\,d\eta }{g(\xi ,\eta )}  -
  \frac{2\,\left( \xi \,f(\eta )+ g(\eta )
       \right) \,\int F(\eta )\,d\eta }{g(\xi ,
     \eta )} +  2\,\int f(\eta )\,d\eta
\end{array}
\]
We introduce the functions:
\begin{equation}\label{eq:FGfgtildeVb3}
\begin{array}{l}
\displaystyle \widetilde{F}(u)=\lambda\,e^{2\,u}  + \nu\, e^u,
\quad
  \displaystyle
\widetilde{G}(v)=\kappa\, e^{2\,v}  +\mu\, e^v\\
\displaystyle \widetilde{f}(u)=\ell\,e^{2\,u}  + n\, e^u, \quad
     \displaystyle
\widetilde{g}(v)=k\, e^{2\,v}  +m\, e^v
\end{array}
\end{equation}
The second integral of motion is:
\[
\begin{array}{rl}
\displaystyle  B=& \displaystyle  p^2_X+p^2_Y - 2 p_X p _Y \,
\frac{\widetilde{F}(X+Y)-\widetilde{G}(X-Y)}{\widetilde{F}(X+Y)+\widetilde{G}(X-Y)}+\\
&\displaystyle   +
4 \frac{\widetilde{f}(X+Y) \widetilde{G}(X-Y)-
\widetilde{g}(X-Y)\widetilde{F}(X+Y)}{\widetilde{F}(X+Y)+\widetilde{G}(X-Y)}
\end{array}
\]
where
\[
X= \ln \, \xi,   \quad p_X= \xi\,p_\xi, \quad Y=\ln\, \eta, \quad p_Y=\eta\, p_\eta
\]
The constants of the Poisson algebra are:
\[
\begin{array}{l}
\alpha=8, \quad\beta=0, \quad \gamma=0, \quad
\delta=0, \quad
 \epsilon=0,\quad
\zeta=32\, \left (
\kappa\,H -k \right) \left(\lambda\,H-\ell\right ),
\\
a=0, \quad d=0, \quad
z=32\, \left( \mu\,H-m\right)\left( \nu\, H-n
\right)
\end{array}
\]
\[
K=64\, \left(\lambda\, H-\ell\right){\left(\mu  H -m\right)}^2
-64 \left( \kappa\, H -k\right) {\left(\nu  H -n \right)}^2
\]

All the above superintegrable systems generally are defined on
manifolds which have neither constant curvature  nor are they
surfaces of revolution. All the known superintegrable systems are
defined on  manifolds of constant curvature or on surfaces of
revolution. Therefore we have proved that there are new
superintegrable systems, which have not yet been studied.

\section{Superintegrable systems corresponding to Koenigs essential forms}
\label{sec:Koenigs}

\begin{table}[p]
\[
\begin{array}{|c|c|c|c|c|c|c|}
\hline \mbox{Class} &    \kappa  &  \lambda &  \mu  & \nu  &
\mbox{\begin{tabular}{c}Essential\\
from \cite{Darboux} \end{tabular}}
&\mbox{\begin{tabular}{c}Classes\\
from \cite{Kress} \end{tabular}}\\
\hline\hline I_1  &16 A_2            &16 A_3            &-A_0                         &4 A_1                       &VII.4  & [3,2]\\
\hline\hline I_2  &A_1               &-8 A_3            &-A_0                         &-2 A_2                      &VII.2&[21,2] \\
\hline\hline I_3  &2(A_2+A_3)        &A_2-A_3           &-2(A_0+A_1)                  &-A_0+A_1                    &VII.1&[111,1] \\
\hline\hline II_1 &                  &                  &                             &                            &&[0,11]      \\
\hline\hline II_2 &2 \sqrt{2} A_1    &16 A_0            &\sqrt{2} A_3                 &4 A_2                       &VII.5 &[3,11]\\
\hline\hline II_3 &2 (A_1+\imag A_0) &2 (\imag A_0-A_1) &-\frac{1}{2} (A_3+\imag A_2) &\frac{1}{2} (A_3-\imag A_2) &VII.3&[21,0] \\
\hline
\end{array}
\]
\caption{\label{tab:Koenigs} \sf Essential forms of Table VII in
Ref. \cite{Darboux} and equivalence classes of ref. \cite{Kress} }
\end{table}

\begin{description}

\item[\fbox{Class I${}_1$} ] Using the coordinate transformation
\[
{\xi} ={\frac{1}{2}\,x^2}, \,  p_{\xi}=\frac{p_x}{x},\, {\eta} =
{\frac{1}{2}\,y^2},\,   p_{\eta}=\frac{p_y}{y}
\]
the metric of the Class I${}_2$ superintegrable systems is reduced
to the metric of the essential form   VII.4 \cite[vol IV,
p.385]{Darboux}, if
\[
\kappa=16 A_2,\quad \lambda=16 A_3,\quad \mu=-A_0,\quad \nu=4A_1
\]
The corresponding superintegrable system  (using the coordinates
of Ref. \cite{Darboux}) is given by the Hamiltonian:
\[
H=\frac{p_x p_y + w(x,y)}{g(x,y)}
\]
\[
\begin{array}{rl}
g(x,y)=&A_0 \left[\frac{1}{(x+y)^2}-\frac{1}{(x-y)^2}\right]+
A_1 \left[{(x+y)^2}-{(x-y)^2}\right]+\\
+&A_2 \left[{(x+y)^4}-{(x-y)^4}\right]
+A_3 \left[{(x+y)^6}-{(x-y)^6}\right]
\end{array}
\]
\[
\begin{array}{rl}
w(x,y)=&a_0 \left[\frac{1}{(x+y)^2}-\frac{1}{(x-y)^2}\right]+
a_1 \left[{(x+y)^2}-{(x-y)^2}\right]+\\
+&a_2 \left[{(x+y)^4}-{(x-y)^4}\right]
+a_3 \left[{(x+y)^6}-{(x-y)^6}\right]
\end{array}
\]
where only three of the constants $a_0,\, a_1,\,a_2,\,a_3$ are
independent, i.e. we can put one among them equal to zero. Using
relations (51) we have that in $x, y$ coordinates the other
integral of motion is:
\[
\begin{array}{rl}
A(x,y)=& \frac{1}{2} p^2_x + \frac{1}{2} p^2_y - {p_x} {p_y}
\frac{\tilde \Phi (x,\, y) - \tilde \Psi (x,\, y)}{\tilde \Phi (x,\, y) + \tilde \Psi (x,\, y)}\\
+&2 \frac{\tilde \phi (x,\, y)\tilde \Psi (x,\, y) - \tilde \psi
(x,\, y) \tilde \Phi (x,\, y)}{\tilde \Phi (x,\, y) + \tilde \Psi
(x,\, y)}
\end{array}
\]
where
\[
\begin{array}{l}
\tilde \Phi (x, y) = A_0 \frac{1}{(x + y)^2} + A_1 (x + y)^2 + A_2
(x + y)^4 + A_3 (x + y)^6 \\
\tilde \Psi (x, y) = - A_0 \frac{1}{(x - y)^2} - A_1 (x - y)^2 -
A_2 (x - y)^4 - A_3 (x - y)^6\\
\tilde \phi (x, y) = a_0 \frac{1}{(x + y)^2} + a_1 (x + y)^2 + a_2
(x + y)^4 + a_3 (x + y)^6 \\
\tilde \psi (x, y) = - a_0 \frac{1}{(x - y)^2} - a_1 (x - y)^2 -
a_2 (x - y)^4 - a_3 (x - y)^6
\end{array}
\]

while using relations (50) we have that in $x, y$ coordinates the
second integral of motion is:

\[
\begin{array}{rl}
B(x,y)=& \frac{1}{x^2} p^2_x + \frac{1}{y^2} p^2_y - 2\,
\frac{ {p_x}{p_y}}{x\,y} \frac{ \Phi (x,\, y) - \Psi (x,\, y)}{\Phi (x,\, y) + \Psi (x,\, y)}\\
+&4 \frac{\phi (x,\, y)\Psi (x,\, y) - \psi (x,\, y) \Phi (x,\,
y)}{x\,\,y\,\,\Phi (x,\, y) + \Psi (x,\, y)}
\end{array}
\]

where

\[
\begin{array}{rl}
\Phi(x,\, y) =& \frac{1}{2} {A_1}\,\left[ {\left( x + y \right)
}^2 - {\left( x - y \right) }^2 \right] + {A_2}\,\left[ {\left( x
+ y \right) }^4 - {\left( x - y \right) }^4 \right] + \\
+&{A_3}\,\Big(\left[ {\left( x + y \right) }^6 - {\left( x - y
\right) }^6 \right] + \\
+& (x+y)^4 (x-y)^2 - (x+y)^2 (x-y)^4 \Big)
\end{array}
\]
\[
\begin{array}{rl}
\Psi(x,\, y) =& {A_0}\,\left[ \frac{1}{(x+y)^2} -
\frac{1}{(x-y)^2} \right] + \frac {1}{2}{A_1}\,\left[ {\left( x +
y \right) }^2 - {\left( x - y \right) }^2 \right] - \\
-&{A_3}\,\Big((x+y)^4 (x-y)^2 - (x+y)^2 (x-y)^4 \Big)
\end{array}
\]
\[
\begin{array}{rl}
\phi(x,\, y) =& \frac{1}{2} {a_1}\,\left[ {\left( x + y \right)
}^2 - {\left( x - y \right) }^2 \right] + {a_2}\,\left[ {\left( x
+ y \right) }^4 - {\left( x - y \right) }^4 \right] + \\
+&{a_3}\,\Big(\left[ {\left( x + y \right) }^6 - {\left( x - y
\right) }^6 \right] + \\
+& (x+y)^4 (x-y)^2 - (x+y)^2 (x-y)^4 \Big)
\end{array}
\]
\[
\begin{array}{rl}
\psi(x,\, y) =& {a_0}\,\left[ \frac{1}{(x+y)^2} -
\frac{1}{(x-y)^2} \right] + \frac {1}{2}{a_1}\,\left[ {\left( x +
y \right) }^2 - {\left( x - y \right) }^2 \right] - \\
-&{a_3}\,\Big((x+y)^4 (x-y)^2 - (x+y)^2 (x-y)^4 \Big)
\end{array}
\]

\item[\fbox{Class I${}_2$} ] Using the coordinate transformation
\[
{\xi} = {- \frac{1}{2} \cos(2 x)}, \quad p_{\xi}= \frac{p_x}{\sin
(2x)}\,, \quad {\eta} = {- \frac{1}{2} \cos(2 y)}\,, \quad
p_{\eta}= \frac{p_y}{\sin (2y)}\,,
\]
the metric of the Class I${}_2$ superintegrable systems is reduced
to the metric of the essential form   VII.4 \cite[vol IV,
p.385]{Darboux}, if
\[
\kappa = A_1, \quad \lambda = -8 A_3,\quad \mu = -A_0,\quad \nu =
-2 A_2
\]
The corresponding superintegrable system  (using the coordinates of ref \cite{Darboux}) is given by the Hamiltonian:
\[
H=\frac{p_x p_y + w(x,y)}{g(x,y)}
\]
where
\[
\begin{array}{rl}
g(x,y)=&\displaystyle A_0 \left[\frac{1}{\sin^2(x+y)}-\frac{1}{\sin^2(x-y)}\right]+
A_1 \left[\frac{1}{\cos^2(x+y)}-\frac{1}{\cos^2(x-y)}\right]+\\
+&A_2 \left[{\cos 2(x+y)}-{\cos 2(x-y)}\right]
+A_3 \left[{\cos 4(x+y)}-{\cos 4(x-y)}\right]
\end{array}
\]
\[
\begin{array}{rl}
w(x,y)=&\displaystyle a_0 \left[\frac{1}{\sin^2(x+y)}-\frac{1}{\sin^2(x-y)}\right]+
a_1 \left[\frac{1}{\cos^2(x+y)}-\frac{1}{\cos^2(x-y)}\right]+\\
+&a_2 \left[{\cos 2(x+y)}-{\cos 2(x-y)}\right]
+a_3 \left[{\cos 4(x+y)}-{\cos 4(x-y)}\right]
\end{array}
\]

where only three of the constants $a_0,\, a_1,\,a_2,\,a_3$ are
independent, i.e. we can put one among them equal to zero. Using
relations (\ref{eq:FGfgtildeVa2}) we have that in $x, y$
coordinates the other integral of motion is:

\[
\begin{array}{rl}
A(x,\,y)=& \frac{1}{4} cot^2(2 x) p^2_x + \frac{1}{4} cot^2(2 y)
p^2_y - \frac{p_x p_y}{2\, tan(2 x)\, tan(2 y)} \frac{\tilde \Phi(x,
y) - \tilde \Psi(x, y)}
{\tilde \Phi(x, y) + \tilde \Psi(x, y)} + \\
+&\frac{1}{tan(2 x)\, tan(2 y)} \frac{\tilde \phi (x,\, y)\tilde
\Psi (x,\, y) - \tilde \psi (x,\, y) \tilde \Phi (x,\, y)}{\tilde
\Phi (x,\, y) + \tilde \Psi (x,\, y)}
\end{array}
\]
where
\[
\begin{array}{rl}
\tilde \Phi(x, y)= A_2 [cos2(x + y)-cos2(x - y)] + A_3 [cos4(x +
y)-cos4(x - y)]
\end{array}
\]
\[
\begin{array}{l} \tilde \Psi(x, y)= {A_0}\,\left[
\frac{1}{sin^2(x+y)} - \frac{1}{sin^2(x-y)} \right] +
{A_1}\,\left[ \frac{1}{cos^2(x+y)} - \frac{1}{cos^2(x-y)} \right]
\end{array}
\]
\[
\begin{array}{rl}
\tilde \phi(x, y)= a_2 [cos2(x + y)-cos2(x - y)] + a_3 [cos4(x +
y)-cos4(x - y)]
\end{array}
\]
\[
\begin{array}{l} \tilde \psi(x, y)= {a_0}\,\left[
\frac{1}{sin^2(x+y)} - \frac{1}{sin^2(x-y)} \right] +
{a_1}\,\left[ \frac{1}{cos^2(x+y)} - \frac{1}{cos^2(x-y)} \right]
\end{array}
\]
while using relations (\ref{eq:FGfgVa2}) we have that in $x, y$
coordinates the second integral of motion is:
\[
\begin{array}{rl}
B(x,\,y)=& \frac{1}{sin^2(2 x)} p^2_x + \frac{1}{sin^2(2 y)} p^2_y -
2\,\frac{p_x p_y}{2\, sin(2 x)\, sin(2 y)} \frac{
\Phi(x, y) - \Psi(x, y)}{\Phi(x, y) + \Psi(x, y)} + \\
+&4\,\frac{1}{sin(2 x)\, sin(2 y)} \frac{\phi (x,\, y)\, \Psi
(x,\, y) - \psi (x,\, y)\, \Phi (x,\, y)}{\Phi (x,\, y) + \Psi
(x,\, y)}
\end{array}
\]
where
\[
\begin{array}{rl}
\Phi(x, y)=&{A_1}\,\left[ \frac{1}{cos^2(x+y)} -
\frac{1}{cos^2(x-y)} \right] + \frac{1}{2}{A_2}\,\left[
cos2(x+y)-cos2(x-y)\right] + \\
+&4\,A_3\,cos^2(x + y)\,cos^2(x - y)\left[
cos2(x+y)-cos2(x-y)\right]
\end{array}
\]
\[
\begin{array}{rl}
\Psi(x, y)=&{A_0}\,\left[ \frac{1}{sin^2(x+y)} -
\frac{1}{sin^2(x-y)} \right] + \frac{1}{2}{A_2}\,\left[
cos2(x+y)-cos2(x-y)\right] - \\
-&4\,A_3\,sin^2(x + y)\,sin^2(x - y)\left[
cos2(x+y)-cos2(x-y)\right]
\end{array}
\]
\[
\begin{array}{rl}
\phi(x, y)=&{a_1}\,\left[ \frac{1}{cos^2(x+y)} -
\frac{1}{cos^2(x-y)} \right] + \frac{1}{2}{a_2}\,\left[
cos2(x+y)-cos2(x-y)\right] + \\
+&4\,a_3\,cos^2(x + y)\,cos^2(x - y)\left[
cos2(x+y)-cos2(x-y)\right]
\end{array}
\]
\[
\begin{array}{rl}
\psi(x, y)=&{a_0}\,\left[ \frac{1}{sin^2(x+y)} -
\frac{1}{sin^2(x-y)} \right] + \frac{1}{2}{a_2}\,\left[
cos2(x+y)-cos2(x-y)\right] - \\
-&4\,a_3\,sin^2(x + y)\,sin^2(x - y)\left[
cos2(x+y)-cos2(x-y)\right]
\end{array}
\]

\item[\fbox{Class I${}_3$} ]
Using the coordinate transformation
\[
{\xi}=\ln\left(\frac{\frac{\wp(x)-\wp(\omega_1)}{\Delta}-1}
{\frac{\wp(x)-\wp(\omega_1)}{\Delta}+1}\right)+\frac{1}{2}
\ln\left(\frac{\frac{\wp(\omega_1)-\wp(\omega_2)}{\Delta}-1}
{\frac{\wp(\omega_1)-\wp(\omega_2)}{\Delta}+1}\right),\quad
p_{\xi}=\frac{\wp(2x)-\wp(\omega_1)}{2 \Delta}p_x
\]
\[
{\eta}=\ln\left(\frac{\frac{\wp(y)-\wp(\omega_1)}{\Delta}-1}
{\frac{\wp(xy)-\wp(\omega_1)}{\Delta}+1}\right)+\frac{1}{2}
\ln\left(\frac{\frac{\wp(\omega_1)-\wp(\omega_2)}{\Delta}-1}
{\frac{\wp(\omega_1)-\wp(\omega_2)}{\Delta}+1}\right),\quad
p_{\eta}=\frac{\wp(2y)-\wp(\omega_1)}{2 \Delta}p_y
\]
where
\[
\Delta^2 = {(\wp(\omega _1)-\wp(\omega _2))}{(\wp(\omega
_1)-\wp(\omega _3))}
\]
the metric of the Class I${}_3$ superintegrable systems is reduced to the metric of the essential form   VII.1 \cite[vol IV, p.385]{Darboux}, if
\[
\kappa = 2 (A_2 + A_3), \quad \lambda = A_2 - A_3,\quad \mu = -2
(A_0 + A_1),\quad \nu = -A_0 + A_1
\]
The corresponding superintegrable system  (using the coordinates of ref \cite{Darboux}) is given by the Hamiltonian:
\[
H=\frac{p_x p_y + w(x,y)}{g(x,y)}
\]
where
\[
\begin{array}{rl}
g(x,y) = &A_0 \left(\wp (x+y)-\wp (x-y)\right)
+  A_1 \left(\wp (x+y+\omega_1)-\wp (x-y + \omega_1)\right)\\
+ & A_2 (\wp (x+y+\omega_2)-\wp (x-y + \omega_2))\\
+ & A_3 (\wp (x+y+\omega_3)-\wp (x-y + \omega_3))\\
\end{array}
\]
\[
\begin{array}{rl}
w(x,y) = &a_0 \left(\wp (x+y)-\wp (x-y)\right)
+  a_1 \left(\wp (x+y+\omega_1)-\wp (x-y + \omega_1)\right)\\
+ & a_2 (\wp (x+y+\omega_2)-\wp (x-y + \omega_2))\\
+ & a_3 (\wp (x+y+\omega_3)-\wp (x-y + \omega_3))\\
\end{array}
\]
where only three of the constants $a_0,\, a_1,\,a_2,\,a_3$ are
independent, i.e. we can put one among them equal to zero. Using
relations (\ref{eq:FGfgVa3}) we have that in $x, y$ coordinates
the other integral of motion is:
\[
\begin{array}{rl}
A(x, y)=&\frac{1}{4 {\Delta}^2} (\wp(2x)-\wp(\omega_1)) p^2_x +
\frac{1}{4 {\Delta}^2} (\wp(2y)-\wp(\omega_1)) p^2_y - \\
-&\frac{\sqrt{\wp(2x)-\wp(\omega_1)}\,
\sqrt{\wp(2y)-\wp(\omega_1)}\,p_x\,p_y}{2\,{\Delta}^2}\,
\frac{\Phi(x,\,y) - \Psi(x,\,y)}{\Phi(x,\,y) + \Psi(x,\,y)} + \\
+&\frac{\sqrt{\wp(2x)-\wp(\omega_1)}\,
\sqrt{\wp(2y)-\wp(\omega_1)}}{{\Delta}^2}\,\frac{\phi(x,\,y)\,\Psi(x,\,y)
- \psi(x,\,y)\,\Phi(x,\,y)}{\Phi(x,\,y) + \Psi(x,\,y)}
\end{array}
\]
where
\[
\begin{array}{rl}
\Phi(x,\,y)=&A_2 (\wp (x+y+\omega_2)-\wp (x-y + \omega_2)) + \\
+&A_3 (\wp (x+y+\omega_3)-\wp (x-y + \omega_3))
\end{array}
\]
\[
\begin{array}{rl}
\Psi(x,\,y)=&A_0 \left(\wp (x+y)-\wp (x-y)\right) + \\
+&A_1 \left(\wp (x+y+\omega_1)-\wp (x-y + \omega_1)\right)
\end{array}
\]
\[
\begin{array}{rl}
\phi(x,\,y)=&a_2 (\wp (x+y+\omega_2)-\wp (x-y + \omega_2)) + \\
+&a_3 (\wp (x+y+\omega_3)-\wp (x-y + \omega_3))
\end{array}
\]
\[
\begin{array}{rl}
\psi(x,\,y)=&a_0 \left(\wp (x+y)-\wp (x-y)\right) + \\
+&a_1 \left(\wp (x+y+\omega_1)-\wp (x-y + \omega_1)\right)
\end{array}
\]
while using relations (\ref{eq:FGfgtildeVa3}) we have that in $x,
y$ coordinates the second integral of motion is:
\[
\begin{array}{rl}
B(x,\,y)=&\frac{1}{\left(\wp (\omega_3)-\wp
(\omega_1)\right)\,\left(\wp (\omega_2)-\wp
(\omega_3)\right)}(\wp(2x)-\wp(\omega_3)) p^2_x + \\
+&\frac{1}{\left(\wp (\omega_3)-\wp (\omega_1)\right)\,\left(\wp
(\omega_2)-\wp (\omega_3)\right)}(\wp(2y)-\wp(\omega_3)) p^2_y -
\\
-&2\,\frac{\sqrt{\wp(2x)-\wp(\omega_3)}\,\sqrt{\wp(2y)-\wp(\omega_3)}\,p_x\,p_y}{\left(\wp
(\omega_3)-\wp (\omega_1)\right)\,\left(\wp (\omega_2)-\wp
(\omega_3)\right)}\,\frac{\tilde \Phi(x,\,y) - \tilde
\Psi(x,\,y)}{\tilde \Phi(x,\,y) + \tilde \Psi(x,\,y)} + \\
+&4\,\frac{\sqrt{\wp(2x)-\wp(\omega_3)}\,\sqrt{\wp(2y)-\wp(\omega_3)}}{\left(\wp
(\omega_3)-\wp (\omega_1)\right)\,\left(\wp (\omega_2)-\wp
(\omega_3)\right)}\,\frac{\tilde \phi(x,\,y)\,\tilde\Psi(x,\,y) -
\tilde \psi(x,\,y)\,\Phi(x,\,y)}{\tilde \Phi(x,\,y) + \tilde
\Psi(x,\,y)}
\end{array}
\]
where
\[
\begin{array}{rl}
\tilde \Phi(x,\,y)=&A_1 \left(\wp (x+y+\omega_1)-\wp (x-y + \omega_1)\right) + \\
+& A_2 (\wp (x+y+\omega_2)-\wp (x-y + \omega_2)) - \\
-&\frac{1}{2} \left( A_0 + A_1 + A_2 + A_3
\right)\frac{\sqrt{\wp(2x)-\wp(\omega_3)}\,\sqrt{\wp(2y)-\wp(\omega_3)}}{\left(\wp
(\omega_3)-\wp (\omega_1)\right)\,\left(\wp (\omega_2)-\wp
(\omega_3)\right)}
\end{array}
\]
\[
\begin{array}{rl}
\tilde \Psi(x,\,y)=&A_0 \left(\wp (x+y)-\wp (x-y)\right) + \\
+& A_3 (\wp (x+y+\omega_3)-\wp (x-y + \omega_3)) + \\
+&\frac{1}{2} \left( A_0 + A_1 + A_2 + A_3
\right)\frac{\sqrt{\wp(2x)-\wp(\omega_3)}\,\sqrt{\wp(2y)-\wp(\omega_3)}}{\left(\wp
(\omega_3)-\wp (\omega_1)\right)\,\left(\wp (\omega_2)-\wp
(\omega_3)\right)}
\end{array}
\]
\[
\begin{array}{rl}
\tilde \phi(x,\,y)=&a_1 \left(\wp (x+y+\omega_1)-\wp (x-y + \omega_1)\right) + \\
+&a_2 (\wp (x+y+\omega_2)-\wp (x-y + \omega_2)) - \\
-&\frac{1}{2} \left( a_0 + a_1 + a_2 + a_3
\right)\frac{\sqrt{\wp(2x)-\wp(\omega_3)}\,\sqrt{\wp(2y)-\wp(\omega_3)}}{\left(\wp
(\omega_3)-\wp (\omega_1)\right)\,\left(\wp (\omega_2)-\wp
(\omega_3)\right)}
\end{array}
\]
\[
\begin{array}{rl}
\tilde \psi(x,\,y)=&a_0 \left(\wp (x+y)-\wp (x-y)\right) + \\
+&a_3 (\wp (x+y+\omega_3)-\wp (x-y + \omega_3)) + \\
+&\frac{1}{2} \left( a_0 + a_1 + a_2 + a_3
\right)\frac{\sqrt{\wp(2x)-\wp(\omega_3)}\,\sqrt{\wp(2y)-\wp(\omega_3)}}{\left(\wp
(\omega_3)-\wp (\omega_1)\right)\,\left(\wp (\omega_2)-\wp
(\omega_3)\right)}
\end{array}
\]

\item[\fbox{Class II${}_1$} ] This case  is not covered by Table VII of Koenigs \cite[Vol. IV, p. 385]{Darboux}. This  class corresponds to the Kress\cite{Kress} equivalence class $[0,11]$  of the nondegenerate superintegrable systems $E_{11},\, E_{20}$  of ref \cite{KaKrPogo01}.

The Hamiltonian is
\[
H= \frac{p_\xi p_\eta   +  k \xi \eta  + \ell \xi +m \eta +n
}{\kappa \xi \eta  + \lambda \xi +\mu \eta +\nu}
\]
the integrals of motion are
\[
\begin{array}{rl}
A=&\displaystyle p_\xi^2- 2 p_\xi p_\eta \frac{\frac{\kappa}{2} \eta^2 + \lambda \eta}{\kappa \xi \eta  +
\lambda \xi +\mu \eta +\nu}+ 2 \left(\frac{k \eta^2}{2}  + \ell \eta\right)-\\
-&\displaystyle 2 \frac{\left(\frac{\kappa}{2} \eta^2 + \lambda
\eta\right)\left(k \xi \eta  + \ell \xi +m \eta +n\right)}{\kappa
\xi \eta + \lambda \xi +\mu \eta +\nu}
\end{array}
\]
\[
\begin{array}{rl}
B=&\displaystyle p_\xi^2 +p_\eta^2 - 2 p_\xi p_\eta \frac{\frac{\kappa}{2} \left(\xi^2+\eta^2 \right) + \lambda \eta+\mu\xi}{\kappa \xi \eta  + \lambda \xi +\mu \eta +\nu}+ 2 \left(\frac{k}{2} \left(\xi^2+\eta^2 \right) + \ell \eta+m\xi\right)-\\
-&\displaystyle 2 \frac{\left(\frac{\kappa}{2} \left(\xi^2+\eta^2
\right) + \lambda \eta+\mu\xi\right)\left(k \xi \eta  + \ell \xi
+m \eta +n\right)}{\kappa \xi \eta  + \lambda \xi +\mu \eta +\nu}
\end{array}
\]

The case VI${}_6$ in Table VI\cite[Vol. IV, p.384]{Darboux} Koenigs  studied separately the cases where $\kappa=0$ and $\kappa\ne0,\; \lambda=\mu=0$.

\item[\fbox{Class II${}_2$} ] Using the coordinate transformation
\[
{\xi} = {\frac{1}{2}\,x^2}, \quad p_{\xi}= \frac{p_x}{x},\quad
 {\eta} = {\frac{1}{2}\,y^2}, \quad p_{\eta}= \frac{p_y}{y}
\]
the metric of the Class II${}_2$ superintegrable systems is reduced to the metric of the essential form   VII.5 \cite[vol IV, p.385]{Darboux}, if
\[
\kappa = 2 \sqrt{2} A_1,\quad \lambda = 16 A_0, \quad \mu =
\sqrt{2} A_3, \quad \nu = 4 A_2
\]
The corresponding superintegrable system  (using the coordinates of ref \cite{Darboux}) is given by the Hamiltonian:
\[
H=\frac{p_x p_y + w(x,y)}{g(x,y)}
\]
where
\[
\begin{array}{rl}
g(x,y)&= A_0 \left[  (x+y)^4-(x-y)^4 \right]+A_1 \left[  (x+y)^3-(x-y)^3 \right]+\\
&+A_2 \left[  (x+y)^2-(x-y)^2 \right]+A_3 \left[  (x+y)-(x-y) \right]
\end{array}
\]

\[
\begin{array}{rl}
w(x,y)&= a_0 \left[  (x+y)^4-(x-y)^4 \right]+a_1 \left[  (x+y)^3-(x-y)^3 \right]+\\
&+a_2 \left[  (x+y)^2-(x-y)^2 \right]+a_3 \left[  (x+y)-(x-y) \right]
\end{array}
\]
where only three of the constants $a_0,\, a_1,\,a_2,\,a_3$ are
independent, i.e. we can put one among them equal to zero. Using
relations (\ref{eq:FGfgtildeVb2}) we have that in $x, y$
coordinates the other integral of motion is:
\[
\begin{array}{rl}
A(x,\,y)=\frac{1}{2} p^2_x + \frac{1}{2} p^2_y -
p_x\,p_y\,\frac{\tilde\Phi(x,\,y) -
\tilde\Psi(x,\,y)}{\tilde\Phi(x,\,y) + \tilde\Psi(x,\,y)} + 2
\frac{\tilde \phi(x,\,y) \tilde \Psi(x,\,y) - \tilde \psi \tilde
\Phi(x,\,y)}{\tilde\Phi(x,\,y) + \tilde\Psi(x,\,y)}
\end{array}
\]
where
\[
\begin{array}{rl}
\tilde\Phi(x,\,y)=A_0 (x+y)^4 + A_1 (x+y)^3 + A_2 (x+y)^2 + A_3
(x+y)
\end{array}
\]
\[
\begin{array}{rl}
\tilde\Psi(x,\,y)=-A_0 (x-y)^4 - A_1 (x-y)^3 - A_2 (x-y)^2 - A_3
(x-y)
\end{array}
\]
\[
\begin{array}{rl}
\tilde\phi(x,\,y)=a_0 (x+y)^4 + a_1 (x+y)^3 + a_2 (x+y)^2 + a_3
(x+y)
\end{array}
\]
\[
\begin{array}{rl}
\tilde\psi(x,\,y)=-a_0 (x-y)^4 - a_1 (x-y)^3 - a_2 (x-y)^2 - a_3
(x-y)
\end{array}
\]
while using relations (\ref{eq:FGfgVb2}) we have that in $x, y$
coordinates the second integral of motion is:
\[
\begin{array}{rl}
B(x,\,y)=&\frac{1}{y^2}\,p^2_y -
2\,\frac{p_x\,p_y}{g(x,\,y)}\,\int\Phi(x) dx -
2\,\frac{w(x,\,y)}{g(x,\,y)}\,\int\Phi(x) dx + \,\int\phi(x) dx
\end{array}
\]
where
\[
\begin{array}{rl}
\Phi(x) = 16 A_0 x + 4 A_1\\
\phi(x) = 16\, a_0 x + 4\, a_1
\end{array}
\]

\item[\fbox{Class II${}_3$} ] Using the coordinate transformation
\[
{\xi} = -{\frac{1}{2}\,cos(2 y)}, \quad p_{\xi}= \frac{1}{sin(2
y)} p_y,\quad
 {\eta} = {e^{2 \imag x}}, \quad p_{\eta}= -\frac{\imag e^{-2 \imag
 x}}{2}\,p_x
\]
the metric of the Class II${}_3$ superintegrable systems is
reduced to the metric of the essential form   VII.3 \cite[vol IV,
p.385]{Darboux}, if
\[
\kappa = 2 (A_1 + \imag A_0), \quad \lambda = 2 (\imag A_0 - A_1),
\quad \mu = - \frac{1}{2} (A_3 + \imag A_2), \quad \nu =
\frac{1}{2} (A_3 - \imag A_2)
\]
where
\[
\begin{array}{rl}
g(x,y)&= A_0 \left[  \sin 4  (x+y)-\sin 4 (x-y) \right]+A_1 \left[  \cos 4 (x+y)-\cos 4 (x-y) \right]+\\
&+A_2 \left[  \sin 2 (x+y)-\sin 2(x-y) \right]+A_3 \left[  \cos 2(x+y)-\sin 2(x-y) \right]
\end{array}
\]
\[
\begin{array}{rl}
w(x,y)&= a_0 \left[  \sin 4  (x+y)-\sin 4 (x-y) \right]+a_1 \left[  \cos 4 (x+y)-\cos 4 (x-y) \right]+\\
&+a_2 \left[  \sin 2 (x+y)-\sin 2(x-y) \right]+a_3 \left[  \cos 2(x+y)-\sin 2(x-y) \right]
\end{array}
\]

where only three of the constants $a_0,\, a_1,\,a_2,\,a_3$ are
independent, i.e. we can put one among them equal to zero. Using
relations (\ref{eq:FGfgtildeVb3}) we have that in $x, y$
coordinates the other integral of motion is:
\[
\begin{array}{rl}
A(x,\,y)=&- \frac{1}{4} p^2_x + \frac{1}{4} cot^2(2 y) p^2_y +
\frac{p_x\,p_y}{2\, \imag \,tan(2 y)} \frac{\tilde\Phi(x,\,y) -
\tilde\Psi(x,\,y)}{\tilde\Phi(x,\,y) + \tilde\Psi(x,\,y)} - \\
-&\frac{1}{2\, \imag \,tan(2 y)} \frac{\tilde\phi(x,\,y) \tilde
\Psi(x,\,y) - \tilde\psi(x,\,y)
\tilde\Phi(x,\,y)}{\tilde\Phi(x,\,y) + \tilde\Psi(x,\,y)}
\end{array}
\]
where
\[
\begin{array}{rl}
\tilde \Phi(x,\, y)=(A_0 + \imag \, A_1)\, e^{4\,\imag\,x} sin(4
y) + (A_2 + \imag \, A_3)\, e^{2\,\imag \, x} sin(2 y)
\end{array}
\]
\[
\begin{array}{rl}
\tilde \Psi(x,\, y)=(A_0 - \imag \, A_1)\, e^{-4\,\imag\,x} sin(4
y) + (A_2 - \imag \, A_3)\, e^{-2\,\imag \, x} sin(2 y)
\end{array}
\]
\[
\begin{array}{rl}
\tilde \phi(x,\, y)=(a_0 + \imag \, a_1)\, e^{4\,\imag\,x} sin(4
y) + (a_2 + \imag \, a_3)\, e^{2\,\imag \, x} sin(2 y)
\end{array}
\]
\[
\begin{array}{rl}
\tilde \psi(x,\, y)=(a_0 - \imag \, a_1)\, e^{-4\,\imag\,x} sin(4
y) + (a_2 - \imag \, a_3)\, e^{-2\,\imag \, x} sin(2 y)
\end{array}
\]
while using relations (\ref{eq:FGfgVb3})we have that in $x, y$
coordinates the second integral of motion is:
\[
\begin{array}{rl}
B(x,\,y)=&\frac{1}{sin^2(2 y)} p^2_y - 2
\frac{p_x\,p_y}{g(x,\,y)} \int{\Phi(x)\,dx} - \\
-&2 \frac{w(x,\,y)}{g(x,\,y)} \int{\Phi(x)\,dx} + 2
\int{\phi(x)\,dx}
\end{array}
\]
where
\[
\begin{array}{rl}
\Phi(x,\,y)=-8 A_0cos(4 x) + 8 A_1 sin(4 x)
\end{array}
\]
\[
\begin{array}{rl}
\phi(x,\,y)=-8 a_0cos(4 x) + 8 a_1 sin(4 x)
\end{array}
\]

\end{description}

\section{Superintegrable potentials on a surface of
revolution with two quadratic integrals of
motion}\label{sec:RevolQQ}

A manifold which is described by a metric of the form
\[
ds^2= g(x+y) dx dy   \quad \mbox{or} \quad ds^2= g(x-y) dx dy
\]
 is called  a surface of revolution.

The above condition is possible only for a specific choice of the
parameters $\kappa,\, \lambda, \, \mu$ and $\nu$. In many cases
the superintegrable systems can be calculated by using the general
forms which are studied in Section \ref{sec:Classification}. The
general forms of these systems by revolution in many instances are
given by the formulas:
\[
H= \frac{p_{\xi}\, p_{\eta}+ f(\xi+\eta)+g(\xi-\eta)}{F(\xi+\eta)}
\quad \mbox{or} \quad H= \frac{p_{\xi}\, p_{\eta}+
f(\xi+\eta)+g(\xi-\eta)}{G(\xi-\eta)}
\]
and
\[
H= \frac{p_{X}\, p_{Y}+ \widetilde{f}(X+Y)+
\widetilde{g}(X-Y)}{\widetilde{F}(X+Y)}\quad \mbox{or}
 \quad H=
\frac{p_{X}\, p_{Y}+ \widetilde{f}(X+Y)+
\widetilde{g}(X-Y)}{\widetilde{G}(X-Y)}
\]
But we must notice that the Liouville or the Lie coordinates are
not always the appropriate ones for concluding whether a surface
is a surface by revolution. Among the parameters $\kappa,\,
\lambda, \, \mu$ and $\nu$ the surfaces of revolution are determined
by two independent parameters.  In Table \ref{tab:RevolutionQQ}
these special values of the parameters $\kappa,\, \lambda, \, \mu$
and $\nu$ are shown. We notice the corresponding potentials given
in references \cite{KaKrMiWin03} and  \cite{KaKrWin02}. This
classification scheme shows that there is the case $R_{11}$ which
is not given in the above references and it is a new not known
superintegrable system. A detailed description of the
superintegrable systems on surfaces of revolution are given
further in the present section.

\begin{table}[p]
\[
\begin{array}{|c|c|c|c|c|c|c|c|}
\hline &\mbox{Class} &    \kappa  &  \lambda &  \mu  & \nu  &
\mbox{\begin{tabular}{c}Potentials\\
by revolution\\
from ref \cite{KaKrMiWin03}\end{tabular}}& \mbox{\begin{tabular}{c}Potentials\\
by revolution\\
from ref \cite{KaKrWin02}\end{tabular}}\\
\hline\hline R_1   & I_1 &0       &0       &\cdot    &\cdot  & 2[A] & \\
\hline       R_2   &     &\cdot   &0       &0        &\cdot  &      & (1) \\
\hline\hline R_3   &I_2  &0       &0       &\cdot    &\cdot  & 2[B] & \\
\hline       R_4   &     &0       &\cdot   &0        &\cdot  & 3[B] & \\
\hline       R_5   &     &\cdot   &0       &\cdot    &0      & 4[A] & \\
\hline       R_6   &     &-\mu    &\cdot   &\cdot    &0      & 2[C] & \\
\hline\hline R_7   &I_3  &0       &0       &\cdot    &\cdot  & 4[B] & \\
\hline       R_8   &     &-\mu    &\nu     &\cdot    &\cdot  & 4[C] & \\
\hline\hline R_9   &II_1 &0       &\mu     &\cdot    &\cdot  &      & (2) \\
\hline       R_{10}&     &\cdot   &0       &0        &\cdot  & 3[A] & \\
\hline\hline R_{11}&II_2 &0       &\cdot   &0        &\cdot  &        &\mbox{new} \\
\hline\hline R_{12}&II_3 &\cdot   &-\kappa &-\nu     &\cdot  & 3[D] & \\
\hline       R_{13}&     &-\mu    &0       &\cdot    &0      & 3[C] & \\
\hline
\end{array}
\]
\caption{ \label{tab:RevolutionQQ} \sf Potentials by revolution with
two quadratic integrals of motion}
\end{table}

\begin{description}

\item[\fbox{$R_1$:  Class I${}_1$  $\kappa=0,\, \lambda=0$} ]  The
form of the Hamiltonian in Liouville coordinates is given by
\[
H=\frac{{p_{\eta} }\,{p_{\xi} }}
   {\nu  + \frac{\mu }{{\left( \xi - \eta \right) }^2}} +
  \frac{k\,\left( \eta  + \xi  \right) +
     4\,\ell\,{\left( \eta  + \xi  \right) }^2 -
     \ell\,{\left( \xi - \eta \right) }^2
     + \frac{m}{{\left( \xi - \eta \right) }^2} + n }
     {\nu  + \frac{\mu }{{\left( \xi - \eta \right) }^2}}
\]
By the coordinate transformation
\[
\xi = \frac{v + \imag u}{2}\,,\, \eta = \frac{v - \imag
u}{2}\,,\,p_{\xi} = p_v - \imag p_u\,,\,p_{\eta} = pv + \imag p_u
\]
and putting $\mu=-1,\, \nu=1,$ the Hamiltonian $2[A]$ of reference
\cite{KaKrMiWin03} is obtained
\[
H = \frac{u^2}{u^2 + 1}\left( p^2_u + p^2_v + k\,v + 4\,\ell \left(
\frac{1}{4}\,u^2 + v^2 \right) - \frac{m + n}{u^2} \right) + n
\]

\item[\fbox{$R_{2}$: Class I${}_1$ $\lambda=0,\, \mu=0$ }]  The
form of the Hamiltonian in Liouville coordinates is given by
\[
H=\frac{{p_{\eta} }\,{p_{\xi} }}{ \kappa \,\left( \eta  + \xi
\right) + \nu} +\frac{k\,\left( \eta  + \xi \right) +
4\,\ell\,{\left( \eta + \xi  \right) }^2  - \ell\,{\left( \xi -
\eta \right) }^2 + \frac{m}{{\left( \xi - \eta \right) }^2} + n }{
\kappa \,\left( \eta + \xi  \right) + \nu }
\]
By the coordinate transformation
\[
\xi = u + \imag v,\,\,\, \eta = u - \imag v\,,\,p_{\xi} =
\frac{1}{2}\, ( p_u - \imag p_v )\,,\,p_{\eta} = \frac{1}{2}\, (
pu + \imag p_v )
\]
and putting $\kappa=1/2,\, \nu=0$ the Hamiltonian $(1)$ of reference
\cite{KaKrWin02} is obtained
\[
H = \frac{p^2_u +p^2_v}{4\,u} + \frac{16\,\ell (4 u^2 + v^2)}{4\,u}
+ \frac{n}{u} - \frac{\frac{m}{4}}{u\,v^2}\,+ 2\,k
\]

\item[\fbox{$R_3$: Class I${}_2$  $\kappa=0,\, \lambda=0,$}]  The
form of the Hamiltonian in Liouville coordinates is given by
\[
H=\frac{{p_{\eta} }\,{p_{\xi} }}{\nu  + \frac{\mu }{{\left( \eta
- \xi \right) }^2}} +
   \frac{\frac{k}{{\left( \eta  + \xi  \right) }^2} +
     \ell\,{\left( \eta  + \xi  \right) }^2  - \ell\,{\left( \xi - \eta \right) }^2
     + \frac{m}{{\left( \xi - \eta \right) }^2}
     + n  }{\nu + \frac{\mu }{{\left( \xi - \eta \right) }^2}}
\]
By the coordinate transformation
\[
\xi = \frac{v + \imag u}{2}\,,\, \eta = \frac{v - \imag
u}{2}\,,\,p_{\xi} = p_v - \imag p_u\,,\,p_{\eta} = pv + \imag p_u
\]
and putting $\mu=-1,\, \nu=1$, the Hamiltonian $2[B]$ of reference
\cite{KaKrMiWin03} is obtained
\[
H = \frac{u^2}{u^2 + 1}\left( p^2_u + p^2_v + \ell\,( u^2 + v^2 ) -
\frac{m + n}{u^2} + \frac{k}{v^2} \right) + n
\]

\item[\fbox{$R_{4}$:  Class I${}_2$ $\kappa=0,\, \mu=0$ }]  The
form of the Hamiltonian in coordinates $(X,Y)$ is given by
\[
H= \frac{p_{X}\, p_{Y}+ 4  \ell e^{2(X+Y)}+n
e^{(X+Y)}+k\,\frac{e^{X+Y}}{(1+e^{X+Y})^2}+m\,\frac{e^{X+Y}}{(-1+e^{X+Y})^2}}{4
\lambda e^{2(X+Y)}+\nu e^{(X+Y)} }
\]
By the coordinate transformation

\[
X= - \frac{1}{2}\,\ln \left(\frac{4}{{\left( \imag v + u
\right)}^2}\right), \, Y= - \frac{1}{2}\,\ln \left(-
\frac{4}{{\left( v + \imag\, u \right)}^2}\right),
\]
\[
p_X=\frac{1}{2} \left( (\imag v + u) p_{u} + (v - \imag u) p_{v}
\right),\,p_Y=-\frac{1}{2} \left( (\imag v - u) p_{u} - (v + \imag
u) p_{v} \right)
\]
and putting $\lambda=1,\,\nu=4$ the Hamiltonian $3[B]$ of
reference \cite{KaKrMiWin03} is obtained
\[
H = \frac{p^2_u + p^2_v + \frac{k}{u^2} - \frac{m}{v^2} + n -
4\,l}{4 + u^2 + v^2}\,+ l
\]
\item[\fbox{$R_5$:  Class I${}_2$  $\lambda=0,\,\nu=0$}]  The form
of the Hamiltonian in coordinates $(X,Y)$ is given by
\[
H= \frac{p_{X}\, p_{Y}+ 4  \ell e^{2(X+Y)}+n
e^{(X+Y)}+k\,\frac{e^{X+Y}}{(1+e^{X+Y})^2}+m\,\frac{e^{X+Y}}{(-1+e^{X+Y})^2}}
{\kappa\,\frac{e^{X+Y}}{(1+e^{X+Y})^2}+\mu\,\frac{e^{X+Y}}{(-1+e^{X+Y})^2}}
\]
By the coordinate transformation
\[
X=\ln\left( \frac{1}{2} (x-\imag y) \right),\,Y=\ln\left(
\frac{1}{2} (x+\imag y) \right)
\]
\[
p_X=\frac{1}{2}\left( (x-\imag y)p_x + (\imag x + y)p_y \right),\,
p_Y=\frac{1}{2}\left( (x+\imag y)p_x + (y-\imag x)p_y \right)
\]
Putting $\kappa=\frac{2-\alpha}{4},\, \mu=\frac{2+\alpha}{4}$, the
Hamiltonian $4[A]$ of reference \cite{KaKrMiWin03} is obtained
\[
H = - \frac{4 x^2 y^2\,\left(p^2_x + p^2_y + n + \frac{1}{4}\,((2 +
\alpha) k + (\alpha - 2) m) \left( \frac{1}{x^2} +
\frac{1}{y^2}\right) + \ell\,(x^2 + y^2) \right)}{y^2 (\alpha -2) +
x^2 (2 + \alpha)} \,+ k + m
\]

\item[\fbox{$R_6$: Class I${}_2$  $\kappa =-\mu,\,\nu = 0$}] The
form of the Hamiltonian in Liouville coordinates is given by
\[
\begin{array}{rl}
H=&\displaystyle \frac{{p_{\eta} }\,{p_{\xi} }} {- \frac{\mu
}{{\left( \eta + \xi \right) }^2} + \lambda \,{\left( \eta + \xi
\right) }^2 - \lambda \,{\left( \xi - \eta \right) }^2 + \frac{\mu
}{{\left(
\xi - \eta \right) }^2} } +\\
 + &  \displaystyle\frac{ \frac{k}{{\left( \eta + \xi
\right) }^2} + \ell\,{\left( \eta  + \xi  \right) }^2 -
\ell\,{\left( \xi - \eta \right) }^2 + \frac{m}{{\left( \xi - \eta
\right) }^2} + n }{- \frac{\mu }{{\left( \eta  + \xi \right) }^2}
+ \lambda \,{\left( \eta + \xi \right) }^2 - \lambda \,{\left( \xi
- \eta \right) }^2 + \frac{\mu }{{\left( \xi - \eta \right) }^2} }
\end{array}
\]
By the coordinate transformation
\[
\xi= u + \imag v,\,\eta = u - \imag v,\,p_{\xi}= \frac{1}{2} (p_u
- \imag p_v),\,p_{\eta}= \frac{1}{2} (p_u + \imag p_v)
\]
and putting $\lambda=\frac{1}{16},\, \mu=-1$, the Hamiltonian
$2[C]$ of reference \cite{KaKrMiWin03} is obtained
\[
H = \frac{p^2_u + p^2_v + 4\,n + \frac{k - 16\,\ell}{u^2} -
\frac{16\,\ell + m}{v^2}}{u^2 + v^2 + \frac{1}{u^2} + \frac{1}{v^2}}
+ 16\,\ell
\]

\item[\fbox{$R_7$: Class I${}_3$  $\kappa = 0,\, \lambda = 0$}]
The form of the Hamiltonian in Liouville coordinates is given by
\[
\begin{array}{rl}
H=&\frac{{\left( e^{\eta  - \xi } - e^{\xi - \eta } \right)
}^2\,{p_{\eta} }\, {p_{\xi} }}{\nu\, \left( e^{\eta  - \xi } +
e^{\xi -
\eta } \right) + \mu }\, + \\
+&\, \frac{{\left( e^{\eta  - \xi } - e^{\xi - \eta } \right) }^2}
{\mu  + \left( e^{\eta  - \xi } + e^{\xi - \eta } \right) \,\nu
}\, \Big( k\frac{e^{2\,\left( \eta + \xi \right) }\,}{{\left( -1 +
e^{2\,\left( \eta  + \xi  \right) } \right) }^2} +\ell
\frac{e^{\eta + \xi }\,\left( 1 + e^{2\,\left( \eta  + \xi
\right) } \right) \,} {{\left( -1 + e^{2\,\left( \eta  + \xi
\right) } \right)
}^2} + \\
+&m \frac{e^{2\,\left( \xi - \eta \right) }\,} {{\left( -1 +
e^{2\,\left( \xi - \eta \right) } \right) }^2} + n\frac{e^{\xi -
\eta }\,\left( 1 + e^{2\,\left( \xi - \eta \right) } \right) \,}
{{\left( -1 + e^{2\,\left(\xi - \eta \right) } \right) }^2} \Big)
\end{array}
\]
By the coordinate transformation
\[
\xi= v + \imag u,\,\eta = v - \imag u,\,p_{\xi}= \frac{1}{2} (p_v
- \imag p_u),\,p_{\eta}= \frac{1}{2} (p_v + \imag p_u)
\]
and putting $\mu=\alpha,\, \nu=1,$ the Hamiltonian $4[B]$ of
reference \cite{KaKrMiWin03} is obtained
\[
H = - \frac{sin^2(2\,u) \left( p^2_u + p^2_v + \frac{\frac{1}{4}\,(k
+ 2\,\ell)}{sinh^2(v)} + \frac{\frac{1}{4}\, (2\,\ell -
k)}{cosh^2(v)}\right) + n\,\alpha - m}{2\,cos(2\,u) + \alpha} + n
\]

\item[\fbox{$R_8$: Class I${}_3$ $\kappa = - \mu,\, \lambda =
\nu$}] The form of the Hamiltonian in Liouville coordinates is
given by
\[
\begin{array}{rl}
H=&\frac{{p_{\eta} }\,{p_{\xi} }} {\frac{\mu + \nu\, \left(e^{\xi
- \eta} + e^{\eta - \xi } \right)} {{\left(e^{\xi - \eta } -
e^{\eta - \xi } \right)}^2} + \frac{-\mu + \nu\, \left(e^{\xi +
\eta} + e^{{-(\xi + \eta) }} \right)\, } {{\left(e^{\xi + \eta} -
e^{-(\xi
+ \eta)} \right)}^2}} + \\
+&\frac{\,k\,\frac{e^{2\,\left( \eta  + \xi \right) }} {{\left( -1
+ e^{2\,\left( \eta  + \xi  \right) } \right) }^2} +
\,\ell\,\frac{e^{\eta + \xi }\,\left( 1 + e^{2\,\left( \eta + \xi
\right) } \right) } {{\left( -1 + e^{2\,\left( \eta  + \xi \right)
} \right) }^2} + \,m\,\frac{e^{2\,\left( -\eta  + \xi \right) }}
{{\left( -1 + e^{2\,\left( -\eta  + \xi  \right) } \right) }^2} +
 \,n\,\frac{e^{-\eta + \xi }\,\left( 1 + e^{2\,\left( -\eta + \xi
\right) } \right)} {{\left( -1 + e^{2\,\left( -\eta + \xi \right)
} \right) }^2}}{\frac{\mu + \nu\, \left(e^{\xi - \eta} + e^{\eta -
\xi } \right)} {{\left(e^{\xi - \eta } - e^{\eta - \xi }
\right)}^2} + \frac{-\mu + \nu\, \left(e^{\xi + \eta} + e^{{-(\xi
+ \eta) }} \right)\, } {{\left(e^{\xi + \eta} - e^{-(\xi + \eta)}
\right)}^2}}
\end{array}
\]
By the coordinate transformation
\[
{{\xi } = {{\mbox{arcsinh}}(\tan(\phi - \imag\,\omega))}},\, {{\eta
} = {{\mbox{arcsinh}}(\tan(\phi + \imag\,\omega))}}
\]
\[
p_\xi = \frac{1}{2}\,\cos^2(\phi - \imag\,\omega)\,\sqrt{1 +
\tan^2(\phi - \imag\,\omega)}\,\,(p_\phi + \imag p_\omega)
\]
\[
p_\eta = \frac{1}{2}\,\cos^2(\phi + \imag\,\omega)\,\sqrt{1 +
\tan^2(\phi + \imag\,\omega)}\,\,(p_\phi - \imag p_\omega)
\]
and putting   $\mu=\alpha,\, \nu=1,$ the Hamiltonian $4[C]$ of
reference \cite{KaKrMiWin03} is obtained
\[
\begin{array}{ll}
H = - \frac{p^2_\phi + p^2_\omega }{\frac{\alpha +
2}{\sinh^2(2\,\omega)} + \frac{\alpha - 2}{\sin^2(2\,\phi)}} -
\frac{\frac{c_1}{\cos^2(\phi)} + \frac{c_2}{\cosh^2(\omega)} +
c_3\,\left(\frac{1}{\sin^2(\phi)} - \frac{1}{\sinh^2(\omega)}
\right)}{\frac{\alpha + 2}{\sinh^2(2\,\omega)} + \frac{\alpha -
2}{\sin^2(2\,\phi)}} + \frac{m + 2\,n - k - 2\,\ell}{2\,\alpha}
\end{array}
\]
where $c_1 = -\frac{m (2 + \alpha) + (k + 2\,\ell)\,(-2 + \alpha)
+ 2\,n\,(2 - 3\,\alpha)}{8\,\alpha}\,,\,c_2 = - \frac{k\,(-2 +
\alpha) + (m + 2\,n)\,(2 + \alpha) - 2\,\ell\,(2 +
3\,\alpha)}{8\,\alpha}$ and $c_3 = \frac{(m + 2\,n)\,(-2 + \alpha)
+ (k+2\,\ell)\,(2 + \alpha)}{8\,\alpha}$

\item[\fbox{$R_9$:  Class II${}_1$ $\kappa = 0,\,\lambda = \mu$}]
The form of the Hamiltonian in Liouville coordinates is given by
\[
H=\frac{{p_{\eta} }\,{p_{\xi} }}{\mu \,\left( \eta  + \xi  \right)
+ \nu } +
  \frac{ k\,\eta \,\xi+ \ell\,\xi+ m\,\eta + n }{\mu \,\left( \eta  + \xi  \right) + \nu }
\]
By the coordinate transformation
\[
\xi= u + \imag v,\,\eta = u - \imag v,\,p_{\xi}= \frac{1}{2} (p_u
- \imag p_v),\,p_{\eta}= \frac{1}{2} (p_u + \imag p_v)
\]
and putting   $\mu=1/2,\, \nu=0,$ the Hamiltonian $(2)$ of
reference \cite{KaKrWin02} is obtained
\[
H = \frac{p^2_u + p^2_v}{4\,u} + \frac{n}{u} + \frac{\imag\,(\ell -
m)\,v}{u} + \frac{k\,(u^2 + v^2)}{u} + \ell + m
\]

\item[\fbox{$R_{10}$:  Class II${}_1$ $\lambda = 0,\,\mu = 0$}]
The form of the Hamiltonian in Liouville coordinates is given by
\[
H=\frac{{p_{\eta} }\,{p_{\xi} }}{\nu  + \eta \,\kappa \,\xi } +
\frac{n + m\,\eta  + \ell\,\xi  + k\,\eta \,\xi }{\nu  + \eta
\,\kappa \,\xi }
\]
by the coordinate transformation
\[
\xi= u + \imag v,\,\eta = u - \imag v,\,p_{\xi}= \frac{1}{2} (p_u
- \imag p_v),\,p_{\eta}= \frac{1}{2} (p_u + \imag p_v)
\]
and putting   $\kappa=\frac{1}{4},\, \nu=1,$ the Hamiltonian
$3[A]$ of reference \cite{KaKrMiWin03} is obtained
\[
H = \frac{p^2_u + p^2_v + 4\,(\ell + m) u + 4\,\imag\,(\ell - m) +
4\,(n - 4\,k)}{4 + u^2 + v^2} + 4\,k
\]

\item[\fbox{$R_{11}$:  Class II${}_2$ $\kappa = 0,\,\mu = 0$}] The
form of the Hamiltonian in Liouville coordinates is given by
\[
H= \frac{{p_{\eta} }\,{p_{\xi} }}{\lambda \,(\eta + \xi) + \nu } +
\frac{n + \frac{m + \ell\,{\sqrt{\eta }}\,\left( \eta  + \xi
\right)  + k\,\left( 3\,\eta  + \xi  \right) }{{\sqrt{\eta
}}}}{\lambda \,(\eta + \xi) + \nu }
\]
by the coordinate transformation
\[
\xi= u + \imag v,\,\eta = u - \imag v,\,p_{\xi}= \frac{1}{2} (p_u
- \imag p_v),\,p_{\eta}= \frac{1}{2} (p_u + \imag p_v)
\]

and putting $\lambda = \frac{1}{2},\, \nu = 0$, we have
\[
H=\frac{p^2_u + p^2_v}{4\,u} + \frac{m + \left( n + 2\,\ell\,u
\right) \, {\sqrt{u - \imag \,v}} + k\,\left( 4\,u - 2\,\imag \,v
\right) }{u\, {\sqrt{u - \imag \,v}}}
\]
For this Hamiltonian the additional integrals of motion have the
form
\[
A=- \frac{\imag }{2}\,{X_1}-\frac{K^2}{2} - \frac{2\,{\sqrt{u -
\imag \,v}}\, \left( \frac{m}{2} + \frac{n\,{\sqrt{u - \imag
\,v}}}{2} - \imag \,k\,v \right) }{u}
\]
\[
B={X_2}-\frac{2\,v\,\left( m\,\left( \imag \,u + \frac{v}{2}
\right) + v\,\left( \frac{n\,{\sqrt{u - \imag \,v}}}{2} - \imag
\,k\,v \right)  \right) }{u\, {\sqrt{u - \imag \,v}}}
\]

where $K,\,X_1,\,X_2$ are the three integrals of the free motion
 of reference\cite{KaKrWin02}.
 \[
 K=p_v, \quad X_1= p_u p_v - \frac{v}{2 u}\left(
 p^2_u+p^2_v\right)
 \; \mbox{and}\;
 X_2=p_v \left(v p_u-u p_v\right) - \frac{v^2}{4 u }\left(
 p^2_u+p^2_v\right)
 \]
This system was not included  in reference\cite{KaKrWin02}.
Therefore is a new superintegrable system, which is studied here
for the first time as far as we known.

 \item[\fbox{$R_{12}$:  Class II${}_3$ $\kappa = - \lambda
,\, \mu =  -\nu$}] The form of the Hamiltonian in $(\xi, \eta)$
coordinates is given by
\[\displaystyle
H=\frac{p_{\xi} p_{\eta} + \ell \xi \eta + \frac{k \xi}{{\eta}^3}
+ \frac{m}{{\eta}^2} + n }{\lambda \xi \eta + \frac{\kappa
\xi}{{\eta}^3} + \frac{\mu}{{\eta}^2} + \nu}
\]
By the coordinate transformation
\[
\xi= 2\,\sqrt{u\, v}\,,\,\eta= \imag
\sqrt{\frac{u}{v}}\,\,,\,p_{\xi}= \frac{1}{2}
\left(\sqrt{\frac{u}{v}}\, p_u + \sqrt{\frac{v}{u}}\, p_v
\right)\,,\,p_{\eta}= - \imag \sqrt{u\,v}\,p_u + \imag
\frac{v^{3/2}}{\sqrt{u}}\,p_v )
\]
and putting   $\kappa = \frac{1}{4}\,,\,\lambda=
-\frac{1}{4}\,,\,\mu = \imag\,,\,\nu= - \imag$ the Hamiltonian
$3[D]$ of reference \cite{KaKrMiWin03} is obtained
\[
H = \frac{u^2\,p^2_u - v^2\,p^2_v +2\,(2\,(\ell - k) + \imag\,n )\,u
+ 2\,(2\,(\ell - k) - \imag\,m)\,v -2\,(k + \ell) }{(u + v) (2 + u
-v)} \,+ 2\,(k - \ell)
\]

 \item[\fbox{$R_{13}$:  Class II${}_3$ $\kappa = - \mu, \lambda  = 0
,\, \nu =  0$}]  
 The form
of the Hamiltonian in $(\xi,\eta)$ coordinates is given by
\[\displaystyle
H= \frac{p_{\xi} p_{\eta} + \ell \xi \eta + \frac{k \xi}{{\eta}^3}
+ \frac{m}{{\eta}^2} + n }{\frac{\kappa \xi}{{\eta}^3} +
\frac{\mu}{{\eta}^2}}
\]
By the coordinate transformation
\[
\xi= \frac{v-u}{\sqrt{u\,v}}\,,\,\eta= \frac{2}{\sqrt{u\, v}}
\]
\[
p_{\xi} = - \frac{u \sqrt{u\,v}}{u + v}\,p_u + \frac{v
\sqrt{u\,v}}{u + v}\,p_v\,,\,p_{\eta} = - \frac{1}{2}
\sqrt{u\,v}\left(u p_u + v p_v \right)
\]
and putting   $\kappa = -4,\,\mu = 4$, the Hamiltonian $3[C]$ of
reference \cite{KaKrMiWin03} is obtained
\[
H = \frac{u^2\,p^2_u - v^2\,p^2_v +\frac{k + m}{2}\,(u + v) +2\,n
\frac{u + v}{u\,v} -4\,\ell \,\frac{u^2 - v^2}{u^2\,v^2} }{(u + v)
(2 + u -v)} - \frac{k}{4}
\]

\end{description}

\section{Superintegrable potentials on a mani\-fold with
curvature zero}\label{sec:QQzero}

Let us consider the manifold corresponding to the  metric in the
Liouville coordinates:
\[
ds^2=g(\xi,\eta) \,d\xi\, d\eta
\]
the curvature is defined by:
\begin{equation} \label{eq:CurvatureZero}
K=- \frac{1}{2 g} \frac{\partial^2 }{\partial \xi \partial \eta}
\ln \,g =0
\end{equation}

The above constraint imposes restrictions on the choice of the
parameters $\kappa,\; \lambda,\; \mu$ and $\nu$. In Table
\ref{tab:Curvature0} we can see the possible choices of the values
of the above parameters.

\begin{table}[p]
\[
\begin{array}{|c|c|c|c|c|c|c|c|c|}
\hline & \mbox{Class} &    \kappa  &  \lambda &  \mu  & \nu  &
\mbox{\begin{tabular}{c}Plane\\
potentials\\
from ref \cite{KaKrPogo01}\end{tabular}}
 &\mbox{\begin{tabular}{c}Drach potentials\\
from ref \cite{Ranada97}\end{tabular}}
 & \mbox{\begin{tabular}{c}Potentials\\
from ref \cite{RaSan99}\end{tabular}} \\
\hline\hline F_1& I_1   &0     &0     &0      &\cdot & E_2    & {(a)}\; (r\ne 0)&V^{\alpha} \\
\hline\hline F_2&I_2   &0     &0     &0      &\cdot & E_1    & {(b)}\; (r\ne 0)&V^{b} \\
\hline       F_3&      &0     &\cdot &0      &0     & E_{16} & {(g)}\; (r\ne 0)&V^{c} \\
\hline\hline F_4&II_1  &\cdot &0     & 0     &0     & E_{20} & {(c)}           &V^{d}  \\
\hline       F_5&      &0     &0     &\cdot  &0     & E_{11} & {(e)}\; (r=0)   & \\
\hline\hline F_6&II_2  &0     &0     &0      &\cdot & E_9    & {(e)}\; (r\ne 0)& \\
\hline       F_7&      &0     &0     &\cdot  &0     & E_{10} &                 & \\
\hline\hline F_8&II_3  &0     &0     &\cdot  &\cdot & E_8    & {(f)}           & \\
\hline       F_9&      &0     &0     &\cdot  &\cdot & E_{7}  & {(i)}\; (r\ne 0)& \\
\hline       F_{10}&   &\cdot &\cdot &0      &0     & E_{17} & {(d)} & \\
\hline       F_{11}&   &\cdot &\cdot &0      &0     & E_{19} & {(h)} \; (r\ne 0)& \\
\hline
\end{array}
\]
\caption{\label{tab:Curvature0} \sf Potentials with curvature zero and
two quadratic integrals of motion}
\end{table}

In this category there are three families of potentials :
\begin{itemize}
\item The potentials on the complex $E$ plane corresponding to the
Hamiltonian:
\[
H=p_x^2+p_y^2+ V(x,y)
\]
these potentials are classified in reference \cite{KaMiPogo00} and
finally the final list of potentials are given in reference
\cite{KaKrPogo01}. In this section we follow the enumeration of
the potentials given by this exhaustive list \cite{KaKrPogo01}

\item The Drach potentials corresponding to the Hamiltonian
\[
H= p_x p_y + V(x,y)
\]
these potentials are classified in reference \cite{Ranada97}. In
this section we follow the enumeration of this reference for the
Drach potentials with quadratic integrals of motion.

\item The potentials on the real hyperbolic plane $H_2$
corresponding to the Hamiltonian:
\[
H=p_x^2-p_y^2+ V(x,y)
\]
these potentials are classified in reference \cite{RaSan99}
\end{itemize}
Generally the condition (\ref{eq:CurvatureZero}) restricts the
choices of the constants $\kappa,\, \lambda,\, \mu$ and $\nu$ in
one independent parameter. These values  characterize the
permitted metrics in the classification given in Section
\ref{sec:Classification}. The permitted choices are given by the
following list

\begin{description}
\item[\fbox{$F_1$:  Class I${}_1$ $\kappa=0,\, \lambda=0, \mu=0$}]
The form of the Hamiltonian in Liouville coordinates is given by
\[
H=\frac{{p_{\eta}}\, {p_{\xi}}}{\nu} +
  \frac{ k\, \left(\xi + \eta \right) + 4\,
  \ell\, {\left( \xi + \eta \right)}^2 -
  \ell\, {\left(\xi -\eta \right)}^2 + \frac{m}{{\left( \xi-\eta
  \right)}^2} + n}{\nu}
\]
 By the coordinate transformation:
\[
 {{\xi } = {x + \imag \,y}},\, {{\eta } = {x -
\imag \,y}} ,\,p_{\xi} = \frac{1}{2} \,(p_x - \imag p_y)
,\,p_{\eta} = \frac{1}{2} \,(p_x + \imag p_y)
\]
and putting $\nu=\frac{1} {4}$, the complex plane Hamiltonian
$E_2$ of reference \cite{KaKrPogo01}  is obtained
\[
H=  p^2_x + p^2_y + 16\,\ell\,\left( 4\,x^2 + y^2 \right) + 8\,k\,x
- \frac{m}{y^2} + 4\, n
\]
Also the potential $V^{\alpha}$ of reference \cite{RaSan99}is
generated.

By the coordinate transformation:
\[
{{\xi } = \frac{x}{r}},\, {{\eta } = y},\, p_{\xi} = r p_x
,\,p_{\eta} = p_y \; \mbox{ where } r \mbox{ is a constant}\] and
putting $\nu=r$, the Drach Hamiltonian $(a)\, (r\ne 0)$ of
reference \cite{Ranada97} is obtained:
\[
H= {p_x}\,{p_y} + \frac{k\, \left(x + r\, y \right)}{r^2} +
  \frac{m\, r}{{\left(x - r\, y \right)}^2} +
  \frac{3\, \ell\, \left(x^2 + r^2\, y^2 \right)}{r^3} + \frac{10\, \ell\, x\,
      y}{r^2} + \frac{n}{r}
\]
This system is generated by the two systems of revolution $R_1$
and $R_2$ see Table \ref{tab:Revolution}. In these cases two
parameters among the $\kappa,\, \lambda, \, \mu$ and $\nu$ are
zero. When a third parameter is zero these systems are degenerated
to the system $F_1$ in Table \ref{tab:Curvature0}.

\item[\fbox{$F_2$:  Class I${}_2$ $\kappa=0,\, \lambda=0,\, \mu=0$
}] The form of the Hamiltonian in Liouville coordinates is given
by
\[
H=\frac{{p_{\eta}}\, {p_{\xi}}}{\nu} +
  \frac{ \frac{k}{( \eta + \xi )^2} +
  \ell\, {\left( \xi + \eta \right)}^2 -
  \ell\, {\left(\xi -\eta \right)}^2 + \frac{m}{{\left( \xi-\eta
  \right)}^2} + n}{\nu}
\]
By the coordinate transformation:
\[
 {{\xi } = {x + \imag \,y}},\, {{\eta } = {x -
\imag \,y}} ,\,p_{\xi} = \frac{1}{2} \,(p_x - \imag p_y)
,\,p_{\eta} = \frac{1}{2} \,(p_x + \imag p_y)
\]
and putting $\nu = \frac{1} {4}$, the complex plane Hamiltonian
$E_1$ of reference \cite{KaKrPogo01} and the Hamiltonian $V^{b}$
of reference \cite{RaSan99} are obtained
\[
H = p^2_x + p^2_y + 16\,\ell\, \left( x^2 + y^2 \right) +
\frac{k}{x^2} -  \frac{m}{y^2} +4\,n
\]
By the coordinate transformation:
\[
{{\xi } = \frac{x}{r}},\, {{\eta } = y},\, p_{\xi} = r p_x
,\,p_{\eta} = p_y \; \mbox{ where } r \mbox{ is a constant}\] and
putting $\nu=r$, the Drach  Hamiltonian ${(b)} (r\ne 0)$ of
reference \cite{Ranada97} is obtained:
\[
H= p_x\,p_y + \frac{4\, \ell\, x\, y}{r^2} +
  \frac{m\, r}{{\left(x - r\, y \right)}^2} +
  \frac{k\, r}{{\left(x + r\, y \right)}^2} + \frac{n}{r}
\]

\item[\fbox{$F_3$:  Class I${}_2$  $\kappa=0,\, \mu=0,\, \nu=0$}]
The form of the Hamiltonian in Liouville coordinates is given by
\[
H=\frac{{p_{\eta} }\,{p_{\xi} }} {4\,\lambda \,\xi\,\eta } +
\frac{\frac{k}{{\left( \eta  + \xi  \right) }^2} + \ell\,{\left(
\eta  + \xi  \right) }^2 - \ell\,{\left( \xi - \eta \right) }^2 +
\frac{m}{{\left( \xi - \eta \right) }^2} + n }
 {4\,\lambda \,\xi\,\eta}
\]
By the coordinate transformation:
\[
{{\xi } = {{\sqrt{\frac{x + \imag \,y}{2}}}}}\,,\,
  {{\eta } = {{\sqrt{\frac{x - \imag \,y}{2}}}}}
\]
\[
p_{\xi} =
  \sqrt{2 (x + \imag y)} \,(p_x - \imag p_y)\,,\,p_{\eta} =
  \sqrt{2 (x - \imag y)} \,(p_x + \imag p_y)
\]
and putting $\lambda =1$, the complex plane  Hamiltonian $E_{16}$
of reference \cite{KaKrPogo01}
\[
H = p^2_x + p^2_y + \,\frac{1}{{\sqrt{x^2 + y^2}}} \, \left( \frac
{n} {2} + \frac {\frac {k} {2}}  {x + {\sqrt{x^2 + y^2}}} + \frac
{\frac {m} {2}} {x - {\sqrt{x^2 + y^2}}} \right) + \ell
\]
and the Hamiltonian $V^{c}$ of reference \cite{RaSan99}
\[
H=p^2_x + p^2_y + \frac{\frac{n}{2}}{\sqrt{x^2 + y^2}} +
\frac{-\frac{k + m}{2}\,x}{y^2\,\sqrt{x^2 + y^2}} +
\frac{\frac{k-m}{2}}{y^2}
\]

are obtained

By the coordinate transformation:
\[
{{\xi } = {{\sqrt{\frac{x}{r}}}}}\,,\,{{\eta } =
{{\sqrt{y}}}}\,,\,p_{\xi} = 2 \sqrt{r x}\,p_x\,,\,p_{\eta} = 2
\sqrt{y}\,p_y
\]
and putting $\lambda =r$, the Drach  Hamiltonian ${(g)} (r\ne 0)$
of reference \cite{Ranada97} is obtained:
\[
H=p_x\,p_y + \frac {\frac {n} {4\, \sqrt {r}}} {\sqrt {x\, y}} +
\frac {\frac{(m - k)} {2} \,r} { \left(  x - r\, y  \right )^2} +
  \frac{\frac{k + m}{4}\, \sqrt{r} (x + r\,
          y)} {\sqrt {x\, y}\, \left (x - r\, y \right )^2} +
  \frac {\ell}{r}
\]

\item[\fbox{$F_4$: Class II${}_1$  $\lambda=0,\, \mu=0,\, \nu=0$}]
The form of the Hamiltonian in Liouville coordinates is given by
\[
H= \frac{{p_{\eta} }\,{p_{\xi} }}{\kappa \,\eta \,\xi } +
  \frac{n + m\,\eta  + \left( \ell + k\,\eta  \right) \,\xi }
   {\kappa \,\eta \,\xi }
\]
By the coordinate transformation:
\[
\begin{array}{ll}
{\xi}  = \frac{\sqrt{x + \sqrt{x^2 + y^2}} + \sqrt{x - \sqrt{x^2 +
y^2}}}{\sqrt{2}}\,,\, {\eta } = \frac{\sqrt{x + \sqrt{x^2 + y^2}}
- \sqrt{x - \sqrt{x^2 + y^2}}}{\sqrt{2}}
\end{array}
\]
\[
\begin{array}{ll}
p_{\xi} = \frac{\sqrt{x + \sqrt{x^2 + y^2}} + \sqrt{x - \sqrt{x^2
+ y^2}}}{\sqrt{2}}\,p_x + &\\
+\frac{-x\left( \sqrt{x - \sqrt{x^2 + y^2}} + \sqrt{x + \sqrt{x^2
+ y^2}}\right) + \sqrt{x^2 + y^2}\,\left( - \sqrt{x - \sqrt{x^2 +
y^2}} + \sqrt{x + \sqrt{x^2 + y^2}}\right)}{\sqrt{2}\,y}\,p_y
\end{array}
\]
\[
\begin{array}{rl}
p_\eta = \frac{\sqrt{2}\,\sqrt{x^2 + y^2}}{\sqrt{x - \sqrt{x^2 +
y^2}} + \sqrt{x + \sqrt{x^2 + y^2}}}\,p_x +
\frac{\imag\,\sqrt{2}\,\sqrt{x^2 + y^2}}{\sqrt{x - \sqrt{x^2 +
y^2}} + \sqrt{x + \sqrt{x^2 + y^2}}}\,p_y
\end{array}
\]
and putting $\kappa = 1$, the complex plane Hamiltonian $E_{20}$
of reference \cite{KaKrPogo01} and the Hamiltonian $V^{d}$ of
reference \cite{RaSan99} are obtained
\[
\begin{array}{rl}
H=p^2_x + p^2_y &+\, \frac {1} {\sqrt{x^2 + y^2}}\, \Big( n + \,+
\frac{\ell + m}{\sqrt{2}}\, \sqrt{x + \sqrt{x^2 + y^2}}\,+
    \\&+\,\frac{\ell - m}{\sqrt{2}} \sqrt{x - \sqrt{x^2 + y^2}}
    \Big)\,+ k
\end{array}
\]
 By the
coordinate transformation:
\[
{{\xi } = {2\,{\sqrt{x}}}}\,,\, {{\eta } =
{2\,{\sqrt{y}}}}\,,\,p_{\xi} = \sqrt{x}\, p_x\,,\,p_{\eta} =
\sqrt{y}\,p_y
\]
and putting $\kappa = \frac {1} {4}$, the Drach  Hamiltonian
${(c)}$ of reference \cite{Ranada97} is obtained:
\[
H=p_x\,p_y + \frac{n}{\sqrt {x\, y}} + \frac {2\,
      m}{\sqrt {x}} + \frac{2\, \ell}{\sqrt{y}} + 4\, k
\]

\item[\fbox{$F_5$:  Class II${}_1$ $\kappa=0,\, \lambda=0, \,
\nu=0$} ] The form of the Hamiltonian in Liouville coordinates is
given by
\[
H= \frac{{p_{\eta} }\,{p_{\xi} }+  k\,\eta \,\xi
+\ell\,\xi+m\,\eta +n}{\eta \,\mu }
\]
By the coordinate transformation:
\[
{{\xi } = {x + \imag \,y}}=z,\quad {{\eta } =
    {2\,{\sqrt{x - \imag \,y}} }}=2 \sqrt{\overline{z}}
\]
\[
p_{\xi} = \frac{1}{2}\,(p_x -\imag\,p_y)\,,\,p_{\eta} =
\frac{1}{2}\,\sqrt{x-\imag\,y}\,(p_x + \imag\,p_y)
\]
and putting $\mu = \frac {1} {8}$, the complex plane Hamiltonian
$E_{11}$ of reference \cite{KaKrPogo01} is obtained
\[
H=p^2_x + p^2_y + 8\,k\,z + 4\,\ell\,\frac{z }{{\sqrt{
        {\overline{z}}}}} + \frac{4\, n   }
   {{\sqrt{{\overline{z}}}}}+8\,m
\]
 Using   the coordinate transformation:
\[
{{\xi } = x},\, {{\eta } = {2\,{\sqrt{y}} }}\,,\,p_{\xi} =
p_x\,,\, p_{\eta} = \sqrt{y}\,p_y
\]

and putting $\mu = \frac {1} {2}$, the Drach Hamiltonian ${(e)}
(r=0)$ of reference \cite{Ranada97} is obtained:
\[
H=p_x\,p_y+ \frac{n }
   {{\sqrt{y}}} + 2\,k\,x +
  \ell\frac{  x}
   {{\sqrt{y}}}+2\,m
\]

\item[\fbox{$F_6$:  Class II${}_2$    $\kappa=0,\, \lambda=0,\,
\mu=0$}] The form of the Hamiltonian in Liouville coordinates is
given by
\[
H=\frac{{p_{\eta} }\,{p_{\xi} }}{\nu }+ \frac{k\,\xi }{{\sqrt{\eta
}}\,\nu }  +
  \frac{\ell\,\xi }{\nu } +
  \frac{3\,k\,{\sqrt{\eta }}}{\nu }  + \frac{\ell\,\eta }{\nu }  +
  \frac{m}{{\sqrt{\eta }}\,\nu } + \frac{n}{\nu }
\]
By the coordinate transformation:
\[
{{\xi } = {x + \imag \,y}},\, {{\eta } = {x - \imag
\,y}}\,,\,p_{\xi} = \frac{1}{2} (p_x - \imag\,p_y)\,,\,p_{\eta} =
\frac{1}{2} (p_x + \imag\,p_y)
\]
and putting $\nu = \frac {1} {4}$\,, the complex plane Hamiltonian
$E_9$ of reference \cite{KaKrPogo01} is obtained
\[
H=p^2_x + p^2_y +
  \frac{8\, k\,\left( 2\,x - \imag \,y \right) }
   {{\sqrt{x - \imag \,y}}} + 8\,\ell\,x +
  \frac{4\, m}{{\sqrt{x - \imag \,y}}}+ 4\, n
\]
By the coordinate transformation:
\[
{{\xi } = \frac{x}{r}}\,,\, {{\eta } = y}\,,\,p_{\xi} = r\,
p_x\,,\,p_{\eta} = p_y
\]
and putting $\nu = r$, the Drach Hamiltonian ${(e)} (r \ne 0)$ of
reference \cite{Ranada97} is obtained:
\[
H=p_x\,p_y + \frac {\frac {m}{r}}{\sqrt {y}} +\frac{\ell } {r^2} \,
(x + r \,
    y) + \frac{\frac {k}{r^2}\, (x + 3 \, r \, y)}{\sqrt {y}} + \frac{n}{r}
\]

\item[\fbox{$F_7$:  Class II${}_2$   $\kappa=0,\, \lambda=0,\,
\nu=0$}] The form of the Hamiltonian in Liouville coordinates is
given by
\[
H=\frac{{p_{\eta} }\,{p_{\xi} }\,{\sqrt{\eta }}}
   {\mu } + \frac{k\, \xi}{\mu} + \frac{\ell \, \sqrt {\eta}\, \xi}{\mu} + \frac{3\,
      k\, \eta}{\mu} + \frac{\ell \, (\eta)^{\frac{3}{2}}}{\mu} + \
\frac{m}{\mu} + \frac{n\, \sqrt{\eta}}{\mu}
\]

Using   the coordinate transformation:
\[
{{\xi } = {x + \imag \,y}}=z\,, \quad {{\eta } = - {\frac{{\left(
x - \imag \,y \right) }^2}{2}}}=-\frac{\overline{z}^2}{2}
\]
\[
p_{\xi} = \frac{1}{2}\,(p_x - \imag\,p_y)\,,\,p_{\eta} = -
\frac{1}{2 (x - \imag y)}\,(p_x - \imag\,p_y)
\]
and putting ${{\mu } = - {\frac{\imag}{4\, \sqrt{2}}}}$\,, the
complex plane Hamiltonian $E_{10}$ of reference \cite{KaKrPogo01}
is obtained
\[
H= p^2_x + p^2_y
 -4 \,n \, \overline{z} + 4\, \sqrt{2}\, \imag \, k\, (z - \frac{3}{2} \, \overline{z}^2 ) -
  4\, \ell\, ( z\, \overline{z} - \frac{1}{2} \, \overline{z}^3) +
  4\,\imag\,\sqrt{2}\,\,m
\]
By the coordinate transformation:
\[
{{\xi } = x}\,,\, {{\eta } = {y^2}}\,,\,p_{\xi} = p_x\,,\,p_{\eta}
= \frac{1}{2 y}\,p_y
\]
and putting $\mu=\frac{1}{2}$, the corresponding Drach Hamiltonian
is obtained:
\[
H= p_x\,p_y + 2\,k\,\left( x + 3\,y^2 \right)+
  2\,l\,y\,\left( x + y^2 \right) +  2\,n\,y + 2\,m
\]
This potential was not included in the list of reference
\cite{Ranada97}.

\item[\fbox{ $F_8$: Class II${}_3$ $\kappa=0,\, \lambda=0$}] The
form of the Hamiltonian in Liouville coordinates is given by
\[
H= \frac{{p_{\eta} }\,{p_{\xi} }}{\nu }+ \frac{k\,\xi }{{\eta
}^3\,\nu } +
  \frac{\ell\,\eta \,\xi }{\nu } +
  \frac{m}{{\eta }^2\,\nu } +\frac{n}{\nu }
\]
Using   the coordinate transformation:
\[
{{\xi } = {x + \imag \,y}}=z\,,\, {{\eta } = {x - \imag
\,y}}=\overline{z}\,,\,p_{\xi} = \frac{1}{2}\,(p_x -
\imag\,p_y)\,,\,p_{\eta} = \frac{1}{2}\,(p_x + \imag\,p_y)
\]
and putting $\mu = 0\,,\,\nu = \frac{1}{4}$\,, the complex plane
Hamiltonian $E_8$ of reference \cite{KaKrPogo01} is obtained
\[
H=p^2_x + p^2_y
    + \frac{4\, k\, z}{\overline{z}^3} + \frac{4\, m}{\overline{z}^2}
    + 4\, \ell \, z\, \overline{z} + 4\, n
\]
By the coordinate transformation:
\[
{{\xi } = x}\,,\, {{\eta } = y}\,,\,p_{\xi} = p_x\,,\,p_{\eta} =
p_y
\]
and putting $\nu = 1$, the Drach Hamiltonians  ${(f)}$ of
reference \cite{Ranada97} is obtained:
\[
H=p_x\,p_y + \ell\,x\,y + \frac{m}{y^2}  + \frac{k\,x}{y^3} + n
\]

\item[\fbox{ $F_9$: Class II${}_3$ $\kappa=0,\, \lambda=0 $}] The
form of the Hamiltonian in $(\xi,\eta)$ coordinates is given by
\[
H= \frac{p_\xi\,p_\eta} {\frac{\mu}{{\eta}^2} + \nu} + \frac{n +
\frac{m}{{\eta}^2} + \left( \frac{k}{{\eta}^3} + \ell\,\eta
\right)\,\xi}{\frac{\mu}{{\eta}^2} + \nu}
\]
By the coordinate transformation
\[
\xi = \frac{1}{2}\,(x + \imag y),\,\eta = \frac{x - \imag \,y -
\sqrt{(x - \imag y)^2 + 16\,\mu\,\nu}}{4\,\nu}\,,
\]
\[
p_\xi = p_x - \imag\,p_y\,,\,p_\eta = - \frac{2\,\nu\,(p_x +
\imag\,p_y)\,\sqrt{(x - \imag y)^2 + 16\,\mu\,\nu}}{x - \imag y -
\sqrt{(x - \imag y)^2 + 16\,\mu\,\nu}}
\]
and putting $\mu = - \frac{c}{4}\,,\, \nu = \frac{c}{4}$ the
complex plane Hamiltonian $E_7$ of reference \cite{KaKrPogo01} is
obtained
\[
\begin{array}{rl}
H = &p^2_x + p^2_y + \frac{\frac{-2 (m + n)}{c}
\overline{w}}{\sqrt{\overline{w}^2 - c^2}} + \frac{- c^2 (k +
\ell)w}{\sqrt{\overline{w}^2 - c^2}\,\left(\overline{w} +
\sqrt{\overline{w}^2 -
c^2}\right) } -\\
-&\frac{4\,k}{c^2}\, w\,\overline{w} + \frac{2\,(m - n)}{c}
\end{array}
\]
where $w = x + \imag y\,,\,\overline{w} = x - \imag y$

 Using   the coordinate transformation:
\[
\xi = y\,,\,\eta = \frac{x + \sqrt{x^2 +
4\,\mu\,\nu}}{2\,\nu}\,,\,p_\xi = p_y\,,\,p_\eta =
\frac{2\,\nu\,\sqrt{x^2 + 4\,\mu\,\nu}}{x + \sqrt{x^2 +
4\,\mu\,\nu}}\,p_x
\]
and putting $\nu = \frac{\sqrt{r}}{2}\,,\,\mu =
\frac{\sqrt{r}}{2}$ the Drach Hamiltonian ${(i)}\; (r\ne 0)$ of
reference \cite{Ranada97} is obtained
\[
H = p_x\,p_y + \frac{\ell - k}{r}x y +
\frac{\frac{n-m}{\sqrt{r}}\,x}{\sqrt{x^2 + r}}+\frac{\frac{k +
\ell}{r}(2 x^2 + r)\, y}{\sqrt{x^2 + r}}
\]

 \item[\fbox{$F_{10}$ Class II${}_3$ $\mu=0,\, \nu=0$}] The
form of the Hamiltonian in Liouville coordinates is given by
\[
H=\frac{{p_{\eta} }\,{p_{\xi} }}{\eta \,\lambda \,\xi }+
\frac{k}{{\eta }^4\,\lambda } +\frac{\ell}{\lambda } +
  \frac{m}{{\eta }^3\,\lambda \,\xi } +
  \frac{n}{\eta \,\lambda \,\xi }
\]
Using   the coordinate transformation:
\[
{{\xi } = {2\,{\sqrt{x - \imag \,y}}}}=2
\sqrt{\overline{z}},\qquad {{\eta } = {2\,{\sqrt{x + \imag
\,y}}}}=2 \sqrt{{z}}
\]
\[
p_\xi = \frac{1}{2}\,\sqrt{x - \imag\,y}\,\,(p_x + \imag\,
p_y)\,,\,\,\,p_\eta = \frac{1}{2}\,\sqrt{x + \imag\,y}\,\,(p_x -
\imag\, p_y)
\]
and putting $\kappa = 0\,,\,\lambda = \frac{1}{16}$\,, the complex
plane Hamiltonian $E_{17}$ of reference \cite{KaKrPogo01} is
obtained
\[
H=p^2_x + p^2_y + \frac{k}{{{{z}}}^2} + 16\, \ell +
  \frac{m}{z{\sqrt{z \overline{z}}}} +
  \frac{4\,n}{\,{\sqrt{{z\overline{z}}}}}
\]

Using   the coordinate transformation:
\[
{{\xi } = {2\,{\sqrt{y}}}},\, {{\eta } =
{2\,{\sqrt{x}}}}\,,\,p_\xi = \sqrt{y}\,p_y\,,\,p_\eta =
\sqrt{x}\,p_x
\]
and putting $\kappa = 0,\,\lambda = \frac{1}{4}$\,, the Drach
Hamiltonians ${(d)}$ of reference \cite{Ranada97} is obtained:
\[
H= p_x\,p_y + \frac{k}{4\,x^2} + 4\,\ell +
  \frac{m}{4\,{\sqrt{y}}\,x^{\frac{3}{2}}} +
  \frac{n}{{\sqrt{y}}\,{\sqrt{x}}}
\]

\item[\fbox{$F_{11}$:  Class II${}_3$ $\mu=0,\, \nu=0$}] The form
of the Hamiltonian in $(\xi,\,\eta)$ coordinates is given by
\[\displaystyle
H = \frac{p_\xi\,p_\eta}{\left( \frac{\kappa}{{\eta}^3} +
\lambda\,\eta \right)\,\xi} + \frac{\left( \frac{k}{{\eta}^3} +
l\,\eta \right)\,\xi + \frac{m}{{\eta}^2} + n }{\left(
\frac{\kappa}{{\eta}^3} + \lambda\,\eta \right)\,\xi}
\]
By the coordinate transformation
\[
\xi = 2\,\sqrt{x + \imag\, y}\,,\,\,\eta = \sqrt{x - \imag\, y +
\sqrt{(x - \imag\, y)^2 -4}}
\]
\[
p_\xi = \frac{1}{2}\,(p_x-\imag\, p_y)\,\sqrt{x + \imag\,
y}\,,\,p_\eta = \frac{(p_x + \imag\, p_y)\,\sqrt{(x - \imag\, y)^2
- 4}}{\sqrt{x - \imag\, y + \sqrt{(x - \imag\, y)^2 - 4} }}
\]
and putting $\kappa = -\,\frac{1}{2}\,,\,\,\lambda = \frac{1}{8}$
we have the Hamiltonian in the complex plane $E_{19}$ of reference
\cite{KaKrPogo01}
\[
H = p^2_x + p^2_y + \frac{(k + 4)
\overline{w}}{\sqrt{{\overline{w}}^2 - 4}} + \frac{-
\frac{m}{\sqrt{2}}}{\sqrt{w (\overline{w} + 2)}} +
\frac{\frac{m}{\sqrt{2}}}{\sqrt{w (\overline{w} - 2)}} +4\,\ell - k
\]

By the coordinate transformation:
\[
\xi = 2\,\sqrt{y}\,,\,\eta = 2\,\sqrt{x + \sqrt{x^2 -
r^2}}\,,\,p_\xi = \sqrt{y}\,\,p_y\,,\,p_\eta = \frac{\sqrt{x^2 -
r^2}}{\sqrt{x + \sqrt{x^2 - r^2}}}\,p_x
\]
and putting $\kappa = - 2 r^2,\,\lambda = \frac{1}{8}$ the Drach
Hamiltonian ${(h)} (r\ne 0)$ of reference \cite{Ranada97} is
obtained
\[
H = {p_x} {p_y} + \frac{\frac{4\,n\,r\,+\,m}{4
\sqrt{2}\,\,r}}{\sqrt{y (x - r)}} + \frac{\frac{4\,n\,r\,-\,m}{4
\sqrt{2}\,\,r}}{\sqrt{y (x + r)}} +
\frac{\left(\frac{16\,\ell\,\,r^2 + k}{4\,r^2}
\right)\,x}{\sqrt{x^2 - r^2}} + \frac{16\,\ell\,r^2 - k}{4 r^2}
\]

\end{description}

\section{Superintegrable potentials on a  manifold with constant
curvature}\label{sec:ConstantCurvQQ}

 Let us consider the manifold
corresponding to the metric in the Liouville coordinates:
\[
ds^2=g(\xi,\eta) \,d\xi\, d\eta
\]
the curvature is defined by:
\begin{equation} \label{eq:CurvatureConstant}
K=- \frac{1}{2 g} \frac{\partial^2 }{\partial \xi \partial \eta}
\ln \,g =\mbox{Constant}
\end{equation}
\begin{table}[p]
\[
\begin{array}{|c|c|c|c|c|c|c|c|}
\hline &\mbox{Class} &    \kappa  &  \lambda &  \mu  & \nu  &
\mbox{\begin{tabular}{c}
Potentials\\
on the sphere
(K=1)\\
from Ref \cite{KaKrPogo01}\end{tabular}}&
\mbox{\begin{tabular}{c}Potentials\\
from Ref \cite{RaSan99}\end{tabular}}\\
\hline\hline C_1& I_1  &0     &0     &1/K  &0    & S_1   &  \\
\hline\hline C_2& I_2  &-1/K  &0     &0    &0    & S_{2} &  \\
\hline       C_3&     &-1/K  &0     &1/K  &0    & S_{4} &  \\
\hline\hline C_4&I_3  &0     &0     &4/K  &0    &       & U^{c} \\
\hline       C_5&    &0     &0     &2/K  &1/K  & S_{9} & U^{a}, U^{b} \\
\hline       C_6&     &-2/K  &-1/K  &2/K  &-1/K & S_{7} & U^{e} \\
\hline       C_7&     &-4/K  &0     &4/K  &0    & S_{8} &  \\
\hline
\end{array}
\]
\caption{\label{tab:CurvatureConstant} \sf Potentials with constant
curvature and two quadratic integrals of motion}
\end{table}

The above constraint imposes restrictions on the choice of the
parameters $\kappa,\; \lambda,\; \mu$ and $\nu$. In Table
\ref{tab:CurvatureConstant} we can see the possible choices of the
values of the above parameters.

\begin{description}
\item[\fbox{$C_1$:  Class I${}_1$ $\kappa=0,\, \lambda=0,\,
\mu=1/K,\, \nu=0$}] The form of the Hamiltonian in Liouville
coordinates is given by
\[
\begin{array}{rl}
H=&K\,{p_{\eta} }\,{p_{\xi} }\,{\left( \xi  -\eta \right) }^2 +\\
+& \displaystyle K\,{\left(\xi - \eta \right) }^2\, \left(
k\,\left( \eta + \xi \right)   + 4\,\ell\,{\left( \eta  + \xi
\right) }^2 - \ell\,{\left( \xi - \eta \right) }^2 +
\frac{m}{{\left( \xi - \eta \right) }^2}  + n \right )
\end{array}
\]
For $K=1$ and using the coordinate transformation:
\[
\begin{array}{ll}
 {\xi} = e^{\imag \,\phi }\,\tan (\frac{\theta }{2})\,,
& p_\xi = \frac{\cot (\frac{\theta }{2})\,
    \left( -\imag \,{p_\phi } +
    {p_\theta }\,\sin\theta \right) }
    {2\,e^{\imag \,\phi }}\\
\eta = -e^{\imag \,\phi }\,
    \cot (\frac{\theta }{2})\,,&
p_\eta = \frac{ \tan (\frac{\theta }{2})\,\left( \imag \,{p_\phi } +
    {p_\theta }\,\sin\theta \right) \,
   }{2\,e^{\imag \,\phi }}
\end{array}
\]
the Hamiltonian $S_1$ of reference \cite{KaKrPogo01} in spherical
coordinates $\theta, \,\phi$ is obtained:
\[
H= p^2_\theta+\frac{p^2_\phi}{\sin^2\theta} -8 k e^{3 \imag \,\phi
}\frac{\cos{\theta}}{\sin^3\theta}+16\,\ell\,e^{4\,\imag \,\phi
}\,
 \frac{ \left( 1 + 2\,\cos (2\,\theta ) \right)}{
  {\sin^4\theta }}+ 4\,n\,\frac{e^{ 2\,\imag  \,\phi }}{
  {\sin^2\theta }}+m
\]

\item[\fbox{ $C_2$: Class I${}_2$ $\kappa=-1/K,\, \lambda=0,\,
\mu=0,\, \nu=0$}] The form of the Hamiltonian in Liouville
coordinates is given by
\[
\begin{array}{rl}
H= &- K\,{p_{\eta} }\,{p_{\xi} }\,{\left( \eta  + \xi  \right) }^2
-\\
-&
  K\,{\left( \eta  + \xi  \right) }^2\,
   \left( \frac{k}{{\left( \eta  + \xi  \right) }^2} +
     \ell\,{\left( \eta  + \xi  \right) }^2  -
     \ell\,{\left( \xi - \eta \right) }^2 + \frac{m}{{\left( \xi - \eta \right) }^2}
      + n \right)
\end{array}
\]
For $K=1$ and using the coordinate transformation:
\[
\begin{array}{ll}
{\xi } = {- \frac{1}{2}\,\imag \,e^{\imag \,\phi }\,\cot
(\frac{\theta }{2})}\,, & {{{p_\xi }} = {\frac{\tan (\frac{\theta
}{2})\,\left( {p_\phi } -\,\imag\, {p_\theta }\,\sin\theta \right)
\,} {e^{\imag \,\phi }}}}\\
{\eta } =  {- \frac{1}{2}\,\imag \,e^{\imag \,\phi }\,\tan
(\frac{\theta }{2})}\,, & {{{p_\eta }} = {\frac{\cot (\frac{\theta
}{2})\, \left( {p_\phi } +\,\imag\, {p_\theta }\,\sin\theta \right)
}{e^{\imag \,\phi }}}}
\end{array}
\]

the Hamiltonian $S_2$ of reference \cite{KaKrPogo01} in spherical
coordinates $\theta, \,\phi$ is obtained:
\[
H=p^2_\theta+\frac{p^2_\phi}{\sin^2\theta} -k -\ell\, e^{4\,\imag
\,\phi }\,\csc^2\theta - m\,\sec^2\theta + n\, e^{2\,\imag \,\phi
}\,\csc^2\theta
\]

\item[\fbox{$C_3$:  Class I${}_2$ $\kappa=-1/K,\, \lambda=0,\,
\mu=1/K,\, \nu=0$}] The form of the Hamiltonian in Liouville
coordinates is given by
\[
H=K\,\frac{{p_{\eta} }\,{p_{\xi} }} {\frac{1}{{\left( \xi - \eta
\right) }^2} - \frac{1}{{\left( \eta  + \xi  \right) }^2}} + K\,
\frac{\frac{k}{{\left( \eta  + \xi \right) }^2} + \ell\,{\left(
\eta  + \xi  \right) }^2 - \ell\,{\left( \xi - \eta \right) }^2 +
\frac{m}{{\left( \xi - \eta \right) }^2} + n }{\frac{1} {{\left(
\xi - \eta \right) }^2} - \frac{1}{{\left( \eta + \xi \right)
}^2}}
\]

For $K=1$ and using the coordinate transformation:

\[
\begin{array}{ll}
{{\xi } = {2\,{\sqrt{-\imag \,e^{\imag \,\phi }\,\cot
(\frac{\theta }{2})}}}}\,, & {{{p_\xi }} = {\frac{-\imag \,
{p_\phi } - \,{p_\theta }\,\sin\theta }{{2\,\sqrt{-\imag
\,e^{\imag \,\phi
}\,\cot( \frac{\theta }{2})}}}}}\\
{{\eta } = {2\,{\sqrt{\imag \,e^{\imag \,\phi }\,\tan
(\frac{\theta }{2})}}}}\,, & {{{p_\eta }} = {\frac{-\imag
\,{p_\phi } + {p_\theta }\,\sin\theta}{2\, {\sqrt{\imag \,e^{\imag
\,\phi }\,\tan(\frac{\theta }{2})}}}}}
\end{array}
\]

the Hamiltonian $S_4$ of reference \cite{KaKrPogo01} in spherical
coordinates $\theta, \,\phi$ is obtained:
\[
H=p^2_\theta\,+\,\frac{p^2_\phi}{\sin^2\theta}\,+\,(m -
k)\,-\,\imag\,(k + m)\, \,\cot\theta
\,-\,64\,\ell\,\frac{e^{2\,\imag \,\phi }}{\sin^2\theta}\,-\,
4\,n\,\frac{e^{\imag \,\phi }}{\sin^2\theta}
\]

\item[\fbox{$C_4$: Class I${}_3$ $\kappa=0,\, \lambda=0, \mu=4/K,
\nu=0$}] The form of the Hamiltonian in Liouville coordinates is
given by
\[
\begin{array}{rl}
H=&\frac{{\left( -1 + e^{2\,\left( -\eta  + \xi  \right) } \right)
}^2\,K\,{p_{\eta} }\,{p_{\xi} }}{4\, e^{2\,\left( -\eta + \xi
\right) }} +\\
+& \frac{{\left( -1 + e^{2\,\left( -\eta  + \xi \right) } \right)
}^2\,K\,\left( \frac{k\, e^{2\,\left( \eta  + \xi \right) }\,
+\,\ell\, {e^{\eta  + \xi }\,\left( 1 + e^{2\,\left( \eta  + \xi
\right) } \right) \,\ell}} {{\left( -1 + e^{2\,\left( \eta  + \xi
\right) } \right)}^2} + \frac{m\,e^{2\,\left( -\eta  + \xi \right)
}+ n\,{e^{-\eta  + \xi }\,\left( 1 + e^{2\,\left( -\eta + \xi
\right) } \right) }}{{\left( -1 + e^{2\,\left( -\eta + \xi \right)
} \right)}^2} \right) }{4\,e^{2\,\left( -\eta + \xi \right) }}
\end{array}
\]

For $K=1$ and using the coordinate transformation

\[
\begin{array}{ll}
{{\xi } = \frac{1}{2}\,{\ln (-\imag \,e^{\imag \,\phi }\,\cot
(\frac{\theta }{2}))}}\,, &{{{p_\xi }} = {-\imag
\,{p_\phi } - {p_\theta }\,\sin\theta}}\\
{{\eta } = \frac{1}{2}\, {\ln (\imag \,e^{\imag \,\phi }\,\tan
(\frac{\theta }{2}))}}\,, & {{{p_\eta }} = {-\imag \,{p_\phi } +
{p_\theta }\,\sin \theta}}
\end{array}
\]

the above Hamiltonian in spherical coordinates $\theta,\, \phi$
has the form

\[
H=p^2_\theta\,+\,\frac{p^2_\phi}{\sin^2\theta}\,+\,\frac{k}{4\,{\sin^2
\theta }\,{\sin^2 \phi }} + \frac{\ell\, cos \phi}{2\,{\sin^2
\theta}\,\sin^2 \phi } + \frac{m}{4} - \frac{n\,\imag }{2}\,\cot
\theta
\]

This is the  Hamiltonian $U^{c}$ of Ref. \cite{RaSan99}  in
spherical coordinates. This potential is missing in the
classification scheme of Ref. \cite{KaKrPogo01}.

 \item[\fbox{$C_5$: Class I${}_3$ $\kappa=0,\,
\lambda=0, \mu=2/K, \nu=1/K$}] The form of the Hamiltonian in
Liouville coordinates is given by
\[
\begin{array}{ll}
H=&K\,\frac{{p_\eta }\,{p_\xi }} {\frac{2\,e^{2\,\left(\xi - \eta
\right) }} {{\left( -1 + e^{2\,\left(\xi - \eta \right) } \right)
}^2}\, +\,\frac{e^{\xi - \eta }\,\left( 1 + e^{2\,\left(\xi - \eta
\right) } \right) }{{\left( -1 + e^{2\,\left( \xi - \eta \right) } \right) }^2}}\, +\\
&+\, K\, \frac{\frac{e^{2\,\left( \eta + \xi \right) }\,k}
{{\left( -1 + e^{2\,\left( \eta  + \xi \right) } \right) }^2} +
\frac{e^{\eta  + \xi }\, \left( 1 + e^{2\,\left( \eta  + \xi
\right) } \right) \,\ell}{{\left( -1 + e^{2\,\left( \eta + \xi
\right) } \right) }^2} + \frac{e^{2\,\left( \xi - \eta \right)
}\,m} {{\left( -1 + e^{2\,\left( \xi - \eta \right) } \right) }^2}
+ \frac{e^{-\eta  + \xi }\, \left( 1 + e^{2\,\left( \xi - \eta
\right) } \right) \,n} {{\left( -1 + e^ {2\,\left( \xi - \eta
\right) } \right) }^2} }{\frac{2\,e^{2\,\left(\xi - \eta \right)
}} {{\left( -1 + e^{2\,\left(\xi - \eta \right) } \right) }^2}\,
+\,\frac{e^{\xi - \eta }\,\left( 1 + e^{2\,\left(\xi - \eta
\right) } \right) }{{\left( -1 + e^{2\,\left( \xi - \eta \right) }
\right) }^2}}
\end{array}
\]
For $K=1$ and using the coordinate transformation:
\[
\begin{array}{ll}
{{\xi } = {\ln (e^{\imag \,\phi }\,\tan (\frac{\theta }{2}))}},
&{{p_\xi }} = \frac{-\imag \,{p_\phi } + {p_\theta }\,\sin \theta }{2}\\
{{\eta } = {\ln ( - e^{\imag \,\phi }\,\cot (\frac{\theta }{2}) )
}}, & {{p_\eta }} = - \frac{\imag \,{p_\phi } + {p_\theta }\,\sin
\theta }{2}
\end{array}
\]

the Hamiltonian $S_9$ of reference \cite{KaKrPogo01} and the
Hamiltonian $U^{b}$ of ref. \cite{RaSan99} in spherical
coordinates $\theta, \,\phi$ are obtained:

\[
\begin{array}{ll}
H=p^2_\theta\,+\,\frac{p^2_\phi}{\sin^2\theta}\,+\\
+\, \frac{m + 2\,n}{4} \,+\, \left( k + 2\,\ell\,\cos (2\,\varphi
) \right) \,{\csc^2\theta}\,
   {\csc^2(2\,\phi )} - \frac{\left( m - 2\,n \right)} {4}\,{\sec^2\theta}
\end{array}
\]

while using the coordinate transformation:
\[
\begin{array}{ll}
{{\xi } = {\frac{-\imag }{4}\,\pi  + \ln (\frac{1 + \imag \,e^{\imag
\,\phi }\,\tan (\frac{\theta }{2})}{1 - \imag \,e^{\imag \,\phi
}\,\tan (\frac{\theta }{2})})}}, \,\,{{\eta } = {\frac{-\imag
}{4}\,\pi  + \ln (\frac{1 - \imag \,e^{\imag \,\phi }\,\cot
(\frac{\theta }{2})}{1 + \imag \,e^{\imag \,\phi }\,\cot
(\frac{\theta }{2})})}},
\end{array}
\]
\[
\begin{array}{ll}
 {{{p_\xi }} = {\frac{-\left( {p_\phi }\,
\left( \cot (\frac{\theta }{2}) + e^{2\,\imag \,\phi }\,\tan
(\frac{\theta }{2}) \right) \right) }{4\, e^{\imag \,\varphi }} -
\frac{\frac{\imag }{2}\,{p_\theta }\, {\cos (\frac{\theta
}{2})}^2\,\left( 1 + e^{2\,\imag \,\phi }\,{\tan (\frac{\theta
}{2})}^2 \right) }{e^{\imag \,\phi }}}},
\end{array}
\]
\[
\begin{array}{ll}
 {{{p_\eta }} = {\frac{\frac{-\imag }{4}\,{p_\theta }\,
 \left( 1 + e^{2\,\imag \,\phi } + \left( -1 + e^{2\,\imag \,\phi }
 \right) \,\cos (\theta ) \right) }{e^{\imag \,\phi }} + \frac{{p\phi }\,
 \left( e^{2\,\imag \,\phi }\,\cot (\frac{\theta }{2}) + \tan (\frac{\theta }{2}) \right) }
 {4\,e^{\imag \,\phi }}}}\\
 \end{array}
\]

we have the Hamiltonian $U^{a}$ of ref. \cite{RaSan99} in
spherical coordinates.

\[
\begin{array}{ll}
H=p^2_\theta\,+\,\frac{p^2_\phi}{\sin^2\theta}\,+\\
+\,\frac{m + 2\,n}{4} + \frac{\left(2\,n -m \right)
\,\csc^2\theta\,\csc^2\phi}{4}\, +\, \frac{k - 2\,l} {2\,{\left(
\cos\theta - \cos\phi\,\sin\theta \right) }^2}\, +\,   \frac{k +
2\,l}{2\,{\left( \cos\theta + \cos\phi\,\sin\theta \right) }^2}
\end{array}
\]

 \item[\fbox{ $C_6$: Class I${}_3$ $\kappa=-2/K,\,
\lambda=-1/K,\, \mu=2/K,\, \nu=-1/K$}] The form of the Hamiltonian
in Liouville coordinates is given by
\[
\begin{array}{rl}
H= &K\,\frac{{p_{\eta} }\,{p_{\xi} }} {\frac{2\,e^{2\,\left( \xi -
\eta \right) }} {{\left( -1 + e^{2\,\left( \xi - \eta  \right) }
\right) }^2} - \frac{e^{\xi - \eta }\, \left( 1 + e^{2\,\left( \xi
- \eta \right) } \right) }{{\left( -1 + e^{2\,\left( \xi - \eta
\right) } \right) }^2} - \frac{2\,e^{2\,\left( \eta  + \xi \right)
}} {{\left( -1 + e^{2\,\left( \eta  + \xi  \right) } \right) }^2}
- \frac{e^{\eta + \xi }\,\left( 1 + e^{2\,\left( \eta + \xi
\right) } \right) }{{\left( -1 + e^{2\,\left( \eta  + \xi \right)
} \right) }^2}} + \\
+&K\,\frac{\frac{e^{2\,\left( \eta  + \xi \right) }\,k} {{\left(
-1 + e^{2\,\left( \eta  + \xi  \right) } \right) }^2} +
\frac{e^{\eta + \xi }\,\left( 1 + e^{2\,\left( \eta + \xi \right)
} \right) \,l}{{\left( -1 + e^{2\,\left( \eta  + \xi \right) }
\right) }^2} + \frac{e^{2\,\left( -\eta  + \xi  \right) }\,m}
{{\left( -1 + e^{2\,\left( \xi - \eta \right) } \right) }^2} +
\frac{e^{\xi - \eta }\, \left( 1 + e^{2\,\left( \xi - \eta \right)
} \right) \,n}{ {\left( -1 + e^{2\,\left( \xi - \eta \right) }
\right) }^2}} {\frac{2\,e^{2\,\left( \xi - \eta \right) }}
{{\left( -1 + e^{2\,\left( \xi - \eta \right) } \right) }^2} -
\frac{e^{\xi - \eta }\, \left( 1 + e^{2\,\left( \xi - \eta \right)
} \right) }{{\left( -1 + e^{2\,\left( \xi - \eta \right) } \right)
}^2} - \frac{2\,e^{2\,\left( \eta  + \xi  \right) }} {{\left( -1 +
e^{2\,\left( \eta  + \xi  \right) } \right) }^2} - \frac{e^{\eta +
\xi }\,\left( 1 + e^{2\,\left( \eta  + \xi  \right) } \right)
}{{\left( -1 + e^{2\,\left( \eta  + \xi  \right) } \right) }^2}}
\end{array}
\]
For $K=1$ and using the coordinate transformation:
\[
\begin{array}{ll}
{{\xi } = {{\mbox{arcsinh}}(-\imag e^{\imag \,\phi }\, \cot
(\frac{\theta }{2}))}}\,, &{{{p_\xi }} = {\frac{ {\sqrt{1 -
e^{2\,\imag \,\phi }\, {\cot (\frac{\theta }{2})}^2}}\, \left( \imag
\,{p_\phi } +{p_\theta }\,\sin (\theta ) \right)}{2\,\imag\,e^{\imag \,\phi }\,
\cot(\frac{\theta}{2})}}}\\
{{\eta } = {{\mbox{arcsinh}}(-\imag \,e^{\imag \,\phi }\,\tan
(\frac{\theta }{2}))}}\,, & {{{p_\eta }} = {\frac{ {\sqrt{1 -
e^{2\,\imag \,\phi }\, {\tan (\frac{\theta }{2})}^2}}\, \left( \imag
\,{p_\phi } -{p_\theta }\,\sin (\theta ) \right)}{2\,\imag\,e^{\imag
\,\phi }\,\tan (\frac{\theta}{2})}}}
\end{array}
\]
the Hamiltonian $S_7$ of reference \cite{KaKrPogo01} and the
Hamiltonian $U^{e}$ of ref. \cite{RaSan99} in spherical
coordinates $\theta, \,\phi$ are obtained:
\[
\begin{array}{rl}
H=&p^2_\theta\,+\,\frac{p^2_\phi}{\sin^2\theta}\,+\,\frac{-k -
2\,l + m - 2\,n}{8} + \frac{\left( k - 2\,l - m -
2\,n \right) \,{\sec (\theta )}^2}{8}\, +\\
&+\frac{\frac{\imag }{8}\,\left( k + 2\,l + m - 2\,n \right)
\,\cos (\phi )\,\sin (\theta )}{{\sqrt{{\cos (\theta )}^2 +
{\sin (\theta )}^2\,{\sin (\phi )}^2}}} + \frac{\left( k - 2\,l +
m + 2\,n \right) \,\sec (\theta )\, \sin (\phi )\,\tan (\theta
)}{8\, {\sqrt{{\cos (\theta )}^2 + {\sin (\theta )}^2\,{\sin (\phi
)}^2}}}
\end{array}
\]

\item[\fbox{ $C_7$: Class I${}_3$ $\kappa=-4/K,\, \lambda=0,\,
\mu=4/K,\, \nu=0$}] The form of the Hamiltonian in Liouville
coordinates is given by
\[
\begin{array}{rl}
H=&K\, \frac{{p_{\eta} }\,{p_{\xi} }} {\frac{4\,e^{2\,\left( \xi -
\eta \right) }}{{\left( -1 + e^{2\,\left( \xi - \eta \right) }
\right) }^2 } - \frac{4\, e^{2\,\left( \eta  + \xi \right)
}}{{\left( -1 + e^{2\,\left( \eta  + \xi  \right) } \right) }^2
}}\, + \\
+& K\, \frac{\frac{e^{2\,\left( \eta  + \xi  \right) }\,k}
{{\left( -1 + e^{2\,\left( \eta  + \xi  \right) } \right) }^2} +
\frac{e^{\eta + \xi }\, \left( 1 + e^{2\,\left( \eta  + \xi
\right) } \right) \,\ell}{{\left( -1 + e^{2\,\left( \eta  + \xi
\right) } \right) }^2} + \frac{e^{2\,\left( \xi - \eta \right)
}\,m} {{\left( -1 + e^{2\,\left( \xi - \eta \right) } \right) }^2}
+ \frac{e^{\xi-\eta }\, \left( 1 + e^{2\,\left( \xi-\eta \right) }
\right) \,n}{{\left( -1 + e^{2\,\left( \xi-\eta \right) } \right)
}^2}}{\frac{4\,e^{2\,\left( \xi - \eta \right) }}{{\left( -1 +
e^{2\,\left( \xi - \eta \right) } \right) }^2 } - \frac{4\,
e^{2\,\left( \eta  + \xi \right) }}{{\left( -1 + e^{2\,\left( \eta
+ \xi  \right) } \right) }^2 }}
\end{array}
\]
For $K=1$ and using the coordinate transformation
\[
{\xi } = \frac{1}{2}\,  \ln \left(\frac{\imag \,( -1 +
             {\sqrt{1 + {\left( \sigma  + \imag \,\tau  \right) }^2}} \,\,) }{\sigma  +
           \imag \,\tau }\right),\,
           {\eta } = \frac{1}{2}\, \ln \left(\frac{\imag \,( 1 +
             {\sqrt{1 + {\left( \sigma  - \imag \,\tau  \right) }^2}} \,\,) }{\sigma  -
           \imag \,\tau }\right)
\]
\[
p_\xi = \sqrt{1 + (\sigma + \imag \tau)^2}\,\,\left((\sigma +
\imag \tau)\,p_\sigma + (-\imag\,\sigma + \tau)\,p_\tau\right)
\]
\[
p_\eta =\imag\, \sqrt{1 + (\sigma - \imag
\tau)^2}\,\,\left((\imag\,\sigma + \tau)\,p_\sigma - (\sigma -
\imag\,\tau)\,p_\tau\right)
\]
the Hamiltonian $S_8$ of reference \cite{KaKrPogo01} in
horospherical coordinates $\sigma, \,\tau$ is obtained.
\[
\begin{array}{rl}
H=&-  \left( {{p\sigma }}^2 + {{p\tau }}^2 \right) \,{\sigma }^2
-\frac{\left( k + m \right) \, \left( -1 + {\sigma }^2 + {\tau }^2
\right) }{8\, {\sqrt{{\sigma }^4 + {\left( -1 + {\tau }^2 \right)
}^2 + 2\, {\sigma }^2\, \left( 1 + {\tau }^2 \right) }}}\,+\\
&+\,\frac{\left(n -\ell \right) \,\left( {\sigma }^2 - \tau  +
{\tau }^2 \right) } {4\,{\sqrt{\left( {\sigma }^2 + {\tau }^2
\right)\,\left( {\sigma }^2 + {\left( -1 + \tau \right) }^2
\right) }}}\,-\frac{\left(\ell + n \right) \, \left( {\sigma }^2 +
\tau + {\tau }^2 \right) }{4\, {\sqrt{\left( {\sigma }^2 + {\tau
}^2 \right) \, \left( {\sigma }^2 + {\left( 1 + \tau \right) }^2
\right) }}}\,+\,\frac{m - k}{8}
\end{array}
\]
\end{description}
We use the inverse transformation

\[
\begin{array}{rl}
{{\sigma }\rightarrow {\frac{\imag \,e^{2\,\eta }}{1 + e^{4\,\eta
}} - \frac{\imag \,e^{2\,\xi }}{1 + e^{4\,\xi }}}} , \, {{\tau
}\rightarrow {- \frac{e^{2\,\eta }}{1 + e^{4\,\eta }} -
\frac{e^{2\,\xi }}{1 + e^{4\,\xi }}}}
\end{array}
\]
for the verification of Poisson brackets. The above
transformations are more appropriate for the corresponding
calculations

\section{Superintegrable systems with a linear and a quadratic
integral}\label{sec:LinearQuadratic}

In the case of a linear integral of motion and a quadratic
integral of motion, there is a Liouville coordinate system where
the Hamiltonian and the linear integral of motion are written as:
\[
I=\left(p_\xi+ p_\eta \right)^2 \quad  \mbox{or} \quad A=\left(
p_\xi- p_\eta\right)^2
\]
and
\[
H=\frac{p_\xi p_\eta}{G(\xi-\eta)}+\frac{g(\xi-\eta)}{G(\xi-\eta)}
\quad  \mbox{or} \quad H=\frac{p_\xi
p_\eta}{F(\xi+\eta)}+\frac{f(\xi+\eta)}{F(\xi+\eta)}
\]
From the forms of the integral of motion, which are given in
Section \ref{sec:Classification} we can find all the possible
subclasses corresponding to a linear and quadratic integral of
motion in Liouville coordinates. In all the above cases we remark
that the potential depends on two parameters among the $k,\,
\ell,\, m$ and $n$. In Table \ref{tab:Linear} we give the possible
cases of superintegrable systems with a linear and a quadratic
integral of motion:

\begin{table}[p]
\[
\begin{array}{|c|c|c|c|c|c|c|c|c|c|c|}
\hline &\mbox{Class} & \kappa & \lambda & \mu & \nu & k & \ell & m
& n \\
\hline\hline  GL_1    & I_1 &0     &0       &\cdot   &\cdot &0     &0    &\cdot   &\cdot \\
\hline        GL_2    &     &\cdot &0       &0       &\cdot &\cdot &0    &0       &\cdot \\
\hline \hline (\simeq GL_1)& I_2 &0  &0     &\cdot   &\cdot &0     &0    &\cdot        &\cdot    \\
\hline        GL_3&     &0         &\cdot   &0       &\cdot &0     &\cdot&0     &\cdot    \\
\hline        GL_4&     &\cdot     &0       &\cdot   &0   &\cdot   &0    &\cdot     &0    \\
\hline \hline GL_5& I_3 &0  &0     &\cdot   &\cdot   &0     &0     &\cdot &\cdot    \\
\hline       (\simeq GL_1)& &\kappa=\mp 2\lambda&\cdot   &\mu=\pm 2\nu &k=\mp 2\ell&\cdot&0 &m=\pm 2 n        &\cdot    \\
\hline\hline  GL_6&II_1 &0  &\lambda=\pm \mu    &\cdot     &\cdot &0     &\ell=\pm m     &\cdot &\cdot     \\
\hline\hline  GL_7&II_2 &0  &\cdot     &0     &\cdot &0     &\cdot     &0 &\cdot   \\
\hline\hline  (\simeq GL_3)&II_3 &0  &\cdot     &0     &\cdot &0     &\cdot     &0 &\cdot   \\
\hline
\end{array}
\]
\caption{\label{tab:Linear} \sf General superintegrable integrable
systems with a linear and a quadratic integral of motion}
\end{table}

In Table \ref{tab:Revolution}, the possible superintegrable
systems which are defined on a surface of revolution with a linear
and a quadratic integral of motion are listed.

\begin{table}[p]
\[
\begin{array}{|c|c|c|c|c|c|c|c|c|c|c|c|}
\hline &\mbox{Class} &\kappa &\lambda &\mu &\nu &k &l &m &n &
\mbox{\begin{tabular}{c}Potentials\\
by revolution\\
from ref \cite{KaKrMiWin03}\end{tabular}}& \mbox{\begin{tabular}{c}Potentials\\
by revolution\\
from ref \cite{KaKrWin02}\end{tabular}}\\
\hline\hline RL_1& I_1  &0     &0     &\cdot &\cdot  &0     &0       &\cdot   &\cdot & 2.2[D] & \\
\hline       RL_2&      &\cdot &0     &0     &\cdot  &\cdot &0       &0       &\cdot &        & (3) \\
\hline\hline RL_3& I_2  &\cdot &0     &\cdot &0      &\cdot &0       &\cdot   &0     & 4.2[D] & \\
\hline       RL_4& II_1 &0     &\cdot &0     &\cdot  &0     &\cdot   &0       &\cdot & 3.2[E] & \\
\hline
\end{array}
\]
\caption{\label{tab:Revolution}\sf Potentials by revolution with a
linear and a quadratic integral of motion}
\end{table}

In Table \ref{tab:Curvature0L}, the possible superintegrable
systems which are defined on a surface of zero curvature with a
linear and a quadratic integral of motion are listed.

\begin{table}[p]
\[
\begin{array}{|c|c|c|c|c|c|c|c|c|c|c|l|}
\hline &\mbox{Class} & \kappa & \lambda & \mu & \nu & k & \ell & m
& n & \mbox{\begin{tabular}{c}potentials\\
from ref \cite{KaKrPogo01}\end{tabular}}\\
\hline\hline FL_1& I_1  &0  &0     &0     &\cdot &0     &0     &\cdot &\cdot & E_6     \\
\hline        FL_2&     &0  &0     &0     &\cdot &\cdot &0     &0     &\cdot & E_5     \\
\hline\hline  FL_3&I_2  &0  &0     &0     &\cdot &0     &\cdot &0     &\cdot & E_3     \\
\hline         FL_4&    &0  &\cdot &0     &0     &0     &\cdot &0     &\cdot & E_{18}  \\
\hline\hline  FL_5&II_1 &0  &0     &0     &\cdot &0     &0     &\cdot &\cdot & E_4     \\
\hline\hline  FL_6&II_2 &0  &0     &0     &\cdot &0     &0     &\cdot &\cdot & E_{13}  \\
\hline\hline  FL_7&II_3 &0  &0     &0     &\cdot &0     &0     &\cdot &\cdot & E_{14}  \\
\hline          FL_8&   &0  &0     &\cdot &\cdot &0     &0     &\cdot &\cdot & E_{12}  \\
\hline
\end{array}
\]
\caption{\label{tab:Curvature0L}\sf Potentials with curvature zero
and with a linear and a quadratic integral of motion}
\end{table}

In Table \ref{tab:CurvatureConstantL}, the possible
superintegrable systems which are defined on a surface of constant
curvature with a linear and a quadratic integral of motion are
listed.

\begin{table}[p]
\[
\begin{array}{|c|c|c|c|c|c|c|c|c|l|}
\hline \mbox{Class} &    \kappa  &  \lambda &  \mu  & \nu  & k &l
&m &n &
\mbox{\begin{tabular}{c}Sphere\\
potentials\\
(K=1)\\
from ref \cite{KaKrPogo01}\end{tabular}}\\
\hline\hline I_1  &0     &0     &1/K  &0     &0     &0 &\cdot &\cdot & S_5 \\
\hline\hline I_2  &-1/K  &0     &0    &0     &\cdot &0 &\cdot &0     & S_3\\
\hline            &-1/K  &0     &1/K  &0     &\cdot &0 &\cdot &0     & S_6 \\
\hline
\end{array}
\]
\caption{\label{tab:CurvatureConstantL} \sf Potentials with constant
curvature and with a linear and a quadratic integral of motion}
\end{table}

\section{Discussion}\label{sec:Discussion}

The findings of this paper are summarized as follows:
\begin{enumerate}
\item  The  superintegrable systems with quadratic integrals of
motion can be classified in  six subclasses. Each subclass depends
on seven  parameters. Four among these parameters ($\kappa,\,
\lambda,\, \mu$ and $\nu$) determine the metric of the manifold,
on which the system is determined. These parameters are
characteristic of the system's "kinetic" energy. The remaining
three parameters  define the  potential (The potential depend on four parameters $k,\; \ell,\; m$ and $n$ but only three among them are independent).
For each subclass, the analytic explicit forms of the manifold
metric, the potential and the integrals of motion are calculated.
Also  the constants of the quadratic Poisson algebra of integrals
of motion are given as functions of the energy and the eight
parameters are given.
\item All the known two dimensional
superintegrable systems are systems defined on surfaces of
constant curvature or on surfaces of revolution. All these systems
are classified in these six classes. Each class is characterized
by the values four parameters $\kappa,\, \lambda,\, \mu$ and
$\nu$, which are determined by the form of the assumed manifold.
If we fix the manifold, let us suppose that the manifold is a
manifold with negative constant curvature, the possible values of
the of the parameters $\kappa,\, \lambda,\, \mu$ and $\nu$ are
calculated for each subclass. Therefore we can "guess" the
existence of the permitted superintegrable systems and to classify
these systems in tables. Using this technique we can classify all
the possible known superintegrable systems and to investigate the
possible missing potentials. With this method a new
superintegrable system was found for the class I superintegrable
systems by revolution, given in ref \cite{KaMiPogo00}.

\item Generally for any values of the parameters $\kappa,\,
\lambda,\, \mu$ and $\nu$ the associated manifolds are neither
surfaces of constant curvature nor surfaces of revolution.
Therefore we have investigated superintegrable systems, which are
not yet known. We believe that all the possible superintegrable
systems with two quadratic integrals of motion are investigated.

\item The two dimensional 'non degenerate' \cite{KaMiPogo00}
superintegrable systems are classified by the values of the
constants of the Darboux equations \cite{RaSan99,KaMiPogo00} and
the constants of the system.
 The relation of these constants with the
constants of the quadratic Poisson algebra is explained.

\item The six classes of superintegrable systems are the equivalence classes of
St\"{a}ckel equivalent systems.

\end{enumerate}

From the above discussion several open problems arise:
\begin{itemize}
\item
The superintegrable systems for the case of
cubic integrals of motion are under investigation
\cite{Ranada97,Tsyg00_TMP,Tsyg00_JPA,KaRo00,GraWin02,Gra03,Gra04}.
The general structure of these systems is recently investigated \cite{KaKrMi05a}
but
 the general form and their classification is not yet known for
  manifolds which carry integrable systems with one third order
  integral of motion.
\item The quantum counterparts of the general six subclasses of
superintegrable systems with quadratic integrals of motion are not
yet known. In Section \ref{sec:Integrable}, the separation of
variables of these systems has been explicitly written. The form
of the separation of  Schroedinger equation can be written in a
Liouville coordinate system. The solutions of the quantum
Schroedinger equation can be calculated. This work is under
current investigation. The form of the Poisson algebra can be
generalized in a quadratic associative algebra, whose  energy
eigenvalues are generally calculated by using deformed oscillator
techniques \cite{BoDasKo93,BoDasKo94,Das00,Das01,Das02}. From the
form of the Poisson algebra, one can be guess that the energy
eigenvalues of these quantum  systems are roots of cubic
polynomials.

\item
The general form of the three dimensional superintegrable systems with
quadratic integrals of motion is not yet known. The
Poisson algebras and the associated quantum counterparts for
the three dimensional superintegrable systems are not yet fully
studied. Recently \cite{KaWiMiPo99,KrKa02} the quantum quadratic algebras
have been written down, which are not generally closed as polynomial algebras.
A systematic calculation of energy
eigenvalues with algebraic methods has  not been  performed yet.
\end{itemize}

\newpage
\appendix
\centerline{\Large \bf Appendix}

\section{Polynomial combinations of integrals}\label{sec:Poly}

Let consider a superintegrable system with two quadratic integrals of motion.
The general forms of the  Hamiltonian and the integrals of motion
in Liouville coordinates are:
\begin{equation}\label{eq:HABpoly}
\begin{array}{l}
H = \frac{p_\xi\,p_\eta}{g(\xi,\,\eta)}+ V(\xi,\eta)\\
A=p^2_\xi+k p^2_\eta -2 c(\xi,\eta) \frac{p_\xi\,p_\eta}{g(\xi,\,\eta)}
+q(\xi,\eta),\qquad k=0\, \mbox{or}  \,1\\
B=a^2(\xi)p^2_\xi+ b^2 (\eta) p^2_\eta - 2 \beta(\xi,\eta) \frac{p_\xi\,p_\eta}{g(\xi,\,\eta)}
+Q(\xi,\eta)
\end{array}
\end{equation}
In this Appendix we consider that the system has quadratic integrals of motion. We assume
that the systems has not any linear integral of motion. That assumption excludes the system $H=p_x^2+p_y^2$, because it possesses two linear integrals of motion.
In this Appendix, we shall prove the following proposition:
\begin{Proposition}\label{prop:Poly}
Let $M$ be an integral of motion, which is a polynomial function
of the momenta of even order. We assume that this integral
contains only  monomials of momenta of even order, i.e.
\begin{equation}\label{eq:PolyInt}
M= \sum\limits_{k +\ell=\mbox{even}}^{2 n}\,
 \alpha_{k,\ell}(\xi, \eta)\; p^{k}_\xi\, p^{\ell}_\eta
\end{equation}
Then $M$ is a polynomial of order $n$ of the integrals $H,\,A,\,B$.
\end{Proposition}
The existence of three integrals of motion implies that the integral $M$ is functionally dependent integral,
i.e. there is some smooth function $\Phi$ (non generally a polynomial one) such that:
\[
\Phi(M,A,B,H)=0  \quad \mbox{or} \quad M= f(A,B,H)
\]
but the function $f(A,B,H)$ is not evident that it is a polynomial one. From (\ref{eq:HABpoly}) we can
see that
\[
\begin{array}{lcl}
p^2_\xi&=&\frac{
\begin{array}{|cc|}
A+2  c H  -q  -  2  c  V & k\\
B+2 \beta H -Q-2  \beta V &  b^{2}(\xi,\eta)
\end{array}}{
\begin{array}{|cc|}
1  &k\\
a^{2}(\xi,\eta) &  b^{2}(\xi,\eta)
\end{array}}\\
p^2_\eta&=&
\frac{
\begin{array}{|cc|}
1   &  A+2  c H  -q  -  2  c  V \\
 a^{2}(\xi,\eta)  &B+2 \beta H -Q-2  \beta V
\end{array}}{
\begin{array}{|cc|}
1  &k\\
a^{2}(\xi,\eta) &  b^{2}(\xi,\eta)
\end{array}}
\\
p_\xi\, p_\eta &=& g\, H - V
\end{array}
\]
The above equations imply that the momentum monomials $p^{k}_\xi\, p^{\ell}_\eta$
 inside the sum sign in (\ref{eq:PolyInt}) can be written as polynomials of the integrals
 $A,B,H$ with coefficients, which depend on $\xi$ and $\eta$. Therefore the integral of motion $M$
 is written:
 \[
M= \sum\limits_{ 0\le i+j+k\le n} c_{i j k}(\xi,\eta)  A^i B^j H^k
 \]
Generally the coefficients $c_{i j k}$ should be constants not
depending on the variables $\xi,\eta$. If these functions are non
constant functions $c_{i j k}(\xi,\eta)$ then we choose a fixed
value of these parameters  $\xi_0,\eta_0$. In general there is an
infinity of trajectories passing through these position values
$\xi_0,\eta_0$. Each trajectory is characterized by a special
value of the integrals $H,A,B$,  therefore $M$ is a polynomial of
fixed constants for an infinity of trajectories. For another pair
of parameters $\xi_1,\eta_1$ we have another choice of the
coefficients in (\ref{eq:PolyInt}) therefore there is a relation
of the form:
\[
\sum\limits_{0 \le i+j+k\le n} \left( c_{i j k}(\xi_1,\eta_1)- c_{i j k}(\xi_0,\eta_0)\right)  A^i B^j H^k=0
\]
The above relations means that the integrals of motion $H,A,B$ are not functionally independent functions,
that is a contradiction to assumption initial regarding the independence of the integrals. Therefore
we have proved that the polynomial expansion (\ref{eq:PolyInt}) of the integral $M$ is indeed unique,
when $M$ is a even polynomial of the momenta.

A direct application of Proposition \ref{prop:Poly} is the following Proposition:
\begin{Proposition}\label{prop:PolyProp}
Let a superintegrable two dimensional system have even qua\-dratic
integrals of motion $H,A,B$. If we put $C=\left\{A,B\right\}$. The
integrals $\left\{A,C\right\}$ and $\left\{B,C\right\}$ can
written as quadratic polynomials of the integrals. The integral
$C^2$ is a cubic polynomial of the integrals $H,A,B$.
\end{Proposition}

The above proposition was taken as a conjecture in the previous work \cite{BoDasKo93}--\cite{Das02}
and \cite{Higgs79}--\cite{Zhedanov92}. Here we prove that this assumption is indeed true. A generalization
of Proposition \ref{prop:PolyProp} is indeed true for superintegrable two dimensional systems with
an integral, which is an odd cubic polynomial in momenta \cite{Tsyg00_JPA,Tsyg00_TMP}.
This Proposition means that the superintegrable two dimensional systems with even quadratic integrals correspond
to a quadratic Poisson algebra, which is characteristic for the superintegrable system.
The situation in three dimensional superintegrable systems is not clear \cite{KaWiMiPo99,KrKa02}

\bigskip

\textbf{Acknowledgements}:  One of the authors (C. D.) would thank W. Miller and J.Kress for their valuable comments and for indicating us the substantial references [14,15,16], which treat the problem of the classification from the point of view of equivalence classes of St\"{a}ckel transform.

\end{document}